\documentclass[fleqn,twoside]{article}
\usepackage{epsfig}
\usepackage{xspace}
\usepackage{longtable}
\usepackage{graphicx}
\usepackage[figuresright]{rotating}

\newcommand{\MeV}{\ensuremath{\mbox{MeV}}\xspace}
\newcommand{\GeVc}{\ensuremath{\mbox{GeV}/c}\xspace}
\newcommand{\MeVc}{\ensuremath{\mbox{MeV}/c}\xspace}
\newcommand{\T}{\ensuremath{\mbox{T}}\xspace}

\newcommand{\mm}{\ensuremath{\mbox{mm}}\xspace}

\newcommand{\mrad}{\ensuremath{\mbox{mrad}}\xspace}
\newcommand{\rad}{\ensuremath{\mbox{rad}}\xspace}

\newcommand{\dedx}{\ensuremath{\mbox{d}E/\mbox{d}x}\xspace}
\newcommand{\pip}{\ensuremath{\pi^+}\xspace}
\newcommand{\pim}{\ensuremath{\pi^-}\xspace}
\newcommand{\piz}{\ensuremath{\pi^0}\xspace}
\newcommand{\bfpip}{\ensuremath{\mathbf {\pi^+}}\xspace}
\newcommand{\bfpim}{\ensuremath{\mathbf {\pi^-}}\xspace}

\newcommand{\evtspill}{\ensuremath{N_{\mathrm{evt}}}\xspace}

\def\be{\begin{equation}}
\def\ee{\end{equation}}
\def\bea{\begin{eqnarray}}
\def\eea{\end{eqnarray}}

\newcommand{\bfGeVc}{\ensuremath{\mathbf {\mbox{\bf GeV}/c}}\xspace}
\begin{document}
\title{\bf Large-angle production of  charged pions by 3~\bfGeVc--12.9~\bfGeVc protons on  
           beryllium, aluminium and lead  targets}

\author{HARP Collaboration}

\maketitle

\begin{abstract}
  Measurements of the  double-differential $\pi^{\pm}$ production
  cross-section    in the range of momentum $100~\MeVc \leq p < 800~\MeVc$ 
  and angle $0.35~\rad \leq \theta  < 2.15~\rad$
  in proton--beryllium, proton--aluminium and proton--lead collisions are presented. 
  The data were taken  with the HARP detector in the T9 beam
  line of the CERN PS.
  The pions were produced by proton beams in a momentum range from
  3~\GeVc to  12.9~\GeVc hitting a target with a thickness of
  5\% of a nuclear interaction length.  
  The tracking and identification of the
  produced particles was performed using a small-radius
  cylindrical time projection chamber (TPC) placed inside a solenoidal
  magnet. 
  Incident particles were identified by an elaborate system of beam
  detectors.
  Results are obtained for the double-differential cross-sections 
  $
  %{{d^2 \sigma^{\pi}}}/{{dpd\Omega }}
  %%{{d^2 \sigma}}/{{dpd\Omega }}
  {{\mathrm{d}^2 \sigma}}/{{\mathrm{d}p\mathrm{d}\theta }}
  $
  at six incident proton beam momenta (3~\GeVc, 5~\GeVc, 8~\GeVc, 
  8.9~\GeVc (Be only), 12~\GeVc and 12.9 \GeVc (Al only)) and 
  compared to previously available data.
\end{abstract}

\begin{center}
(To be submitted to The European Physical Journal C)
\end{center}
\clearpage
\thispagestyle{plain}
\begin{center}
{\large HARP collaboration}\\
\newcommand{\afdoct}{{3}\xspace}
\vspace{0.1cm}
{\small
%\begin{center}
M.G.~Catanesi, 
%M.T.~Muciaccia, 
E.~Radicioni%,
%S.~Simone
\\ 
{\bf Universit\`{a} degli Studi e Sezione INFN, Bari, Italy} %done
\\
R.~Edgecock, 
M.~Ellis$^{1}$,          %, OK
%%%DROP: S.~Robbins$^{2,3}$,      %, OK
F.J.P.~Soler$^{2}$
\\
{\bf Rutherford Appleton Laboratory, Chilton, Didcot, UK} %done
\\
C.~G\"{o}\ss ling %,
%M.~Mass  % one paper (same as roma iii student)
\\
{\bf Institut f\"{u}r Physik, Universit\"{a}t Dortmund, Germany} %done
\\
%I.~Boyko,          % not member of HARP
S.~Bunyatov, 
%G.~Chelkov,        % pending explicit request
%%A.~Chukanov,      % removed by Boris (only shifts)
%%D.~Dedovitch,     % pending explicit request
%A.~Elagin,         % not member of HARP
%%M.~Gostkin,       % pending explicit request 
%%A.~Guskov,        % pending explicit request, 
%D.~Khartchenko,    % removed by Boris (only shifts)
%%O.~Klimov,        % removed by Boris (only shifts) 
%S.~Kotov,          % (only shifts) 
%%I.~Krasin,        % techn.paper only (technician)
A.~Krasnoperov, 
%Z.~Kroumchtein,    % pending explicit request
%%D.~Kustov,        % removed by Boris (only shifts) 
%D.~Naumov,         % late request
%Y.~Nefedov,        % pending explicit request
%K.~Nikolaev,       % not member of HARP
B.~Popov$^3$, 
%I.~Potrap,,        % (only shifts) 
V.~Serdiouk,        % TPC only - propose to keep on this paper
V.~Tereschenko %, 
%A.~Zhemchugov      % pending explicit request
\\
{\bf Joint Institute for Nuclear Research, JINR Dubna, Russia} %later
\\
E.~Di~Capua, 
G.~Vidal--Sitjes$^{4}$  % now at ..
%tech: V. Carassiti
%tech: F. Evangelisti
\\
{\bf Universit\`{a} degli Studi e Sezione INFN, Ferrara, Italy}  %done
\\
A.~Artamonov$^5$,   % and ITEP, Moscow
%%%DROP: P.~Arce$^8$,        % and univ. 
%F.~Dydak, 
S.~Giani, 
S.~Gilardoni,       %$^3$, %supported by DOCT 
P.~Gorbunov$^{5}$,  %,                new footnote
A.~Grant,  
A.~Grossheim$^{7}$, %$^1$, %supported by DOCT 
%%%DROP: P.~Gruber$^{11}$,    %supported by DOCT 
V.~Ivanchenko$^{8}$,  %,                new footnote 
%J.-C.~Legrand,     % only technical paper, or first TPC paper?
A.~Kayis-Topaksu$^{9}$,
%L.~Linssen,  %% asked to be removed
J.~Panman, 
I.~Papadopoulos,  
%%%DROP: J.~Pasternak, %$^{1,4}$,  %supported by DOCT 
E.~Tcherniaev, 
I.~Tsukerman$^5$,   % and ITEP, Moscow
R.~Veenhof, 
C.~Wiebusch$^{10}$,    % now at ..
%J.~Wotschack,
P.~Zucchelli$^{6,11}$ %on leave of absence from INFN-Ferrara
\\
{\bf CERN, Geneva, Switzerland} 
\\
A.~Blondel, 
S.~Borghi$^{12}$,  % new footnote or CERN
%%%DROP: M.~Campanelli,       % ???????????????
M.C.~Morone$^{13}$, 
G.~Prior$^{14}$,   %supported by DOCT 
R.~Schroeter
\\
{\bf Section de Physique, Universit\'{e} de Gen\`{e}ve, Switzerland} %done
\\
%%%DROP: R.~Engel,
C.~Meurer
\\
{\bf Institut f\"{u}r Physik, Forschungszentrum Karlsruhe, Germany}
\\
\newcommand{\afkyot}{{19}\xspace}
%%%DROP: I.~Kato$^{10,\afkyot}$ %,                new footnote
%T.~Nakaya$^{\afkyot}$,
%K.~Nishikawa$^{\afkyot}$,
%S.~Ueda$^{\afkyot}$
%%%DROP: \\
%%%DROP: {\bf University of Kyoto, Japan} % 
%%%DROP: \\
%V.~Ableev, % one or few papers
%C.~Cavion, 
U.~Gastaldi%, 
%M.~Placentino
\\
{\bf Laboratori Nazionali di Legnaro dell' INFN, Legnaro, Italy} %done
\\
\newcommand{\aflanl}{{15}\xspace}
G.~B.~Mills$^{\aflanl}$  
\\
{\bf Los Alamos National Laboratory, Los Alamos, USA} % 
\\
J.S.~Graulich$^{16}$, 
G.~Gr\'{e}goire 
\\
{\bf Institut de Physique Nucl\'{e}aire, UCL, Louvain-la-Neuve,
  Belgium} %ok
\\
M.~Bonesini,
%%M.~Calvi,         % one physics paper
%%%DROP: A.~De~Min,          % ??????
F.~Ferri           % ?
%%%DROP: M.~Paganoni,        % ?????? 
%%%DROP: F.~Paleari          % ?
% for technical paper:  Francesco Chignoli 
\\
{\bf Universit\`{a} degli Studi e Sezione INFN Milano Bicocca, Milano, Italy} %done
\\
%S.~Gninenko, 
M.~Kirsanov
%Yu.~Musienko, 
%A.~Poljarush, 
%A.~Toropin
%V.Postoev % for technical paper
\\
{\bf Institute for Nuclear Research, Moscow, Russia} %done (?)
\\
A. Bagulya, 
%V.~Chechin, 
% acknowledgements, tech: V.~Ermilova$^{\dagger}$, 
V.~Grichine,  % also supported Geneva
N.~Polukhina%, 
%N.~Starkov
\\
{\bf P. N. Lebedev Institute of Physics (FIAN), Russian Academy of
Sciences, Moscow, Russia} %done
\\
V.~Palladino
\\
{\bf Universit\`{a} ``Federico II'' e Sezione INFN, Napoli, Italy} % ok 
\\
\newcommand{\afclmb}{{15}\xspace}
L.~Coney$^{\afclmb}$, 
D.~Schmitz$^{\afclmb}$
\\
{\bf Columbia University, New York, USA} % 
\\
G.~Barr, 
A.~De~Santo$^{17}$%%%DROP: , % now at ..
%%%DROP: C.~Pattison, 
%%%DROP: K.~Zuber$^{23}$  % now at ..
\\
{\bf Nuclear and Astrophysics Laboratory, University of Oxford, UK} % ok
\\
%M.~Baldo~Ceolin,  preferes not to sign
% tech: G.~Barichello, 
F.~Bobisut, 
D.~Gibin,
A.~Guglielmi, 
%M.~Laveder, 
%A.~Menegolli, 
M.~Mezzetto
% tech: A. Pepato
%, M.~Vascon
\\
{\bf Universit\`{a} degli Studi e Sezione INFN, Padova, Italy} % done
\\
J.~Dumarchez%%%DROP: , 
%S.~Troquereau, % one paper
%%%DROP: F.~Vannucci 
\\
{\bf LPNHE, Universit\'{e}s de Paris VI et VII, Paris, France} % done
\\
%%V.~Ammosov,        %pending explicit request 
%V.~Gapienko,  %shifts?  reduce for next paper
%%V.~Koreshev,        %pending explicit request 
%%A.~Semak,   % ???,        %pending explicit request
%Yu.~Sviridov,  %shifts?  reduce for next paper
%tech: E.~Usenko, now at INR Troitsk, and LA paper (RPC)
%%V.~Zaets    % ???,        %pending explicit request
%%\\
%%{\bf Institute for High Energy Physics, Protvino, Russia} %done
%%\\
U.~Dore
\\
{\bf Universit\`{a} ``La Sapienza'' e Sezione INFN Roma I, Roma,
  Italy} % ok 
\\
D.~Orestano, 
%M.~Pasquali,  %student worked on calorimeter - one paper
F.~Pastore, 
A.~Tonazzo, 
L.~Tortora
% tech: Alfredo Iaciofano, Marco Lobello, Franco Marinilli
\\
{\bf Universit\`{a} degli Studi e Sezione INFN Roma III, Roma, Italy}
% done
\\
C.~Booth, 
%%%DROP: C.~Buttar$^{4}$,  %"now at the University of Glasgow".
%%%DROP: P.~Hodgson, 
L.~Howlett
%tech:  R. Nicholson
\\
{\bf Dept. of Physics, University of Sheffield, UK} %done
\\
M.~Bogomilov, 
% tech:  K.Burin
M.~Chizhov, 
D.~Kolev, 
% tech: P.~Petev, I.~Rusinov, 
R.~Tsenov
\\
{\bf Faculty of Physics, St. Kliment Ohridski University, Sofia,
  Bulgaria} %done
\\
%G.~Maneva, 
S.~Piperov, 
%S.~Stoykova, 
P.~Temnikov
\\
{\bf Institute for Nuclear Research and Nuclear Energy, 
Academy of Sciences, Sofia, Bulgaria} % done
\\
M.~Apollonio, 
P.~Chimenti,  % also supported Geneva
G.~Giannini%%%DROP: , 
%%%DROP: G.~Santin$^{24}$  % presently (now) at ESA / ESTEC, Noordwijk, The Netherlands , also supported Geneva, remove after one
\\
{\bf Universit\`{a} degli Studi e Sezione INFN, Trieste, Italy} % done
\\
%Y.~Hayato$^{\afkek}$, 
%A.~Ichikawa$^{\afkek}$, 
%T.~Kobayashi$^{\afkek}$
%\\
%{\bf KEK, Tsukuba, Japan} %
%\\
J.~Burguet--Castell, 
A.~Cervera--Villanueva, 
J.J.~G\'{o}mez--Cadenas, % also supported Geneva
J. Mart\'{i}n--Albo,
P.~Novella,
M.~Sorel%%%DROP: ,
%%%DROP: A.~Tornero
\\
{\bf  Instituto de F\'{i}sica Corpuscular, IFIC, CSIC and Universidad de Valencia,
Spain} % done
}
\end{center}
\thispagestyle{plain}
\vfill
%\newpage
%\vspace{1cm}
\rule{0.3\textwidth}{0.4mm}
\newline
\newpage
$^{~1}${Now at FNAL, Batavia, Illinois, USA.}
\newline
%%$^{~2}$Jointly appointed by Nuclear and Astrophysics Laboratory,
%%            University of Oxford, UK.
%%\newline
%%$^{~3}${Now at Codian Ltd., Langley, Slough, UK.}
%Now at Bergische Universit\"{a}t Wuppertal, Germany.}
%%\newline
$^{~2}${Now at University of Glasgow, UK.}
\newline
$^{~3}${Also supported by LPNHE, Paris, France.}
\newline
%$^{~6}${Supported by the CERN Doctoral Student Programme.}
%\newline
%
$^{~4}${Now at Imperial College, University of London, UK.}
\newline
$^{~5}${ITEP, Moscow, Russian Federation.}
%%\newline
%%$^{~8}${Permanently at Instituto de F\'{\i}sica de Cantabria,
%%            Univ. de Cantabria, Santander, Spain.} 
\newline
$^{~6}${Now at SpinX Technologies, Geneva, Switzerland.}
\newline
$^{~7}${Now at TRIUMF, Vancouver, Canada.}
%%\newline
%%$^{11}${Now at University of St. Gallen, Switzerland.}
\newline
$^{~8}${On leave of absence from Ecoanalitica, Moscow State University,
Moscow, Russia.}
%short: EMSU, 119899, Moscow, Russia 
%the Budker Institute for Nuclear Physics, Novosibirsk, Russia.
\newline
$^{~9}${Now at \c{C}ukurova University, Adana, Turkey.}
\newline
$^{10}${Now at III Phys. Inst. B, RWTH Aachen, Aachen, Germany.}
\newline
$^{11}$On leave of absence from INFN, Sezione di Ferrara, Italy.
\newline
$^{12}${Now at CERN, Geneva, Switzerland.}
\newline
$^{13}${Now at University of Rome Tor Vergata, Italy.}
\newline
$^{14}${Now at Lawrence Berkeley National Laboratory, Berkeley, California, USA.}
%%\newline
%%$^{19}${K2K Collaboration.}
\newline
$^{15}${MiniBooNE Collaboration.}
\newline
$^{16}${Now at Section de Physique, Universit\'{e} de Gen\`{e}ve, Switzerland, Switzerland.}
\newline
$^{17}${Now at Royal Holloway, University of London, UK.}
%%\newline
%%$^{23}${Now at University of Sussex, Brighton, UK.}
%\newline
%$^{18}${Now at ESA/ESTEC, Noordwijk, The Netherlands.}
%

\clearpage
%\tableofcontents
%\clearpage
%\listoffigures
%\clearpage
%\listoftables
%\clearpage

\section{Introduction}

 %+++ short intro HARP goals +++
The HARP experiment~\cite{harp-prop} 
makes use of a large-acceptance spectrometer for
 systematic study of the hadron
production on a large range of target nuclei for beam momenta from 1.5 to 15~\GeVc. 
The main motivations are to measure pion yields for a quantitative
design of the proton driver of a future neutrino factory~\cite{ref:nufact}, 
to provide measurements to allow substantially improved calculations of
the atmospheric neutrino
flux~\cite{Battistoni,Stanev,Gaisser,Engel,Honda} to be made
and to measure particle yields as input for the flux
calculation of accelerator neutrino experiments, 
such as K2K~\cite{ref:k2k,ref:k2kfinal},
MiniBooNE~\cite{ref:miniboone} and SciBooNE~\cite{ref:sciboone}. 

 %+++++++++++++++++++++++++++physics motivation intro++++++++++++++

Measurements of the double-differential cross-section, 
$
%{{d^2 \sigma^{\pi}}}/{{dpd\Omega }}
{{\mathrm{d}^2 \sigma^{\pi}}}/{{\mathrm{d}p\mathrm{d}\theta }}
$
%of positive and negative pion production for 
for $\pi^{\pm}$ production at large angles by
protons of 3~\GeVc, 5~\GeVc, 8~\GeVc, 8.9~\GeVc (Be only),  12~\GeVc 
and 12.9~\GeVc (Al only) momentum impinging
on a thin beryllium, aluminium or lead target of 5\% nuclear interaction length
($\lambda_{\mathrm{I}}$) are presented. 
These measurements are of special interest for target materials used in conventional
accelerator neutrino beams (Be, Al) and in neutrino factory designs (Pb).

In this energy range and for these nuclear targets,  only very sparse data 
sets are available 
from previous experiments, usually with  large uncertainties 
\cite{ref:piroue,ref:boyarinov}, aside what has been published 
in references \cite{ref:E910}, \cite{ref:shibata}.

Results for other nuclei, such as Be, Al 
for pions produced in the forward direction and  C, Cu, Sn, Ta for
pion production at large angles  are presented in 
different HARP publications \cite{ref:bePaper,ref:alPaper,
ref:harp:cacotin,ref:harp:tantalum}.
HARP is the first experiment to provide a large data set taken with
many different targets, full particle identification and large detector 
acceptance down to low secondary momentum ($\simeq 200$ MeV/c).
This paper completes the range of solid target materials for which HARP
data are available.
The combination of the data sets make it possible to perform systematic
comparisons of hadron production models with measurements at different
incoming beam momenta over a large range of target atomic number $A$.
%+++++++++++intro part of detector description++++++++++++++++++
%

Data were taken in the T9 beam of the CERN PS.
About $3.1 \times 10^{5}, 5.1 \times 10^{5}, 1.6 \times 10^{5}$ 
well-reconstructed secondary pion 
tracks for the beryllium, aluminium and lead targets were selected
from
1.6, 2.3 and 0.9 millions of
 incoming protons,
which gave an interaction trigger in the Large Angle spectrometer.

%
%+++++++++++short description of analysis++++++++++++++++++
%

The analysis proceeds by selecting tracks in the Time Projection
Chamber (TPC) in events with incident beam protons.  
Momentum and polar angle measurements and particle identification are
based on the measurements of track position and energy deposition in
the TPC.
An unfolding method is used to correct for experimental resolution,
efficiency and acceptance and to obtain the double-differential pion
production cross-sections.  The method allows a full error evaluation to
be made.
The analysis follows closely the  methods used for the determination of
$\pi^{\pm}$ production by
protons on a tantalum target which are fully described 
in Ref.~\cite{ref:harp:tantalum} and will be only briefly outlined here.
A comparison with available data is presented. 

\section{Experimental apparatus and data analysis}
\label{sec:apparatus}
\begin{figure}[tbp]
  \begin{center}
    \hspace{0mm} \epsfig{file=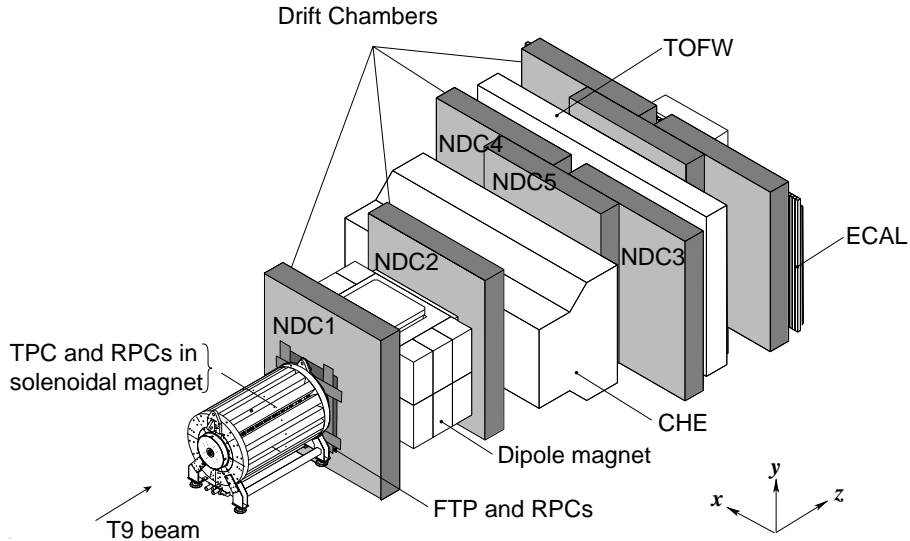,width=12cm}
  \end{center}
\caption{Schematic layout of the HARP detector. 
The convention for the coordinate system is shown in the lower-right
corner. 
The three most downstream (unlabelled) drift chamber modules are only partly
equipped with electronics and are not used for tracking.
}
\label{fig:harp}
\end{figure}

 The HARP detector is shown in Fig.~\ref{fig:harp} and is fully
 described in reference \cite{ref:harpTech}.
The forward spectrometer, mainly used in the particle production 
analysis for the conventional
neutrino beams and atmospheric neutrino flux, comprises a dipole magnet,
 large planar drift chambers 
(NDC)~\cite{NOMAD_NIM_DC}, a time-of-flight wall (TOFW) \cite{ref:tofPaper}, 
a threshold Cherenkov counter
(CHE) and an electro-magnetic calorimer (ECAL).
% together with 
% the convention used for the coordinate system.
%
In the large-angle region of particle production a cylindrical TPC with a 
radius of 408~\mm (active region)
is  positioned inside a solenoidal magnet with a field of 0.7~\T. 
The TPC detector was designed to measure and identify tracks in the
angular region from 0.25 to 2.5~\rad with respect to the beam axis.
The target is placed inside the inner field cage (IFC) of the TPC such that,
in addition to particles produced in the forward direction, 
backward-going tracks can be measured.
The targets have a nominal thickness of
5\%~$\lambda_{\mathrm{I}}$ and a cylindrical shape with a nominal
diameter of 30~\mm and each of the three targets have a purity above 99.95\%. 
%The beryllium target has a thickness of 20.46~mm
% and a density of 1.848~g/cm$^3$,
%the aluminium target has a thickness of 19.80~mm
% and a density of 2.69~g/cm$^3$ and 
%the lead target has a thickness of 8.37~mm
% and a density of 11.34~g/cm$^3$.
The Be, Al and Pb targets have a thickness of 20.46~mm, 19.80~mm and
8.37~mm with a measured variation of less than $\pm 0.02~\mm$, $\pm 0.07~\mm$
and $\pm 0.08~\mm$, respectively.  

The TPC is used
for tracking, momentum determination and the measurement of the
energy deposition \dedx for particle identification~\cite{ref:tpc:ieee}.
A set of resistive plate chambers (RPC) form a barrel inside the solenoid 
around the TPC to measure the arrival time of the secondary
particles~\cite{ref:rpc}. 
%The momentum of produced particles is obtained from the curvature of
%their trajectories in the magnetic field.
%The emission angle is given by the direction of the trajectory near
%the interaction point.
Charged particle identification (PID) can be achieved by measuring the 
ionization per unit length in the gas (\dedx) as a function of the total
momentum of the particle. 
%This  measurement is obtained with the TPC~\cite{ref:tpc:ieee}. 
Additional PID can be performed through a time-of-flight 
measurement with the RPCs.
%~\cite{ref:rpc,ref:barr:rpc,ref:ieee:rpc}.

% 
 
%In addition to the data taken with the thin carbon target of
%5\% nuclear interaction length ($\lambda_{\mathrm{I}}$),
%runs were also taken with an empty target holder, a 
%thin 2\%~$\lambda_{\mathrm{I}}$ target and a
%long 100\%~$\lambda_{\mathrm{I}}$ target.
%Data taken with a liquid hydrogen target at 3~\GeVc, 5~\GeVc and
%8~\GeVc incident beam momentum together with cosmic-ray data were used 
%to provide an absolute calibration of the efficiency, momentum scale and
%resolution of the detector. 
%Moreover, tracks produced in runs with Pb, Ta, Sn and Cu targets in
%the same period and with the same beam settings were used for
%the calibration of the detector, event reconstruction and analysis
%procedures (see reference \cite{ref:harp:tantalum} for further details). 

The momentum of the T9 beam is known with a precision of
the order of 1\%~\cite{ref:t9}. 
The absolute normalization of the number of incident protons was
performed using a total of 1,148,120 incident proton triggers. 
These are triggers where the same selection on the beam particle was
applied but no selection on the interaction was performed.
The rate of this trigger was down-scaled by a factor 64.
A cross-check of the absolute normalization was provided by counting
tracks in the forward spectrometer.

%A detailed description of the HARP apparatus is
%given in Ref.~\cite{ref:harpTech}. 
%In this analysis the detector components
%of the large-angle spectrometer and the beam instrumentation are employed
%and are briefly summarized in the following.
%
 Beam instrumentation provides identification of the incoming
 particle, the determination of the time when it hits the target, 
 and the impact point and direction of the beam particle
 on the target. 
 It is based on a set of four multi-wire proportional chambers (MWPC)
 to measure position and direction of the incoming beam particles 
 and time-of-flight detectors and 
 N$_2$-filled Cherenkov counters to identify incoming particles.  
 Several trigger detectors are installed to select events with an
 interaction and to define the normalization.

The beam of positive particles used for this measurement contains mainly 
positrons, pions and protons, with small components of kaons and
deuterons and heavier ions.
%Its composition depends on the selected beam momentum.
The proton fraction in the incoming beam varies from 35\% at
3 GeV/c to 92\% at 12 GeV/c. 
The length of the accelerator spill is 400~ms with a typical intensity
of 15~000 beam particles per spill.

In addition to the usual need for calibration of the detector, a number of
%difficulties 
hardware shortfalls, discovered mainly after the end of data-taking,
had to be overcome to use the TPC data reliably in the analysis.
The TPC is affected by a relatively large number of dead or noisy 
pads and static and dynamic distortions of the reconstructed trajectories.
Static distortions are caused by the inhomogeneity of the electric field,
due to an accidental mismatch between the inner and outer field cage 
(powered by two distinct HV supplies) and 
other sources.
% a partial "transparency" of the cathode wire grid. 
Dynamic distortions are caused instead by the
build-up of ion-charge density in the drift volume during the 400~ms 
long beam spill. 
All these effects were fully studied and available corrections are described 
in detail in Ref.~\cite{ref:harp:tantalum}. While methods to correct the dynamic 
distortions of the TPC tracks are being developed, a practical approach has been
followed in the present analysis. Only the events corresponding to the
early part of the spill, where the effects of the dynamic distortions are
still small, have been used\footnote{This translates into a cut on the maximum 
number of events ($N_{evt}$) to be retained.}. 
The time interval between spills 
is large enough to drain all charges in the TPC related to the effect of the beam.
The combined effect of the distortions on the kinematic quantities used
in the analysis has been studied in detail and only that part of the data
for which the systematic errors can be assessed with physical 
benchmarks was used, as explained in \cite{ref:harp:tantalum}. More than 
%$35 \%$ 
$30 \%$ 
of the recorded data can thus be used in the current analysis.

The absolute momentum scale is determined by using elastic scattering 
events off a hydrogen target. The angle of the forward scattered particle is
used to predict the momentum of the recoil proton, to be compared to the one
 measured by the TPC. To study the stability of this measurement protons
are selected in a narrow band with a relatively large \dedx where \dedx 
depends strongly on momentum. 
The average momentum for the selected protons remains stable within $3 \%$
as a function of time-in-spill over the part of the spill used in this
analysis. 
%%\section{Data selection and analysis}
%%\label{sec:selection}
Only a short outline of the data analysis is presented here, for more details
see Ref. \cite{ref:harp:tantalum}.

The analysis proceeds by first selecting a beam proton hitting the
target, not accompanied by other tracks. 
Then an event is required to give a large angle interaction (LAI) trigger  to be
retained. 
After the event selection the sample of tracks to be used for analysis
is defined.
At least twelve space points in the TPC out of a maximum of twenty are
required to consider a track. This cut ensures a good measurement of the
track parameters and the \dedx. For the selected tracks a cut on 
$ d_{0}^{'}$ (the distance of closest approach 
to the extrapolated trajectory 
of the incoming beam particle in the plane perpendicular to the beam
direction) 
and $ z^{'}_{0}$ (the z-coordinate where the distance of the
secondary track and the beam track is minimal) is applied.
Finally, only tracks with $100~\MeVc \leq p \leq 800~\MeVc$ and $p_{T}\geq 55~\MeVc$
are accepted. 

Table~\ref{tab:events} shows the number of events and the number of $\pi^{\pm}$ 
selected in the p--Be, p--Al and  p--Pb analysis.
The total number of events taken by the data acquisition (``Total DAQ
events'') includes trigger of all types as well as calibration events. 
The number of accepted events for the analysis (``Accepted protons with 
LAI'') is obtained from incoming protons 
in coincidence with a large angle trigger. 
The large difference between the two numbers
is due to the relatively large fraction of pions in the
beam and to the larger number of triggers taken for the measurements
with the forward dipole spectrometer.
These data will be the subject of other publications. 
The fraction of data used for the analysis (``Fraction of triggers used'')
after a cut on the maximum event number to be retained in the spill
(``$N_{evt}$ cut'') to avoid dynamic distortion corrections is then reported.
Finally, the rows ``Negative particles'', ``Positive particles '',
``$\bfpim$ selected with PID'' and ``$\bfpip$ selected with PID'' show
the number of accepted tracks with negative and positive charge and the ones
passing in addition the pion PID criteria, respectively.
%To give an impression of the complexity of the events, one can define an
%`average multiplicity' as the ratio of the number of tracks with at
%least twelve hits in the TPC (regardless of their momentum, angle or
%spatial position) and the number of events accepted by the selection
%criteria with at least one such track. 
%The average multiplicity obtained according to this definition is
%reported in Table~\ref{tab:events}.
%With this definition, the average multiplicity is 2.2, 2.6, 3.1 and 3.4
%in the 3~\GeVc, 5~\GeVc, 8~\GeVc and 12~\GeVc beams, respectively.
 
% beryllium (lead)
\begin{table}[tbp!] 
\caption{Total number of events and tracks used in the beryllium,
  aluminium and lead 
  5\%~$\lambda_{\mathrm{I}}$ target data sets, and the number of
  protons on target as calculated from the pre-scaled trigger count. 
 For each entry the first line shows beryllium target data, the second
  line -- aluminium target data and the third (last) line -- lead target data.} 
\label{tab:events}
{\small
\begin{center}
\begin{tabular}{llrrrrrr} \hline
%\bf{Data set}          &         &\bf{3 \bfGeVc}&\bf{5 \bfGeVc}&\bf{8 \bfGeVc}&\bf{8.9 \bfGeVc} &\bf{12 \bfGeVc} &\bf{12.9 \bfGeVc}\\ \hline
\bf{Data set (\bfGeVc)}          &         &\bf{3}&\bf{5}&\bf{8}&\bf{8.9} &\bf{12} &\bf{12.9}\\ \hline
    Total DAQ events     &  (Be)     & 1409710 & 1705362      & 2010031      &  3969685        &  1288354       &   --            \\
                         & (Al)     & 1586331 & 1094308      & 1706919      &   --            &  619021        &  6644256        \\
                         & (Pb)     & 1299264 & 2234984      & 1949950      &   --            &  630417        &  --              \\
  Acc. protons with LAI  &           & 77223   &  182423      &  365500      &  692150         &  300939        &   --             \\
                         &           & 69794   &  120948      &  341687      &   --            &  71554         &  1715323         \\
                         &           & 79188   &  207004      &  415982      &   --            &  188134        &  --              \\
  Fraction of triggers used   &      & 35\%   &  33\%       &   36\%      &  41\%          &    41\%       &   --              \\
    (\evtspill cut)           &      & 48\%   &  40\%       &   33\%      &  --             &  42\%        &   35\%            \\   
                              &      & 36\%   &  32\%       &   36\%      &  --             &   27\%        &   --             \\
  \bf{$\bfpim$ selected with PID} &  & 3120    &   11168      &  29337       &  63887          &  29506         &   --              \\
                                  &  & 3882    &   9233       &  27809       &  --             &  19290         &   168229          \\
                                  &  & 2347    &   11842      &  42576       &  --             &  18092         &   --               \\
 \bf{$\bfpip$ selected with PID}  &  & 5520    &   15331      &  37049       &  78727          &  35136         &   --               \\ 
                                  &  & 6396    &   13045      &  35991       &  --             &  23440         &   203924           \\
                                  &  & 3203    &   13318      &  46150       &  --             &  19040         &   --           \\ \hline
\end{tabular}
\end{center}
}
\end{table}
% counters of target.py used:
% #11
% #41 x64
% #21
% #22
% nevt
% #23
% #23/#22
% #72
% mult (mean of trk_mult1 in ..target.aida)
% #84
% #87
% #91
% #92
% #88
% #89

\section{Experimental results}
\label{sec:results}

The double-differential cross-section for the production of a particle of 
type $\alpha$ can be expressed in the laboratory system as:

\begin{equation}
{\frac{{\mathrm{d}^2 \sigma_{\alpha}}}{{\mathrm{d}p_i \mathrm{d}\theta_j }}} =
\frac{1}{{N_{\mathrm{pot}} }}\frac{A}{{N_A \rho \cdot t}}
 \sum_{i',j',\alpha'} M_{ij\alpha i'j' \alpha'}^{-1} \cdot
{N_{i'j'}^{\alpha'} } 
\ ,
\label{eq:cross}
\end{equation}

where $\frac{{\mathrm{d}^2 \sigma_{\alpha}}}{{\mathrm{d}p_i \mathrm{d}\theta_j }}$
is expressed in bins of true momentum ($p_i$), angle ($\theta_j$) and
particle type ($\alpha$).
The factor  $\frac{A}{{N_A \rho \cdot t}}$ in Eq.~\ref{eq:cross}
is the inverse of the number of target nuclei per unit area
($A$ is the atomic mass,
$N_A$ is the Avogadro number, $\rho$ and $t$ are the target density
and thickness)\footnote{We do not make a correction for the attenuation
of the proton beam in the target, so that  the
cross-sections are strictly valid for a $\lambda_{\mathrm{I}}=5\%$ target.}.

%% The terms on the right-hand side of the equation are as follows.

%\begin{itemize}

%\item 
The `raw yield' $N_{i'j'}^{\alpha'}$ 
is the number of particles of observed type $\alpha'$ in bins of reconstructed
momentum ($p_{i'}$) and  angle ($\theta_{j'}$). 
These particles must satisfy the event, track and PID 
selection criteria.
Although, thanks to the stringent PID selection,  the background from
misidentified protons in the pion sample is small, the pion and proton
raw yields ($N_{i'j'}^{\alpha'}$, for 
$\alpha'=\pim, \pip, \mathrm{p}$) have been measured simultaneously. 
It is thus possible to correct for the small remaining proton
background in the pion data without prior assumptions concerning the
proton production cross-section.
%Figure~\ref{fig:raw-yield} shows the $p$ and \tht distribution for the
%momentum bins and angular bins chosen in the analysis.

The matrix $ M_{ij\alpha i'j' \alpha'}^{-1}$ 
corrects for the  efficiency and resolution of the detector. 
It unfolds the true variables $ij\alpha$ from the reconstructed
variables $i'j'\alpha'$  with a Bayesian technique~\cite{dagostini} 
and corrects  
the observed number of particles to take into account effects such as 
trigger efficiency, reconstruction efficiency, acceptance, absorption,
pion decay, tertiary production, 
PID efficiency, PID misidentification and electron background. 
The method used to correct for the various effects is  described in
more detail in Ref.~\cite{ref:harp:tantalum}.

In order to predict the population of the migration matrix element 
$M_{ij\alpha i'j'\alpha'}$, the resolution, efficiency
and acceptance of the detector are obtained from the Monte Carlo.
This is accurate provided that the Monte Carlo
simulation describes these quantities correctly. 
Where some deviations
from the control samples measured from the data are found, 
the data are used to introduce (small) corrections to the
Monte Carlo. 
Using the unfolding approach, possible known biases in the measurements
are taken into account automatically as long as they are described by
the Monte Carlo.
In the experiment simulation, which is based on the GEANT4
toolkit~\cite{ref:geant4}, the materials in the beam-line and the 
detector are accurately described as well as
the relevant features of the detector response and 
the digitization process.
%The experiment simulation is based on the GEANT4
%toolkit~\cite{ref:geant4}.
%The materials in the beam line and the detector are accurately
%reproduced in this simulation, as well as the relevant features of the
%detector response and the digitization process. 
In general, the Monte Carlo
simulation compares well with the data, as shown in Ref.~\cite{ref:harp:tantalum}.

The result is normalized to the number of incident protons on target
$N_{\mathrm{pot}}$. 
The absolute normalization of the result is calculated in the first
instance relative to the number of incident beam particles accepted by
the selection. 
After unfolding, the factor  $\frac{A}{{N_A \rho \cdot t}}$ is applied.
The beam normalization using down-scaled incident proton triggers 
has uncertainties smaller than 2\%
  for all beam momentum settings.

The background due to interactions of the primary
protons outside the target (called `Empty target background') is
measured using data taken without the target mounted in the target
holder.
Owing to the selection criteria which only accept events from the
target region and the good definition of the interaction point this
background is negligible ($< 10^{-5}$).
The background of
interactions of the primary proton outside the target can be suppressed
for large angle tracks measured in the TPC owing to the good resolution
in $z$.  This is contrary to the situation in the forward spectrometer
where an interaction in the target cannot be distinguished from an
interaction in upstream or downstream
material~\cite{ref:alPaper,ref:bePaper}.

The effects of these uncertainties on the final results are estimated
by repeating the analysis with the relevant input modified within the
estimated uncertainty intervals.
In many cases this procedure requires the construction of a set of
different migration matrices.
The correlations of the variations between the cross-section bins are
evaluated and expressed in the covariance matrix.
Each systematic error source is represented by its own covariance
matrix.
The sum of these matrices describes the total systematic error.

\subsection{Cross-section measurements}

%%%%%%%%%%%%%%%%%%%%%%%%%%%%

\begin{figure}[btp]
\begin{center}
%\begin{sideways}
\epsfig{figure=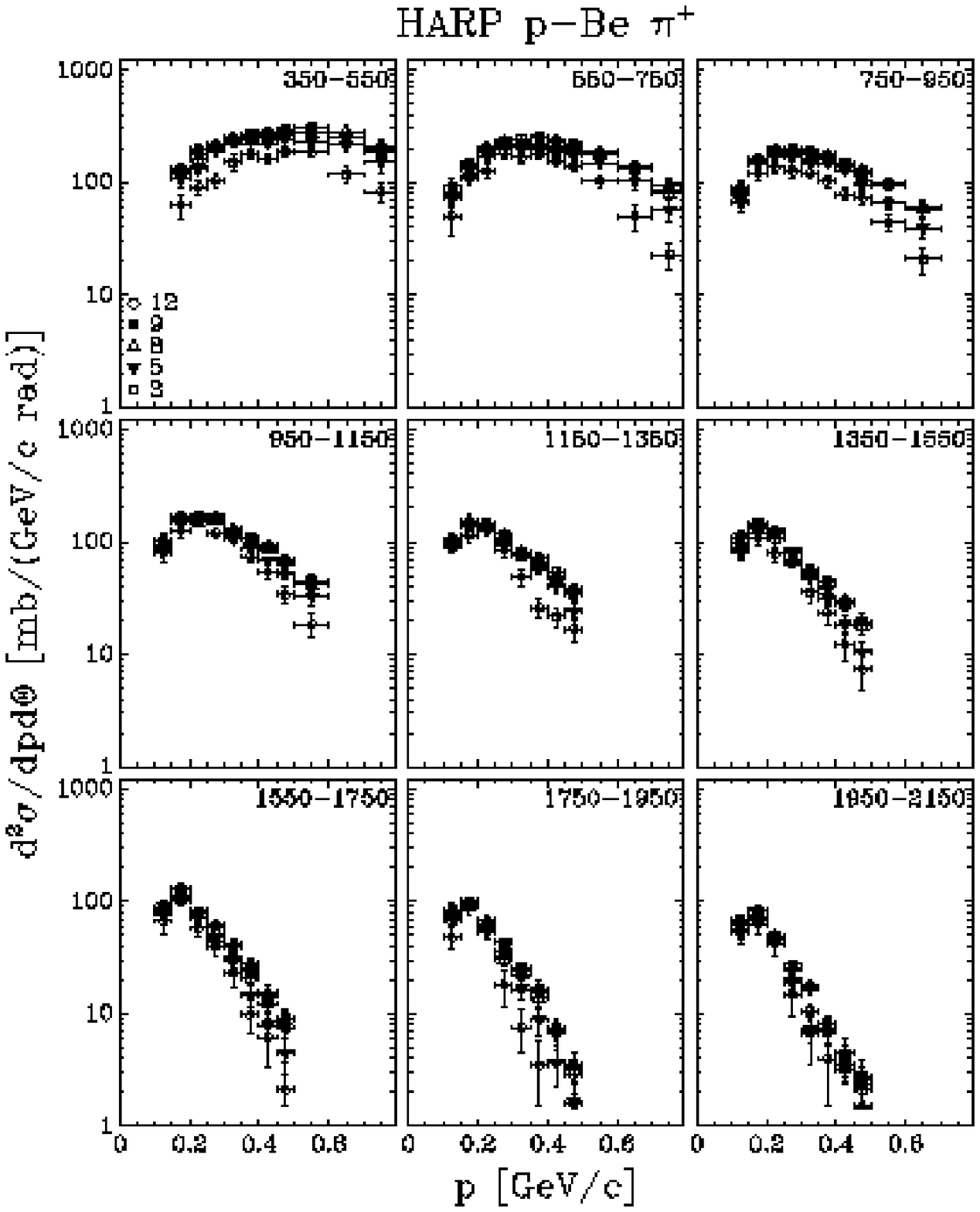,width=0.49\textwidth} 
\epsfig{figure=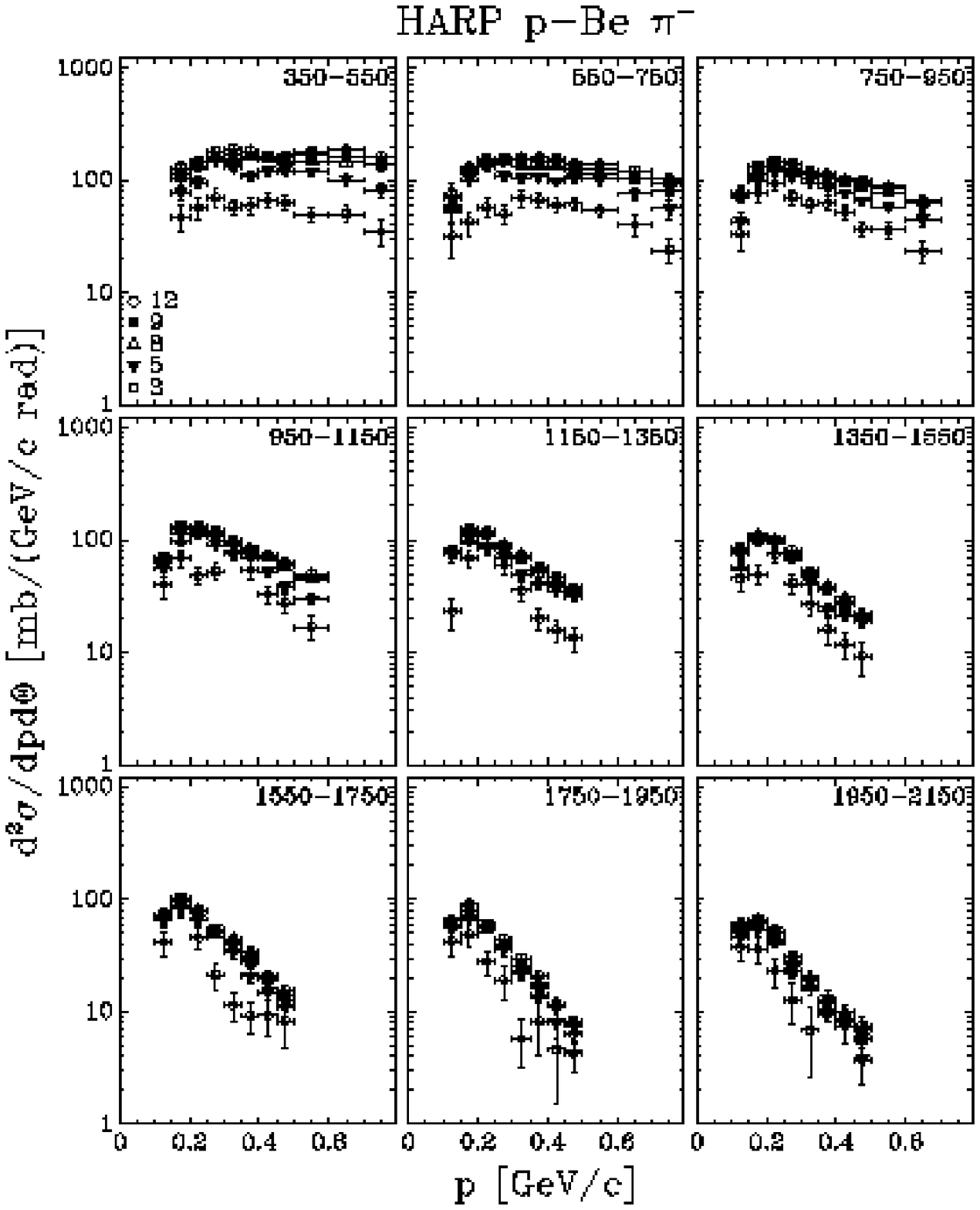,width=0.49\textwidth} 
%\end{sideways}
\caption{
Double-differential cross-sections for \pip production (top) and  \pim
 production (bottom) in
p--Be interactions as a function of momentum displayed in different
angular bins (shown in \mrad in the panels).
%The results are given for four incident beam momenta (filled triangles:
%3~\GeVc; open triangles: 5~\GeVc; filled rectangles: 8~\GeVc; open
%circles: 12~\GeVc). 
The error bars represent the combination of statistical and systematic
 uncertainties. 
}
\label{fig:xs-p-th-pbeam-be}
\end{center}
\end{figure}
\begin{figure}[tbp]
\begin{center}
%\begin{sideways}
\epsfig{figure=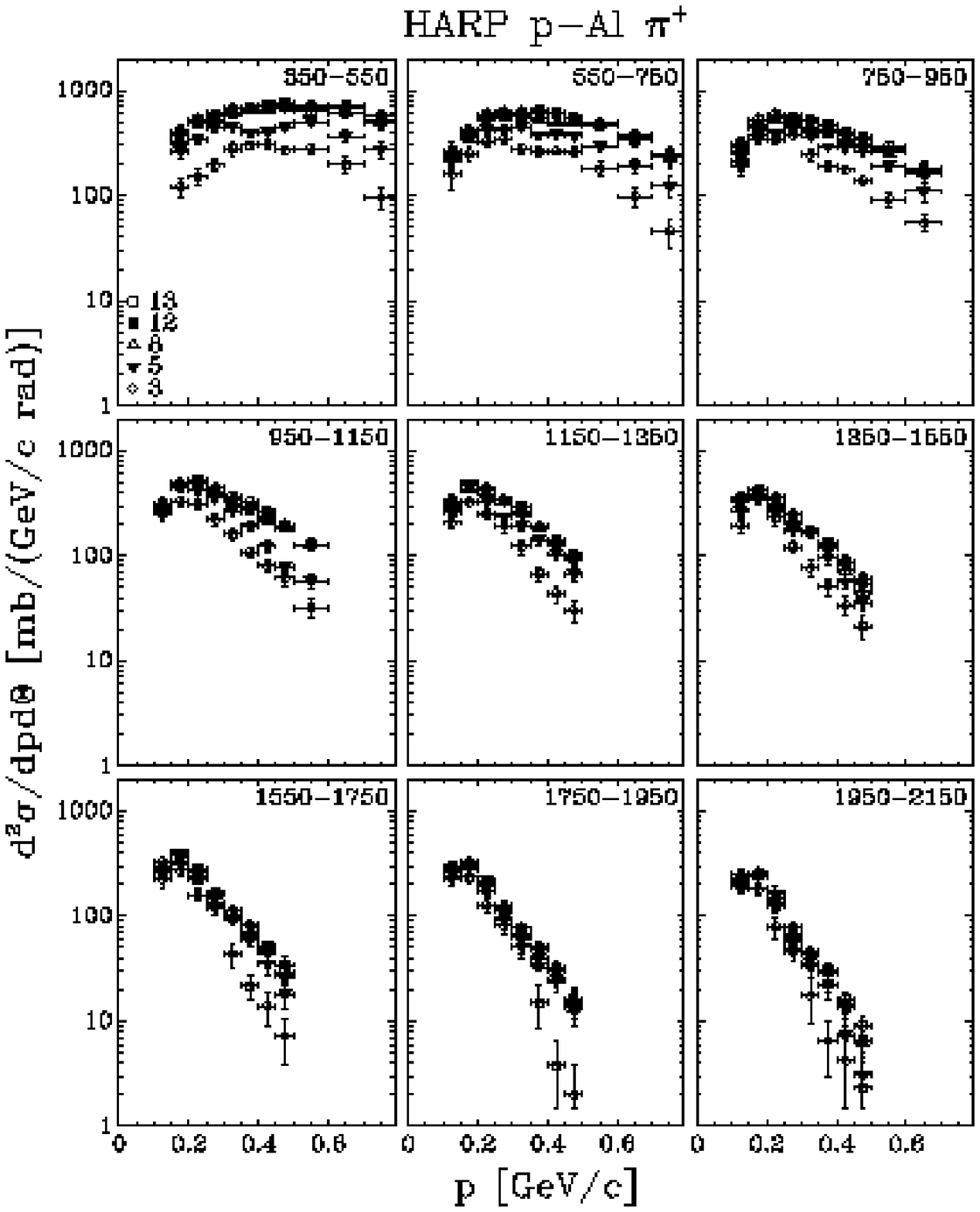,width=0.49\textwidth} 
\epsfig{figure=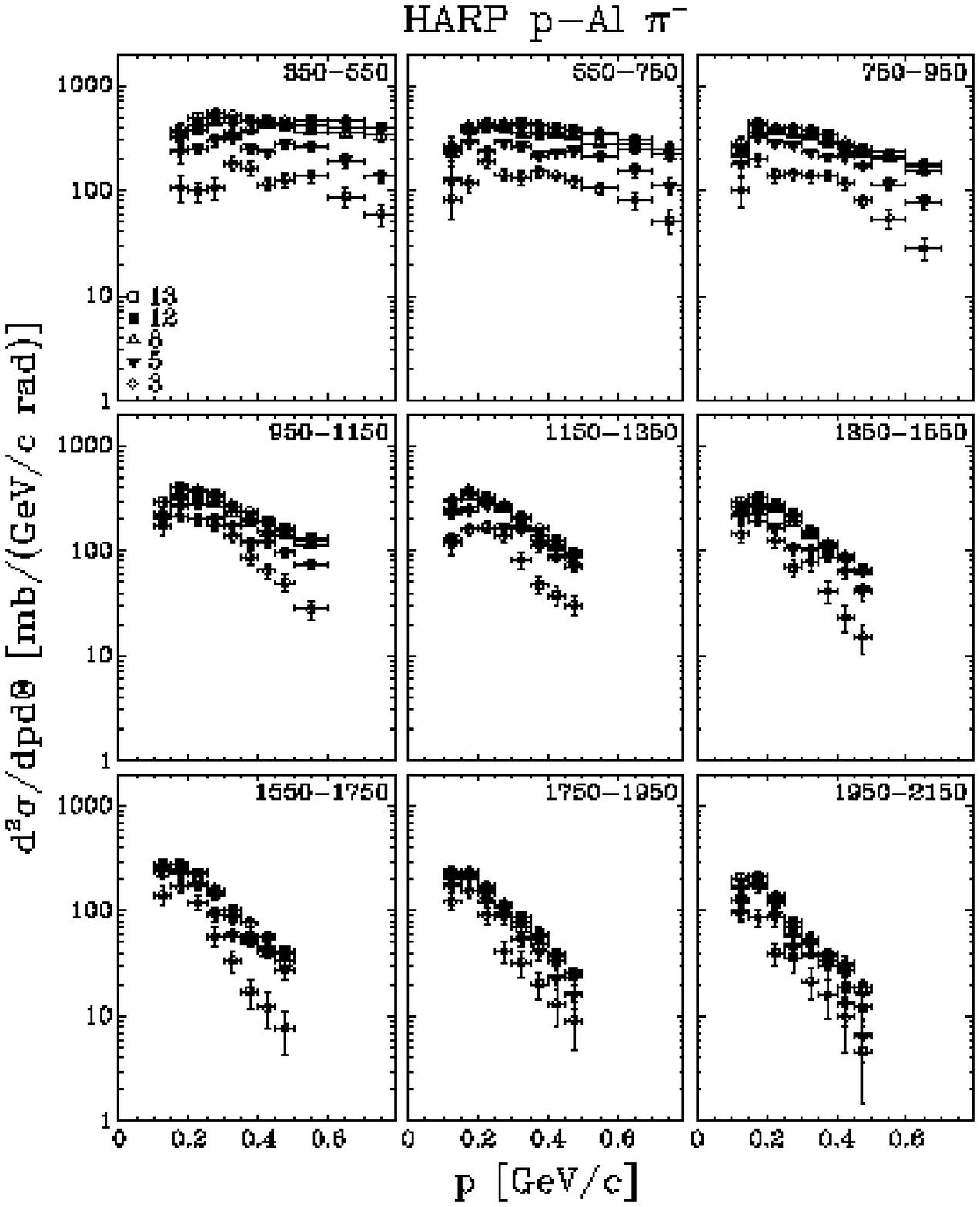,width=0.49\textwidth} 
%\end{sideways}
\caption{
Double-differential cross-sections for \pip production (top) and  \pim
 production (bottom) in
p--Al interactions as a function of momentum displayed in different
angular bins (shown in \mrad in the panels).
%The results are given for four incident beam momenta (filled triangles:
%3~\GeVc; open triangles: 5~\GeVc; filled rectangles: 8~\GeVc; open
%circles: 12~\GeVc). 
The error bars represent the combination of statistical and systematic
 uncertainties. 
In the figure, the symbol legend 13 (9) refers to 12.9 (8.9) \GeVc nominal
beam momentum.
}
\label{fig:xs-p-th-pbeam-al}
\end{center}
\end{figure}
\begin{figure}[tbp]
\begin{center}
%\begin{sideways}
\epsfig{figure=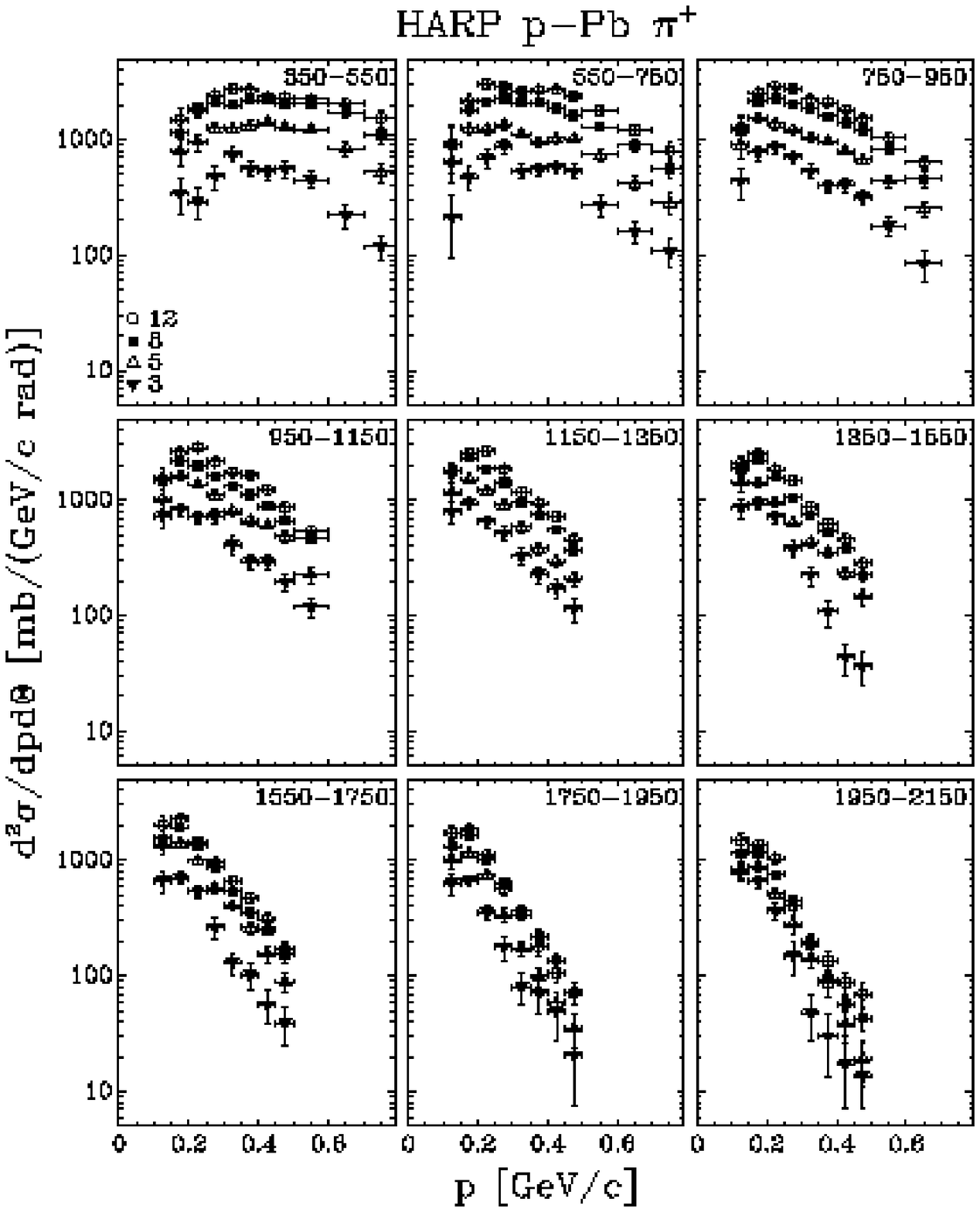,width=0.49\textwidth} 
\epsfig{figure=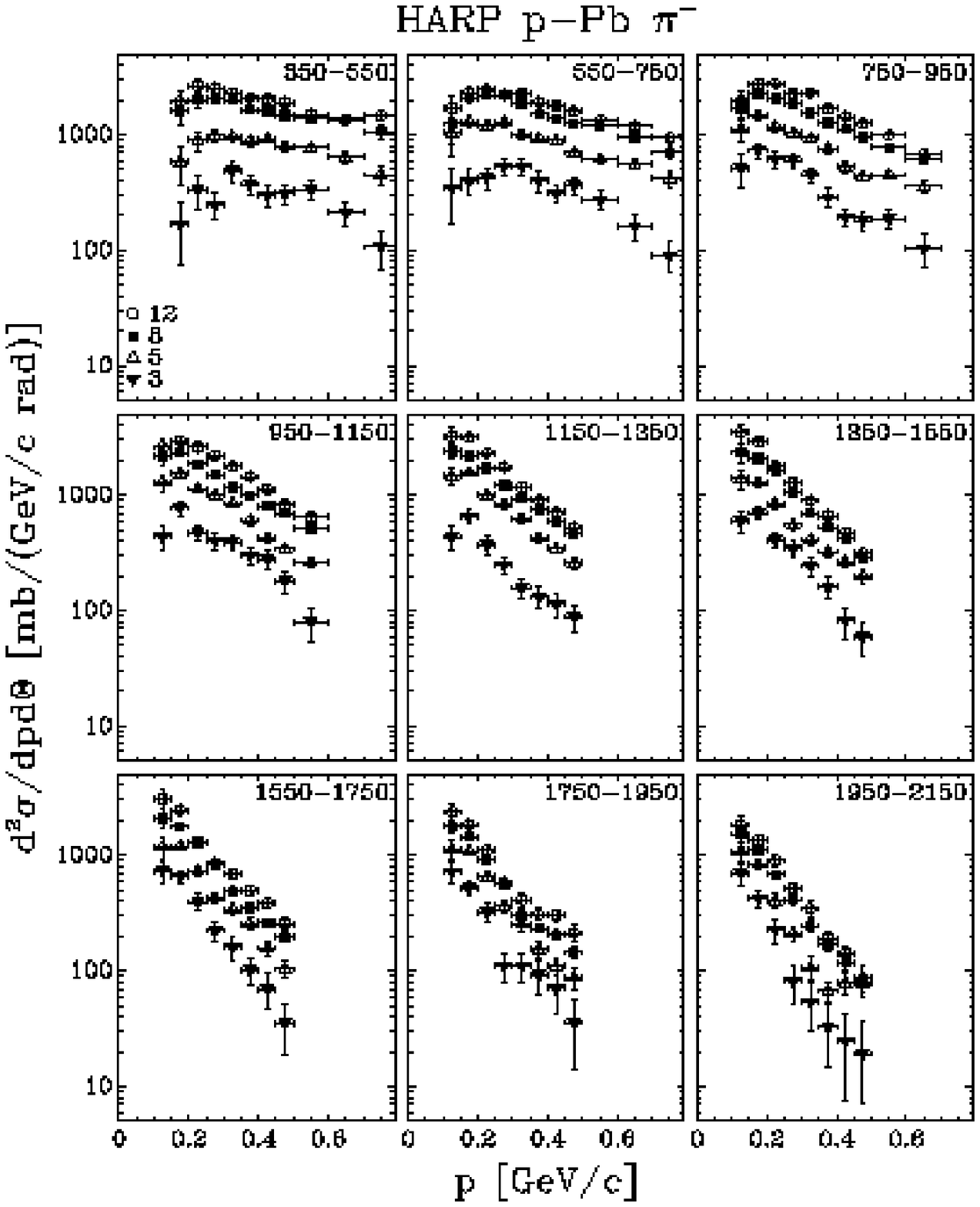,width=0.49\textwidth} 
%\end{sideways}
\caption{
Double-differential cross-sections for \pip production (top) and  \pim
 production (bottom) in
p--Pb interactions as a function of momentum displayed in different
angular bins (shown in \mrad in the panels).
%The results are given for four incident beam momenta (filled triangles:
%3~\GeVc; open triangles: 5~\GeVc; filled rectangles: 8~\GeVc; open
%circles: 12~\GeVc). 
The error bars represent the combination of statistical and systematic
 uncertainties. 
}
\label{fig:xs-p-th-pbeam-pb}
\end{center}
\end{figure}

The measured double-differential cross-sections for the 
production of \pip and \pim in the laboratory system as a function of
the momentum and the polar angle for each incident beam momentum are
shown in Fig.~\ref{fig:xs-p-th-pbeam-be}, \ref{fig:xs-p-th-pbeam-al} and
\ref{fig:xs-p-th-pbeam-pb} for Be, Al and Pb, respectively.
%Figures~\ref{fig:xs-p-th-pbeam-plus} and \ref{fig:xs-p-th-pbeam-minus}
%show 
%the measurement of the double-differential cross-section for
%the production of $\pi^+$
%(Fig.~\ref{fig:xs-p-th-pbeam-plus}) and $\pi^-$
%(Fig.~\ref{fig:xs-p-th-pbeam-minus}) in the laboratory
%system as a function of 
%the momentum and the polar angle for each incident beam momentum. 
The error bars  shown are the
square-roots of the diagonal elements in the covariance matrix,
where statistical and systematic uncertainties are combined
in quadrature.
Correlations cannot be shown in the figures.
The correlation of the statistical errors (introduced by the unfolding
procedure) are typically smaller than 20\% for adjacent momentum bins and
even smaller for adjacent angular bins.
The correlations of the systematic errors are larger, typically 80\% for
adjacent bins.
The overall scale error  is not shown.
The latter error is 2\% for Be and Al and 3\% for Pb due to the
larger variation in the measured thickness of the lead target.
The results of this analysis are also tabulated in Appendix A.
% A discussion of the error evaluation is given below. 
%The measurements for the different beam momenta are overlaid in the
%same figure.

\begin{figure}[tbp]
  \epsfig{figure=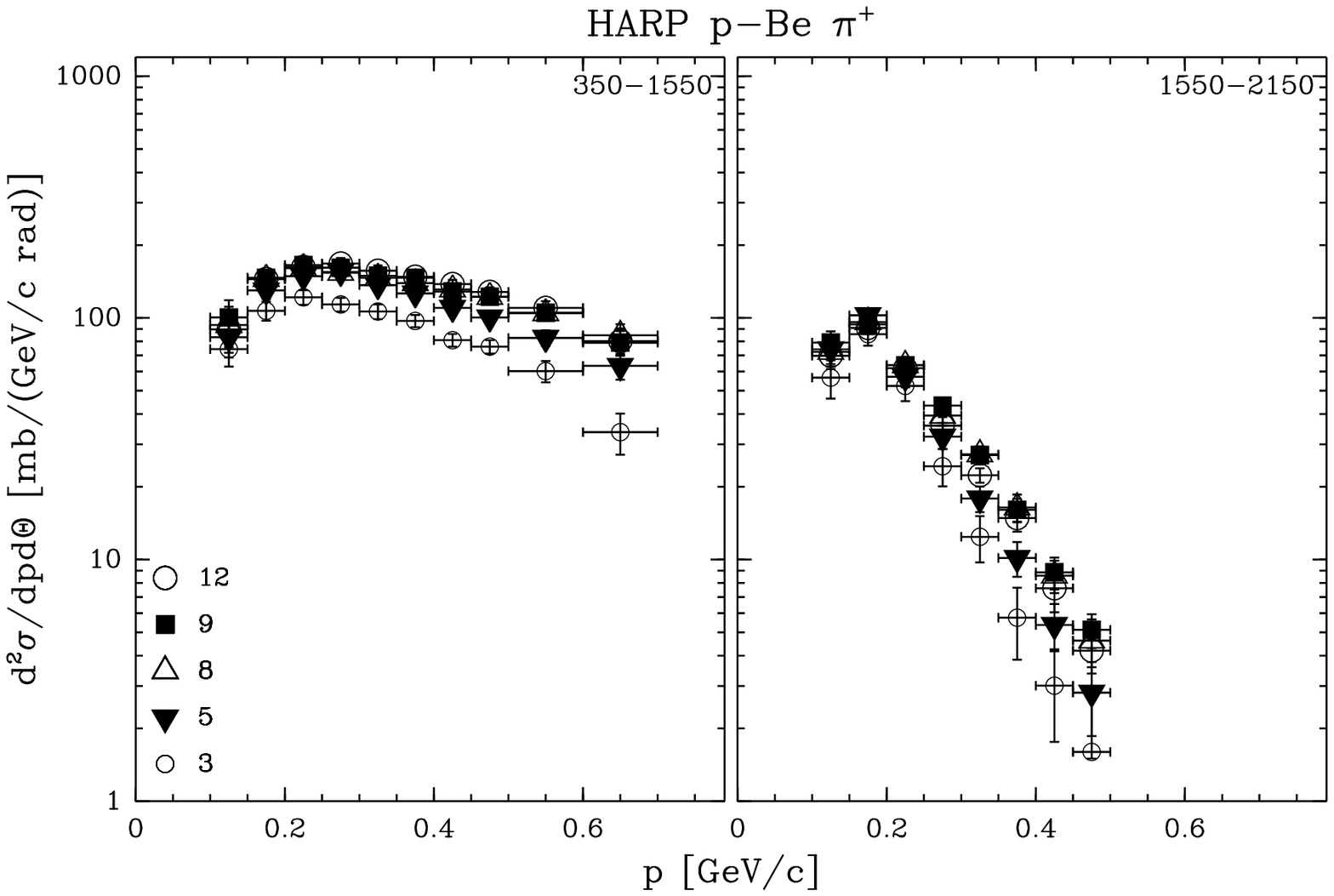,width=0.49\textwidth}
  ~
  \epsfig{figure=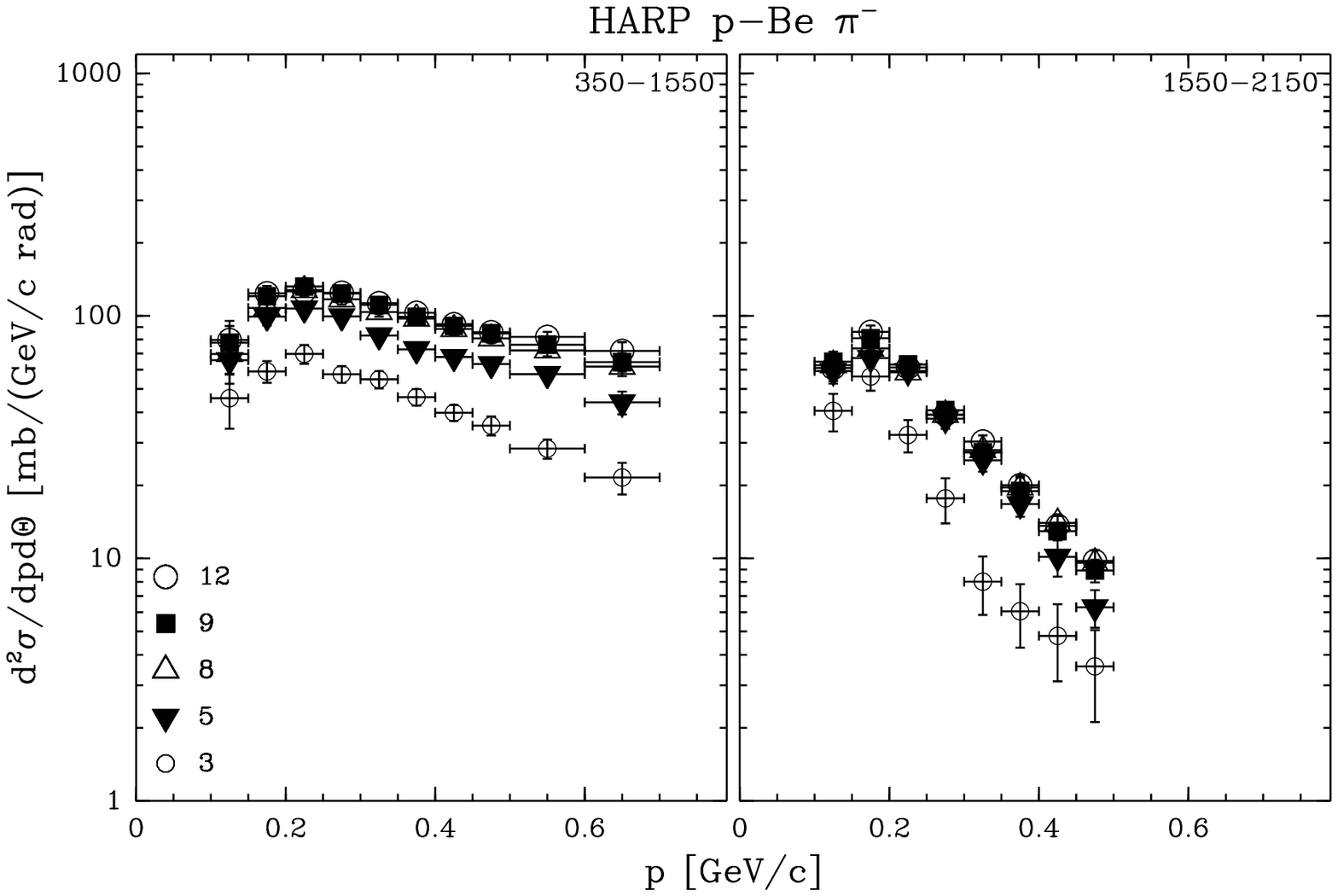,width=0.49\textwidth}
  \epsfig{figure=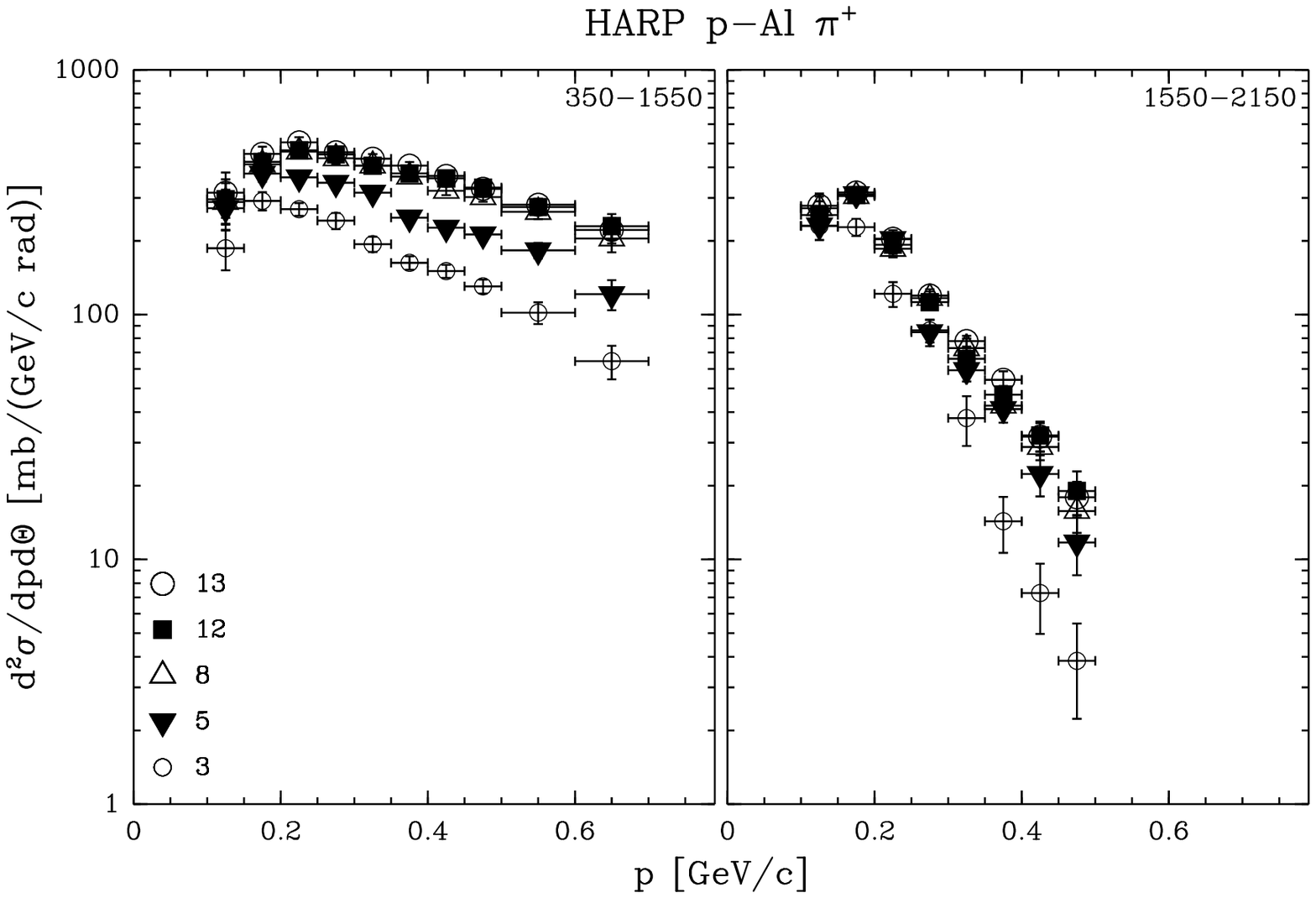,width=0.49\textwidth}
  ~
  \epsfig{figure=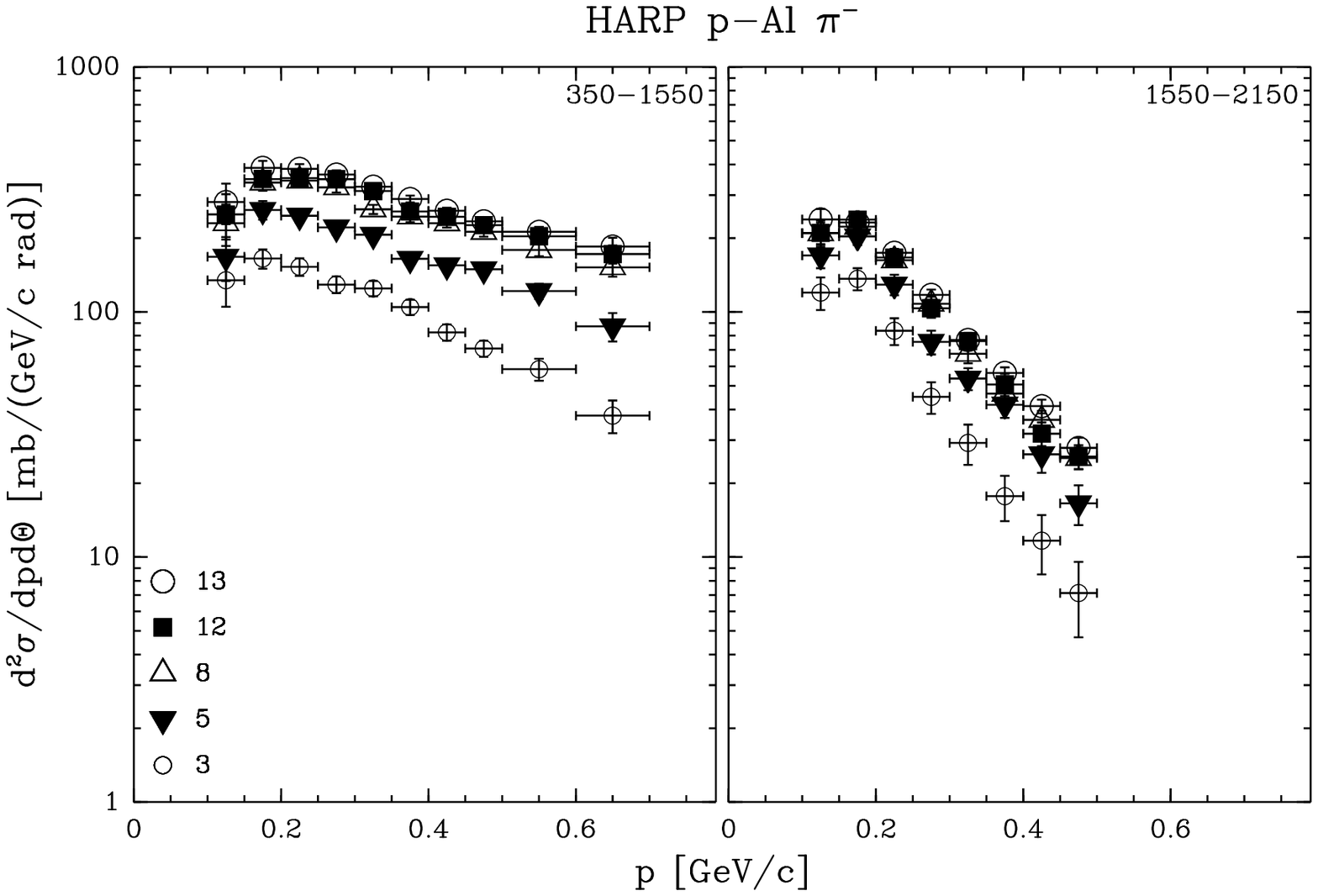,width=0.49\textwidth}
  \epsfig{figure=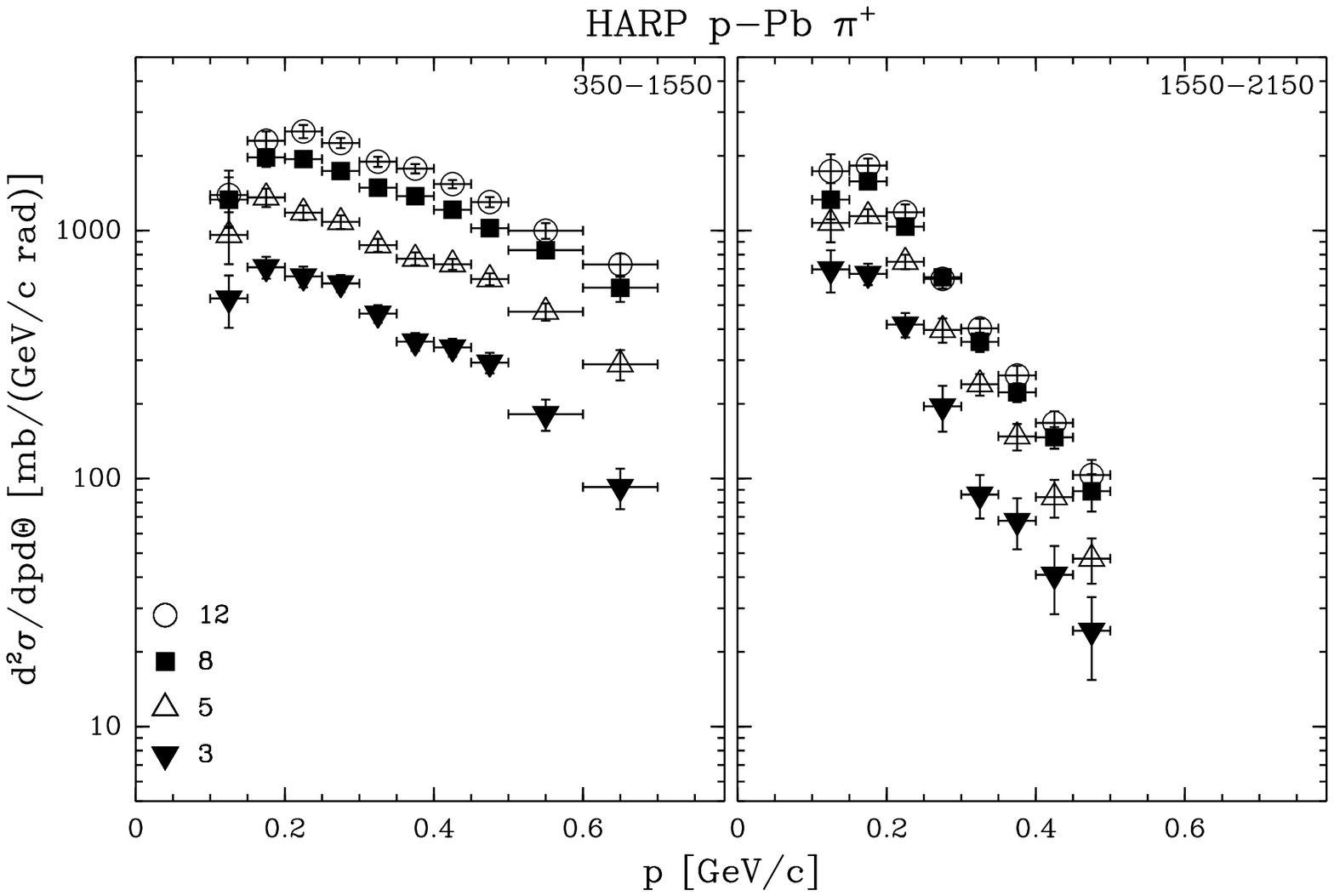,width=0.49\textwidth}
  ~
  \epsfig{figure=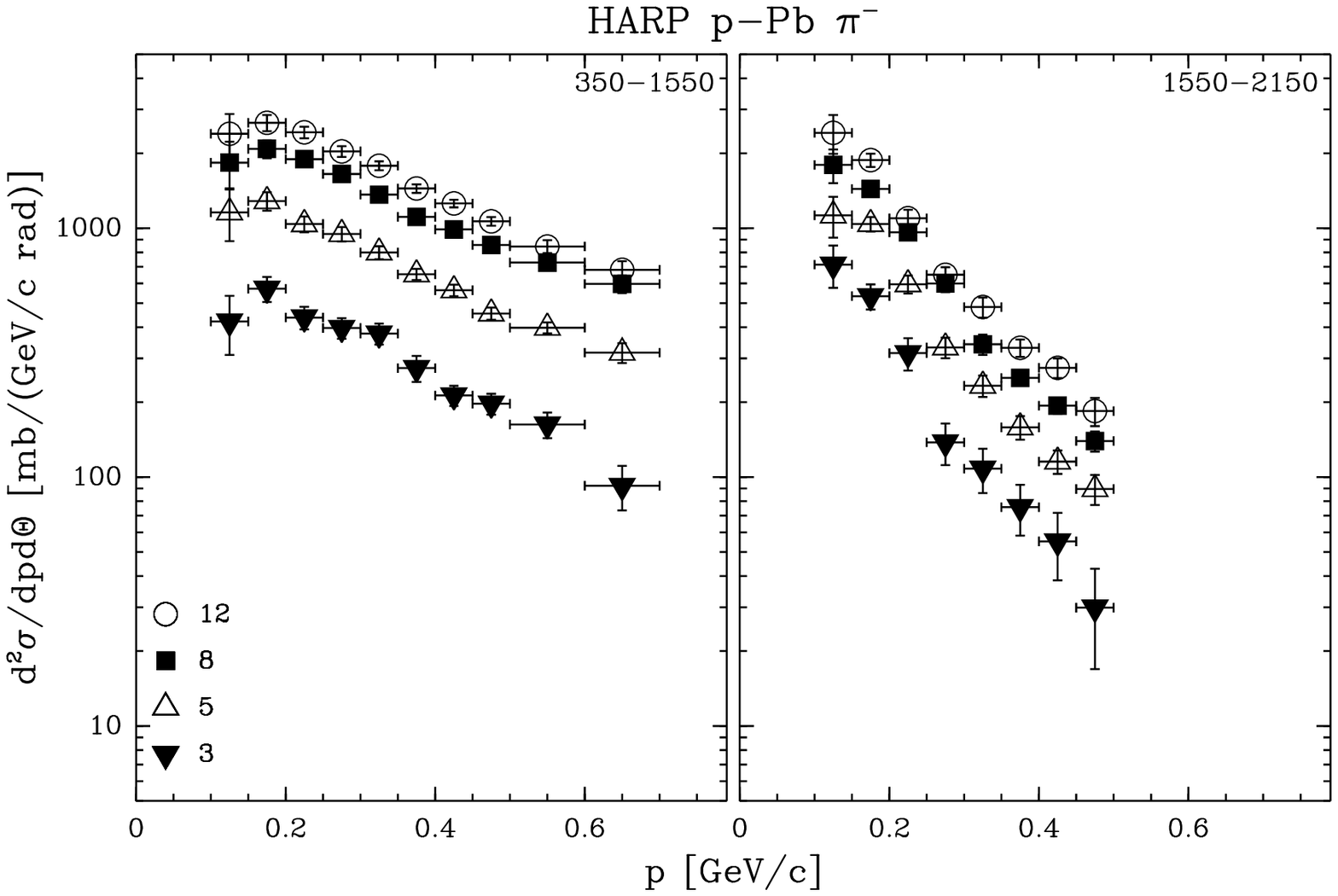,width=0.49\textwidth}
\caption{
Double-differential cross-sections for \pip  and \pim production in 
p--Be (top), p--Al (middle) and p--Pb (bottom) interactions as a function of momentum averaged over the
angular region covered by this experiment (shown in mrad).
The left panel of each pair shows forward production 
(350~\mrad  $\le~\mrad \theta <$ 1550~\mrad), while
the right panel of each pair shows backward production 
(1550~\mrad  $\le \theta <$ 2150~\mrad).
%The results are given for four incident beam momenta (filled triangles:
%3~\GeVc; open triangles: 5~\GeVc; filled rectangles: 8~\GeVc; open
%circles: 12~\GeVc).
The error bars obtained after summing the bins of the
 double-differential cross-sections take into account the correlations
 of the statistical and systematic uncertainties. 
In the figure, the symbol legend 13 (9) refers to 12.9 (8.9) \GeVc nominal
beam momentum.
}
\label{fig:xs-p-pbeam}
\end{figure}

To better visualize the dependence on the incoming beam momentum, the
same data averaged over the angular range (for the
forward going and backward going tracks) covered by the analysis are shown
separately for \pip and \pim in Fig.~\ref{fig:xs-p-pbeam}.
The spectrum of pions produced in the backward direction is much
steeper than that in the forward direction.
The increase of the pion yield per proton is visible in 
addition to a change of spectrum towards higher momentum of the
secondaries produced by higher momentum beams in the forward
direction. 
This dependence is much weaker for Be than for Pb.
%%%%%%%%%%%%%%%%%%%%%%%%%%%%%%%%%%%%%%%%%%%%%%%%%%%%%

\begin{figure}[tbp]
\begin{center}
  \epsfig{figure=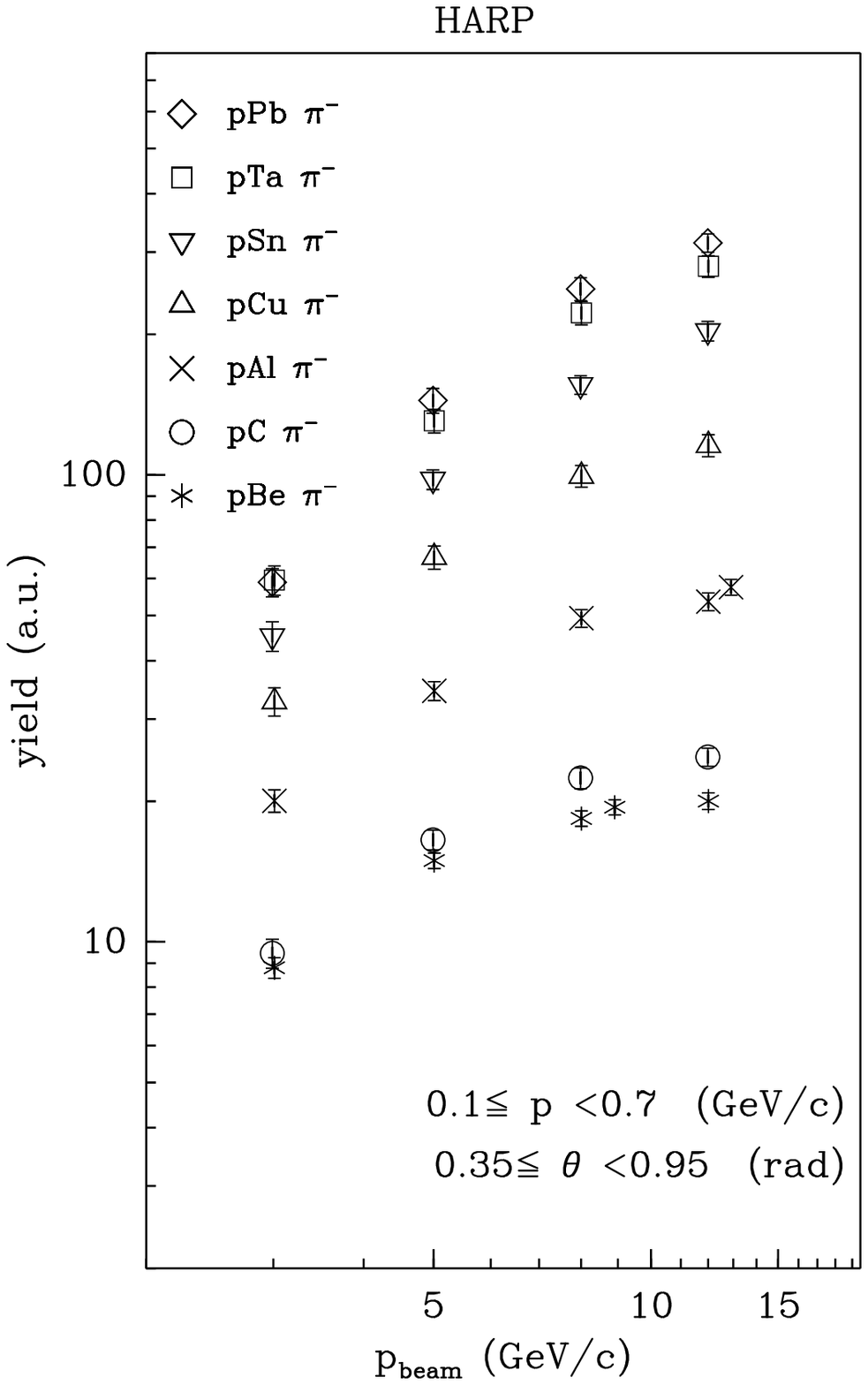,width=0.425\textwidth}
 ~
  \epsfig{figure=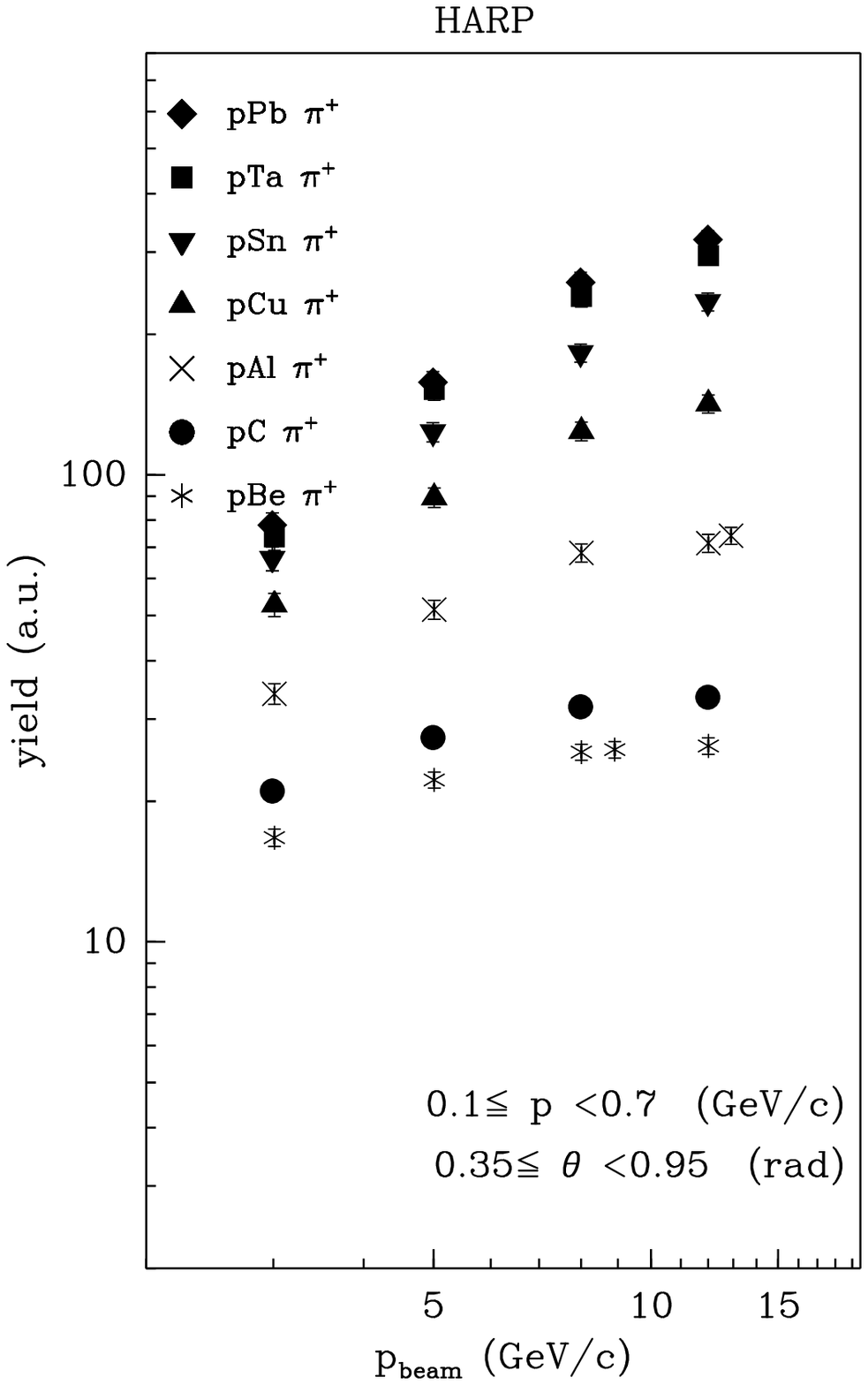,width=0.425\textwidth}
\end{center}
\caption{
 The dependence on the beam momentum of the \pim (left) and \pip (right) 
  production yields
 in p--Be, p--C, p--Al, p--Cu, p--Sn, p--Ta, p--Pb
 interactions integrated over the forward angular region 
 ($0.350~\rad \leq \theta < 0.950~\rad$) and momentum ($100~\MeVc \leq p < 700~\MeVc$).
 The results are given in arbitrary units, with a consistent scale
 between the left and right panel.
%Although the units are indicated as ``arbitrary'',
%for the largest region, the yield is expressed as 
%${{\mathrm{d}^2 \sigma}}/{{\mathrm{d}p\mathrm{d}\Omega }}$ in
%mb/(\GeVc~sr).  
Data points for different target nuclei and equal momenta are slightly
 shifted horizontally with respect to each other to increase the visibility.
}
\label{fig:xs-trend}
\end{figure}

The dependence of the integrated pion yields on the incident beam
momentum is shown in Fig.~\ref{fig:xs-trend} and compared with 
previous results obtained with the same apparatus. 
The \pip and \pim yields are integrated over the region 
$0.350~\rad \leq \theta < 0.950~\rad$ and $100~\MeVc \leq p < 700~\MeVc$.
Whereas the beam energy dependence of the yields in the
p--C, p--Be  data differs clearly from the dependence in the p--Ta,
p--Pb data one can
observe that the p--Al, p--Cu and p--Sn data display a
smooth transition between them.
The dependence in the p--C, p--Be data is much more flat with a saturation of
the yield between 8~\GeVc and 12~\GeVc
with the p--Al, p--Cu and p--Sn showing an intermediate behaviour.
%% Also the \pip and \pim production yields exhibit a different behaviour.

\begin{figure}[tbp]
\begin{center}
  \epsfig{figure=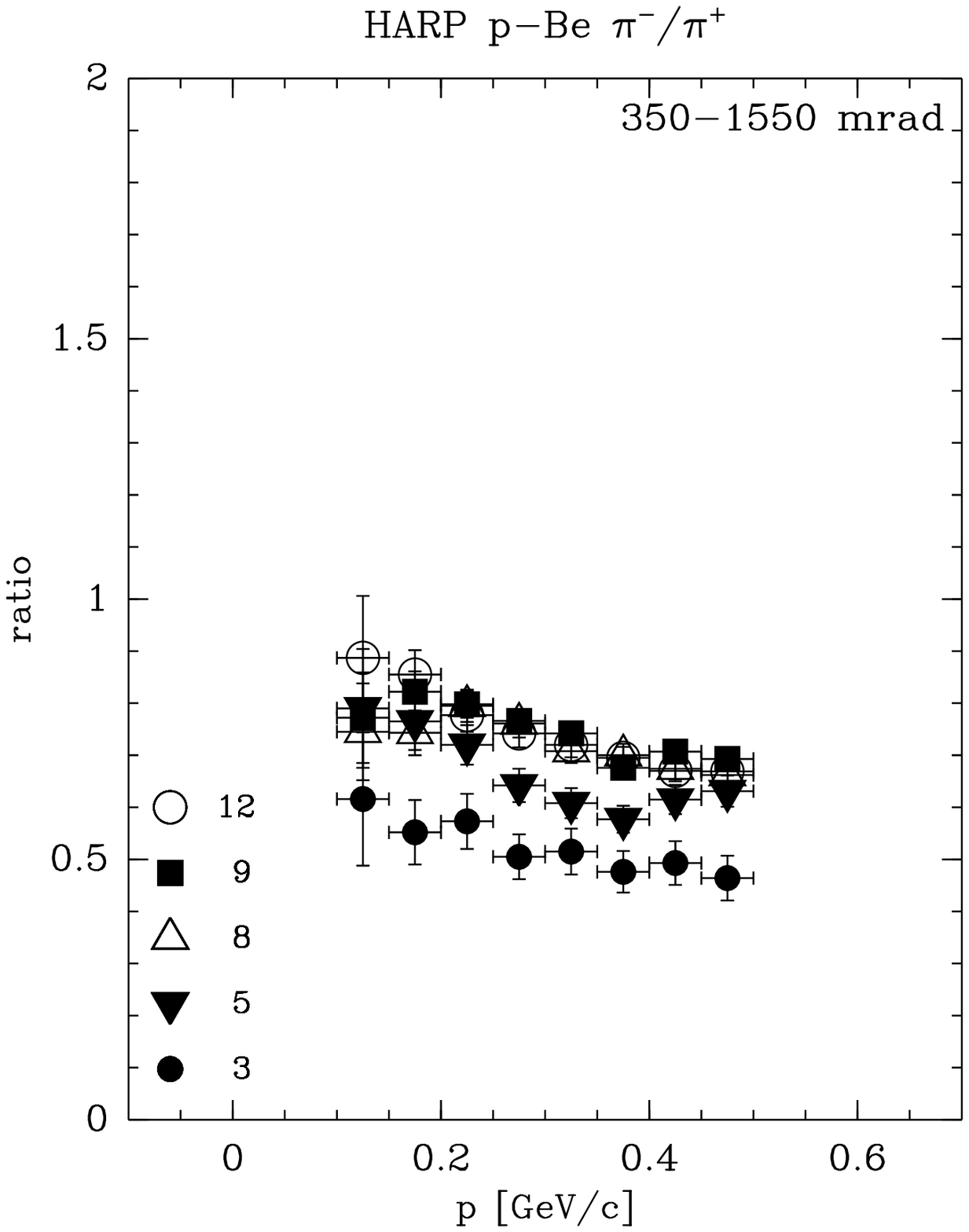,width=0.30\textwidth}
 ~ 
  \epsfig{figure=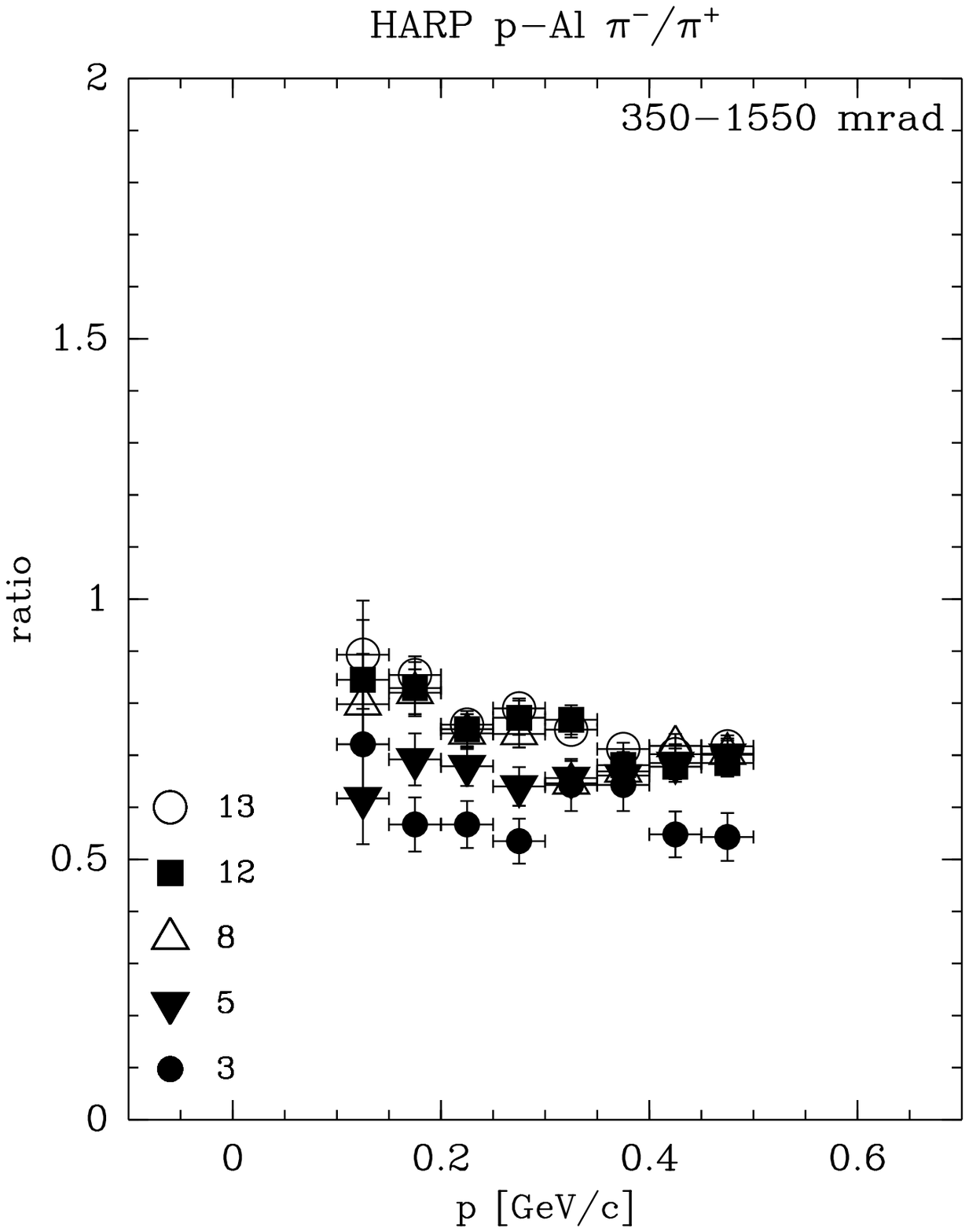,width=0.30\textwidth}
 ~
  \epsfig{figure=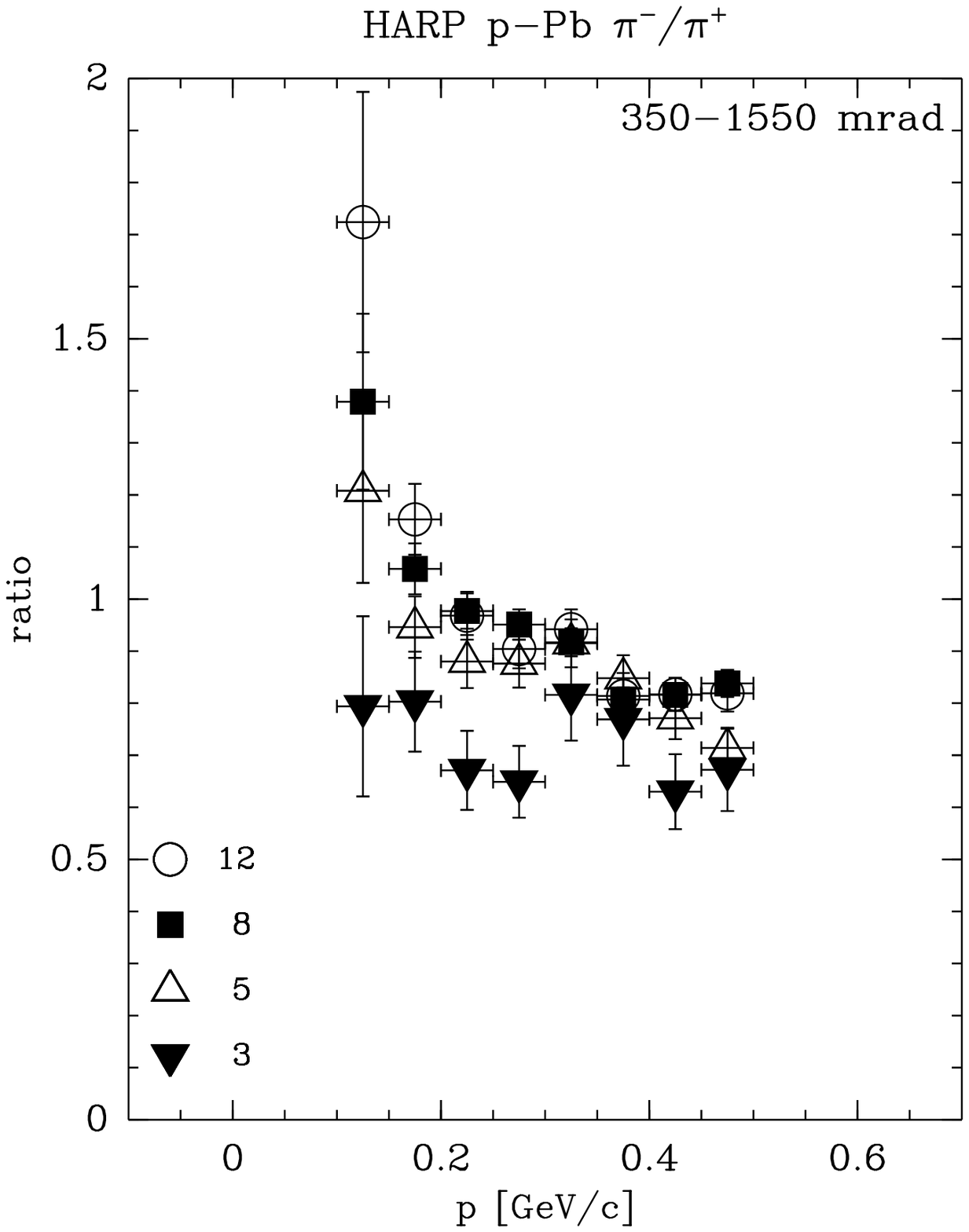,width=0.30\textwidth}
\end{center}
\caption{
The ratio of the differential cross-sections for \pim and \pip
 production in 
p--Be (left panel), p--Al (middle panel)  and p--Pb (right panel)  interactions as a function of secondary momentum integrated over the
forward angular region (shown in mrad).
In the figure, the symbol legend 13 (9) refers to 12.9 (8.9) \GeVc nominal
beam momentum.
%The results are given for four incident beam momenta (filled triangles:
%3~\GeVc; open triangles: 5~\GeVc; filled rectangles: 8~\GeVc; open
%circles: 12~\GeVc). 
}
\label{fig:xs-ratio}
\end{figure}

The integrated \pim/\pip ratio in the forward direction is displayed in
Fig.~\ref{fig:xs-ratio} as a function of secondary momentum. 
In the covered  part of the momentum range in most bins more \pip's are 
produced than \pim's.
The \pim/\pip ratio 
has features similar to the ones observed in our p--C~\cite{ref:harp:cacotin}
 and p--Ta data~\cite{ref:harp:tantalum}.
In the p--Pb data the ratio is closer to unity than 
for the p--Be, p--Al data.
In the lead data a similar effect is observed as in the previously
published tantalum data, namely that the number of \pip's produced 
is smaller than the number of \pim's in the lowest momentum bin
(100~\MeVc--150~\MeVc) for the 8~\GeVc and 12~\GeVc incoming beam
momenta. 
A similar effect was seen by E910 in their p--Au data~\cite{ref:E910}. 
Lower-$A$ targets do not show this behaviour.

\begin{figure}[tbp]
\begin{center}
  \epsfig{figure=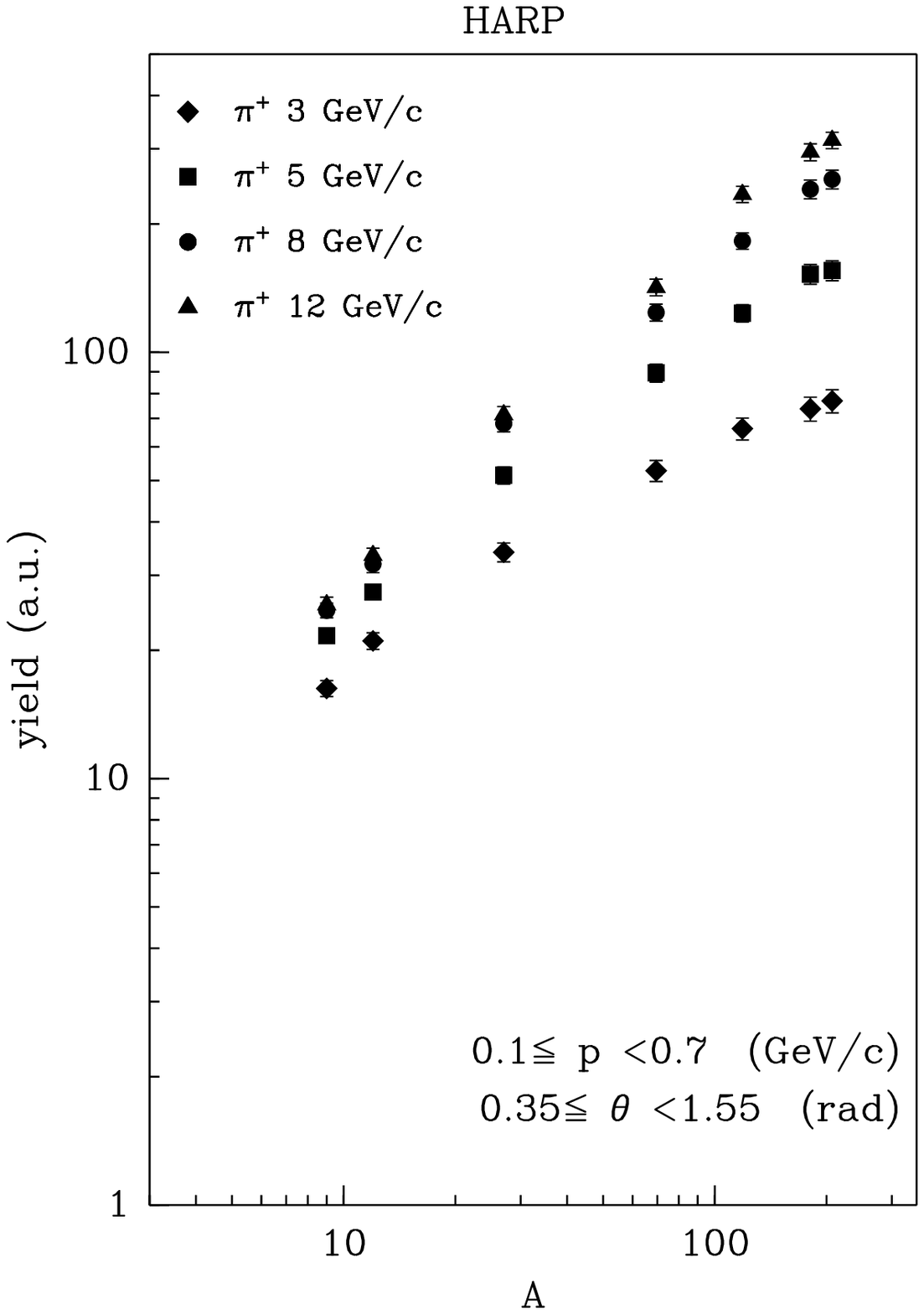,width=0.425\textwidth}
 ~
  \epsfig{figure=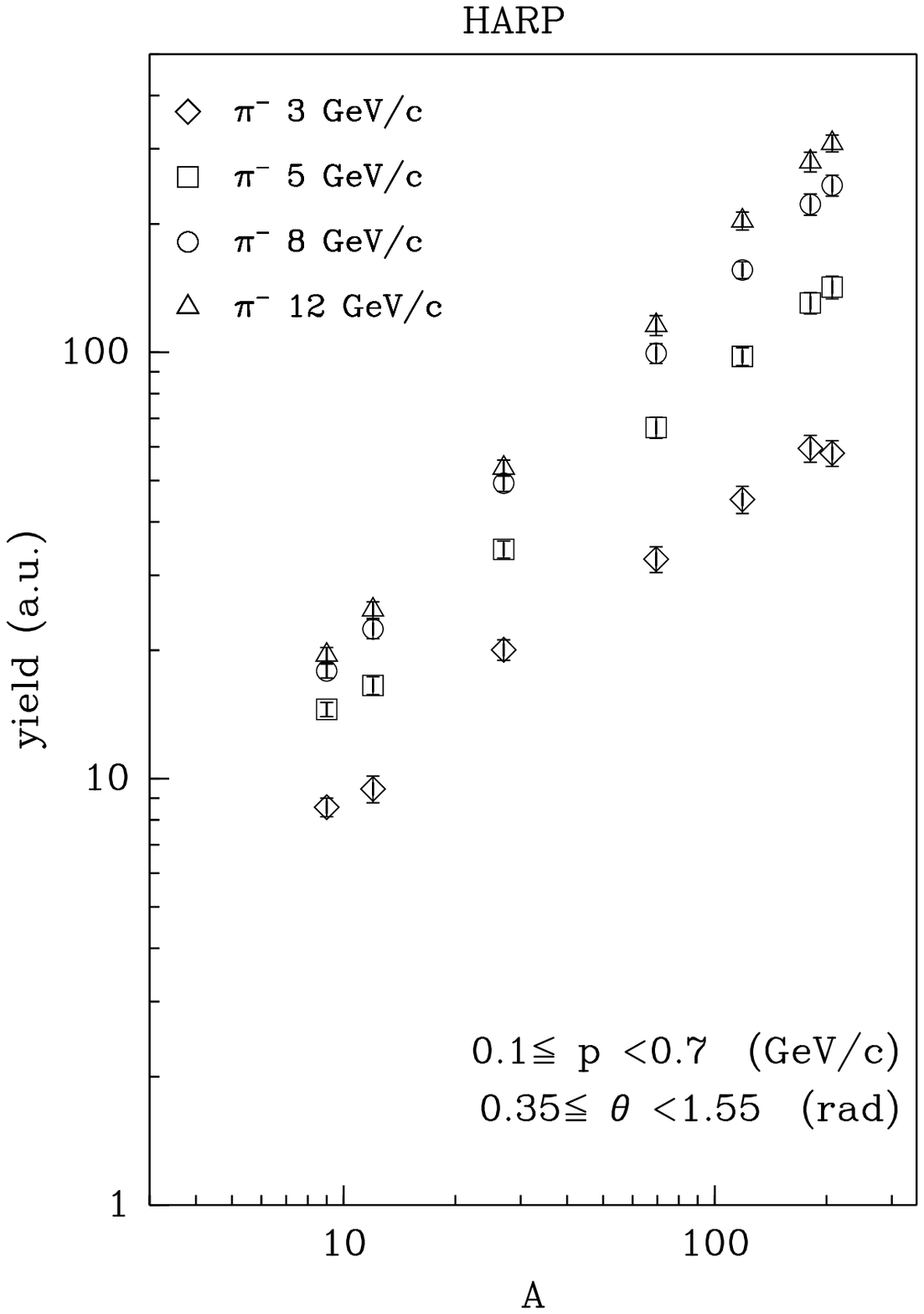,width=0.425\textwidth}
\end{center}
\caption{
 The dependence on the atomic number $A$ of the pion production yields
 in p--Be, p--C, p--Al, p--Cu, p--Sn, p--Ta, p--Pb
 interactions integrated over the forward angular region 
 ($0.350~\rad \leq \theta < 1.550~\rad$) and momentum ($100~\MeV/c \leq p < 700~\MeV/c$).
 The results are given in arbitrary units, with a consistent scale
 between the left and right panel.
 The vertical scale used in this figure is consistent with the one in
 Fig.~\ref{fig:xs-trend}. 
}
\label{fig:xs-a-dep}
\end{figure}

The dependence of the integrated pion yields on the atomic number $A$ is
shown in Fig.~\ref{fig:xs-a-dep} combining the results with the p--Ta 
data (Ref.~\cite{ref:harp:tantalum}), 
the p--C data  
and the p--Cu, p--Sn data (Ref.~\cite{ref:harp:cacotin})
taken with the same apparatus and analysed
using the same methods.  
The \pip yields integrated over the region 
$0.350~\rad \leq \theta < 1.550~\rad$ and $100~\MeVc \leq p < 700~\MeVc$ are
shown in the left panel and the \pim data integrated over the same region
in the right panel for four different beam momenta.
One observes a smooth behaviour of the integrated yields.
The $A$-dependence is slightly different for \pim and \pip production,
the latter saturating earlier, especially at lower beam momenta.

%
% desribe the A-dependence here.
%

The experimental uncertainties are summarized in
Table~\ref{tab:errors-3}.   
One observes that only for the 3~\GeVc beam  the statistical error is 
similar in magnitude to the systematic error, while the statistical
error is negligible for the 8~\GeVc and 12~\GeVc beam settings.
The statistical error is calculated by error propagation as part of the
unfolding procedure. 
It takes into account that the unfolding matrix is obtained from the
data themselves\footnote{The migration matrix is calculated without
prior knowledge of the cross-sections, while the unfolding procedure
determined the unfolding matrix from the migration matrix and the
distributions found in the data.} and hence contributes also to the
statistical error. 
This procedure almost doubles the statistical error, but avoids an important 
systematic error which would otherwise be introduced by assuming a
cross-section model {\em a priori} to calculate the corrections. 

The largest systematic error corresponds to the uncertainty in the
absolute momentum scale, which was estimated to be around 3\% using elastic
scattering~\cite{ref:harp:tantalum}.
At low momentum in the relatively small angle forward
direction the uncertainty in the subtraction of the electron and
positron background due to \piz production is dominant ($\sim$6\%--10\%).
This uncertainty is split between the variation in the shape of the
\piz spectrum and the normalization using the identified electrons. 
The target region definition  and
the uncertainty in the PID efficiency and background from tertiaries
(particles produced in secondary interactions)
are of similar size and are not negligible ($\sim 2-3 \%$).
Relatively small errors are introduced by the uncertainties in
the absorption correction, absolute knowledge of the angular and the
momentum resolution.
The correction for tertiaries is relatively large at low momenta and large 
angles ($\sim 3-5 \%$). 
As expected, this region is most affected by this component.
The errors are quoted for the positive pion data.
Owing to the similarity of the spectra the errors are very similar
for the negative pions.

As already mentioned above, the overall normalization has an uncertainty
of 2\% for Be and Al and 3\% for Pb, and is not reported in the table.
It is mainly due to the uncertainty in the efficiency that beam protons
counted in the normalization actually hit the target, with smaller
components from the target thickness and density and beam particle
counting procedure. 

% copper 
\begin{table}[tbp!] 
\begin{center}
\caption{Experimental uncertainties for the beryllium, aluminium and lead analyses. The numbers
  represent the uncertainty in percent of 
  the cross-section integrated over the angle and momentum region indicated. 
The overall normalization has an uncertainty
 of 2\% for Be and Al and 3\% for Pb, and is not reported in the table. 
%Systematic errors 
%  for beryllium (aluminium) at 8.9 (12.9) GeV/c are assumed similar to the ones at 8 (12) GeV/c.
} 
\label{tab:errors-3}
\vspace{2mm}
\small{
\begin{tabular}{ l l rrr | rrr | rr} \hline
\bf{p (\GeVc) }&\multicolumn{4}{c|}{0.1 -- 0.3}
                            &\multicolumn{3}{c|}{0.3 -- 0.5}
                            &\multicolumn{2}{c}{0.5 -- 0.7} \\
\hline
\bf{Angle}& &350--&950--&1550--
            &350--&950--&1550--
                       &350--&950-- \\
\bf{(\mrad)}  &  &950&1550&2150
                       &950&1550&2150
                       &950&1550 \\
%3
\hline
\bf{3 \GeVc }&&&&&&&&\\
\hline
\bf{Total syst.} & (Be)    &11.6 & 4.6  & 2.6 & 3.6 & 7.5  & 10.5  &9.6 &16.4 \\
                 & (Al)    &9.9 &  5.1  & 3.5 & 3.9 & 7.7  & 12.9  &10.0&14.4   \\                 
                 & (Pb)    &  11.7 &  6.8  &  6.3 &  3.7 &  5.9 & 6.4 & 11.0 & 14.7  \\
                          
\bf{Statistics}  & (Be)        & 4.5 & 3.8 & 4.9 & 3.5  & 5.9 &16.0 & 4.7 &13.0 \\
                 & (Al)        & 4.0 & 3.3 & 4.2 & 3.2  & 5.1 & 11.3&   4.2&   9.9  \\
                 & (Pb)        & 5.8      &  4.6     &  5.5     & 4.8       &  6.8     & 14.1      &   6.7    & 13.4  \\ 
%5
\hline
\bf{5 \GeVc }&&&&&&&&\\
\hline
\bf{Total syst.}  & (Be)   &11.0  & 4.7  & 2.7  & 3.9  & 4.8  & 7.8  & 6.5  &11.4 \\
                  & (Al)   &9.9   & 5.0  & 3.5  & 3.9  & 5.0  & 6.8  & 8.5  &12.2 \\    
                  & (Pb)   & 12.4       & 7.4       & 6.3       & 4.0       & 5.2       &  6.8      &  6.8      & 11.3 \\ 
\bf{Statistics        } & (Be)  & 2.2  & 1.8  & 2.3  & 1.4  & 2.2  & 4.5  & 1.7  & 3.5  \\
                        & (Al)  & 2.7  & 2.4  & 3.0  & 2.1  & 3.2  & 6.3  & 2.5  & 5.2  \\
                        & (Pb)  &  2.8      &  2.3      &    2.8    &   2.2     &  2.9      & 5.2       & 2.7       &  4.6       \\ 
%8
\hline
\bf{8 \GeVc }&&&&&&&&\\
\hline
\bf{Total syst.}    & (Be) &10.8  & 5.2  & 3.1  & 5.1  & 3.9  & 6.8  & 6.7  &11.3 \\
                    & (Al) & 9.6   &5.2   &3.6  & 4.1  & 4.3  & 7.2  & 7.3  &11.4  \\
                    & (Pb) &  11.1      &    7.4    &  6.4      &   4.2     &  4.7      &  6.6      &  6.6      & 9.3      \\ 
\bf{Statistics}     & (Be)      & 1.7  & 1.4  & 1.8  & 1.1  & 1.7  & 3.2  & 1.2 & 2.5 \\
                    & (Al)      & 1.6  & 1.5  & 1.9  & 1.2  & 1.9  & 3.6  & 1.4 &  2.7 \\
                    & (Pb)      &   1.5     &   1.3     &   1.6     &   1.1     &   1.6     &  2.7      &  1.3      & 2.2 \\          
\hline
\bf{8.9 \GeVc }&&&&&&&&\\
\hline
\bf{Total syst.}   & (Be) & 8.8 &  4.4 &  2.7 &  4.5 &  4.0 &  7.8 &  7.4 & 12.3 \\
\bf{Statistics}    & (Be)  & 1.1 &  1.0 &  1.3 &  0.8 &  1.2 &  2.4 &  0.9 &  1.8 \\
%12
\hline
\bf{12 \GeVc }&&&&&&&&\\
\hline
\bf{Total syst.}   & (Be) &10.7 & 5.1  & 2.9  & 4.3 & 3.8  & 6.7  & 6.6  &11.1 \\
                   & (Al) &9.8  & 5.8  & 4.1  & 3.9 &  4.5 &  6.8 &  7.5 & 10.8  \\
                   & (Pb) & 10.8      &  7.1      &   6.7     &    2.9   &    4.4    &  5.6      &  7.1      &  9.1 \\  
\bf{Statistics}    & (Be)  & 1.6  & 1.4  & 1.7  & 1.0  & 1.5  & 2.9  & 1.1  & 2.2 \\
                   & (Al)  & 1.9  & 1.9  & 2.4  & 1.5  & 2.2  & 4.4  & 1.6  & 3.2  \\
                   & (Pb)  &  2.1      & 1.9       & 2.4       &  1.7      &  2.4       &  4.3     &  2.0  & 3.4 \\    
\hline
\bf{12.9 \GeVc }&&&&&&&&\\
\hline
\bf{Total syst.}   & (Al) &9.7 &  5.4 &  3.8 &  3.7 &  4.1 &  6.7 &  7.3 & 11.1 \\
\bf{Statistics}    & (Al)  & 0.7 &  0.6 &  0.8 &  0.5 &  0.8 &  1.4 &  0.6 &  1.0 \\
\end{tabular}
}
\end{center}
\end{table}

The Pb data presented in this paper are particularly relevant for the
design of the input stage of future neutrino factories.
This experiment covers the full momentum range of
interest for production angles above 0.35~\rad. 
When one defines the effective coverage of the kinematic range as the
fraction of muons transported by the input stage of a
neutrino factory design originating from decays for which the pion
production cross-section is within the kinematic range measured by the
present experiment then one evaluates this 
effective coverage to be close to 70\%~\cite{ref:fernow}, 
using a particular model for pion
production at an incoming beam momentum of 10.9~\GeVc~\cite{ref:brooks}
for the ISS input stage~\cite{ref:iss}.  

\begin{figure}[tbp]
  \begin{center}
  \epsfig{figure=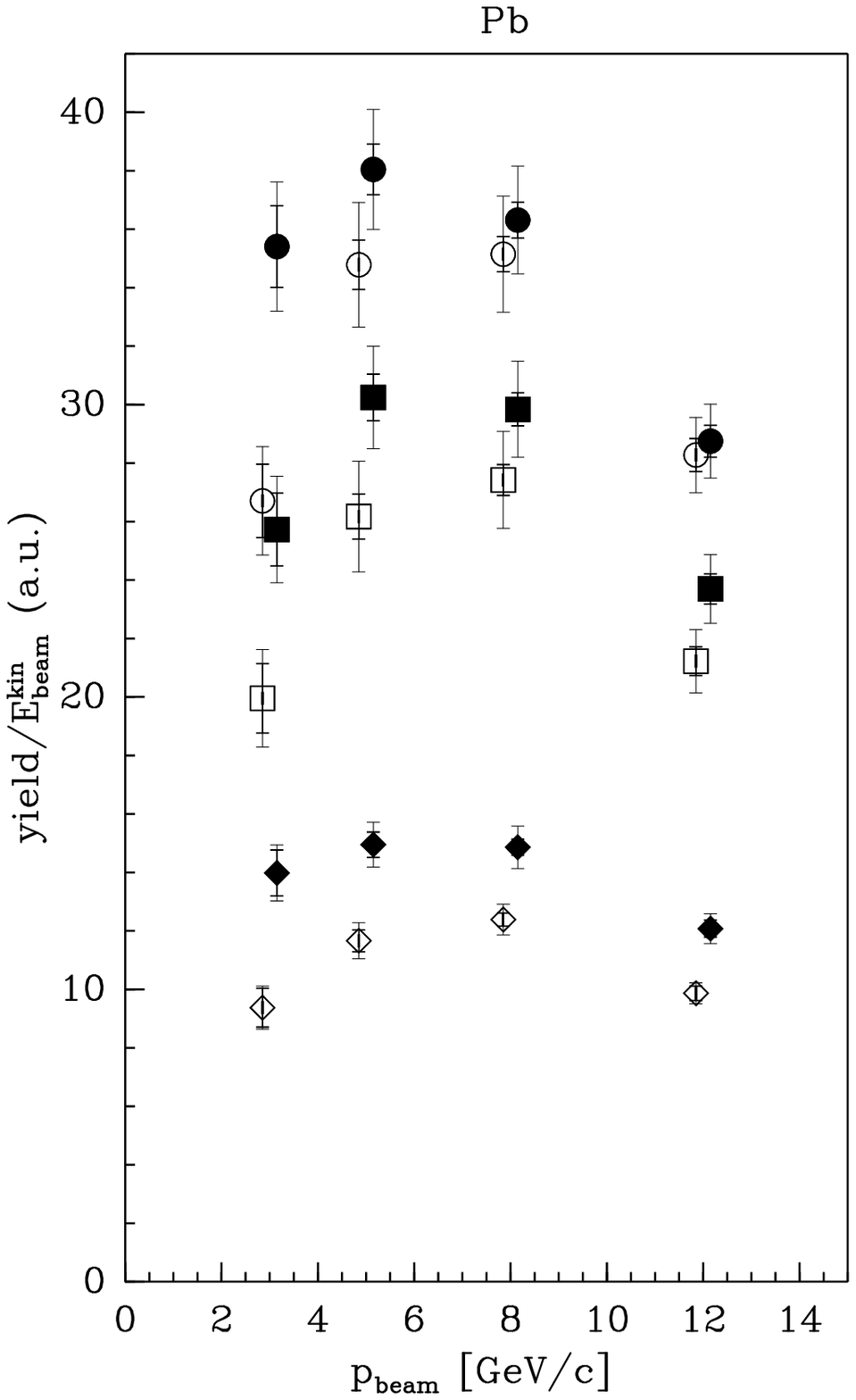,width=0.4\textwidth} 
  \epsfig{figure=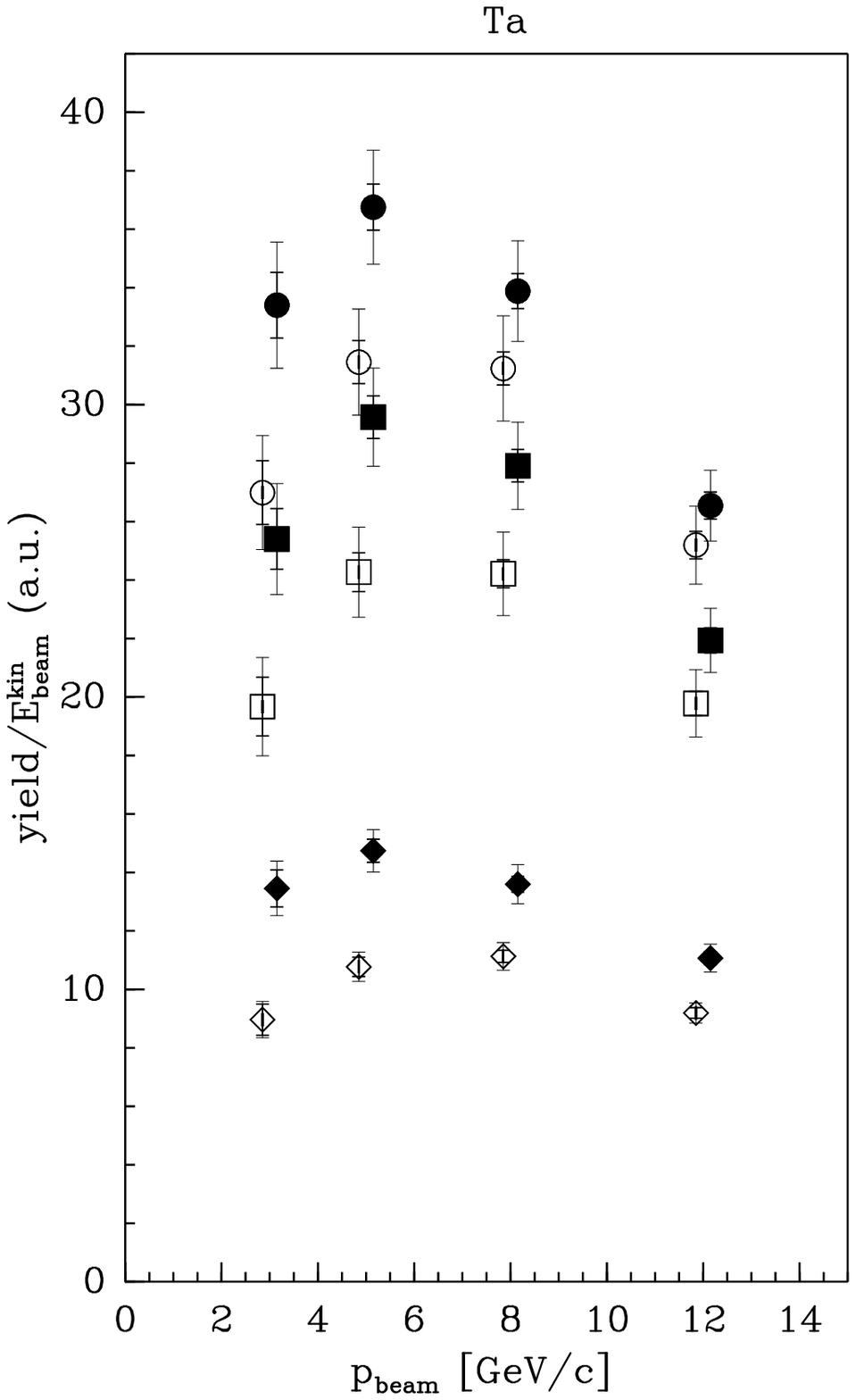,width=0.4\textwidth} 
  \end{center}
\caption{
The integrated yields
normalized to the kinetic energy of the proton of the \pip (closed
symbols) and \pim (open symbols) yield as a function of incident proton 
beam momentum. 
Shown are the yields for lead (left) and tantalum (right). 
The circles indicate the integral over the full HARP acceptance (in the
 forward direction), the
squares are integrated over $0.35 \ \rad <\theta< 0.95 \ \rad$, while the diamonds
are calculated for the restricted angular range and
$250~\MeVc<p<500~\MeVc$.
The full error bar shows the overall (systematic and statistical)
error, while the inner error bar shows the error relevant for the
point--to-point comparison.  
For the latter error only the uncorrelated systematic uncertainties
were added to the statistical error.
}
\label{fig:nufact-yield}
\end{figure}

The \pip and \pim production cross-sections were
integrated over the full HARP kinematic range in the forward hemisphere
($100~\MeVc<p<700~\MeVc$ and $0.35 <\theta< 1.55$).
The results are shown in Fig.~\ref{fig:nufact-yield}.
The integrated
yields normalized to the kinetic energy of the incoming beam particles
are shown for Pb in the left panel and compared with the Ta data in the
right panel.  
The outer error bars indicate the total statistical and systematic errors.
If one compares the \pip and \pim rates for a given beam momentum or
if one compares the rates at a different beam momentum the relative
systematic error is reduced by about a factor two.
The relative uncertainties are shown as inner error bar.
It is shown that in our kinematic coverage the optimum yield is
between 5~\GeVc and 8~\GeVc.
To show the trend the rates within restricted ranges are also given: a
restricted angular range ($0.35 <\theta< 0.95$)  and a range further
restricted in momentum ($250~\MeVc<p<500~\MeVc$).
The latter range may be most representative for the neutrino factory.
One notes that the Pb and Ta data yield the same conclusions.
Although the units are indicated as ``arbitrary'',
for the largest region, the yield is expressed as 
${{\mathrm{d}^2 \sigma}}/{{\mathrm{d}p\mathrm{d}\Omega }}$ in
mb/(\GeVc~sr). 
For the
other regions the same normalization is chosen, but now scaled with the
relative bin size to show visually the correct ratio of number of pions
produced in these kinematic regions. 

Of course this analysis only gives a simplified picture of the results.
One should note that the best result can be obtained by using the
full information of the double-differential cross-section and
by developing designs optimized specifically for each single beam
momentum. 
Then these optimized designs can be compared.

\subsection{Comparisons with earlier data}
\label{sec:compare}

\begin{figure}[btp]
\begin{center}
   \begin{minipage}[b]{0.47\textwidth}
    \begin{center}
     \epsfig{figure=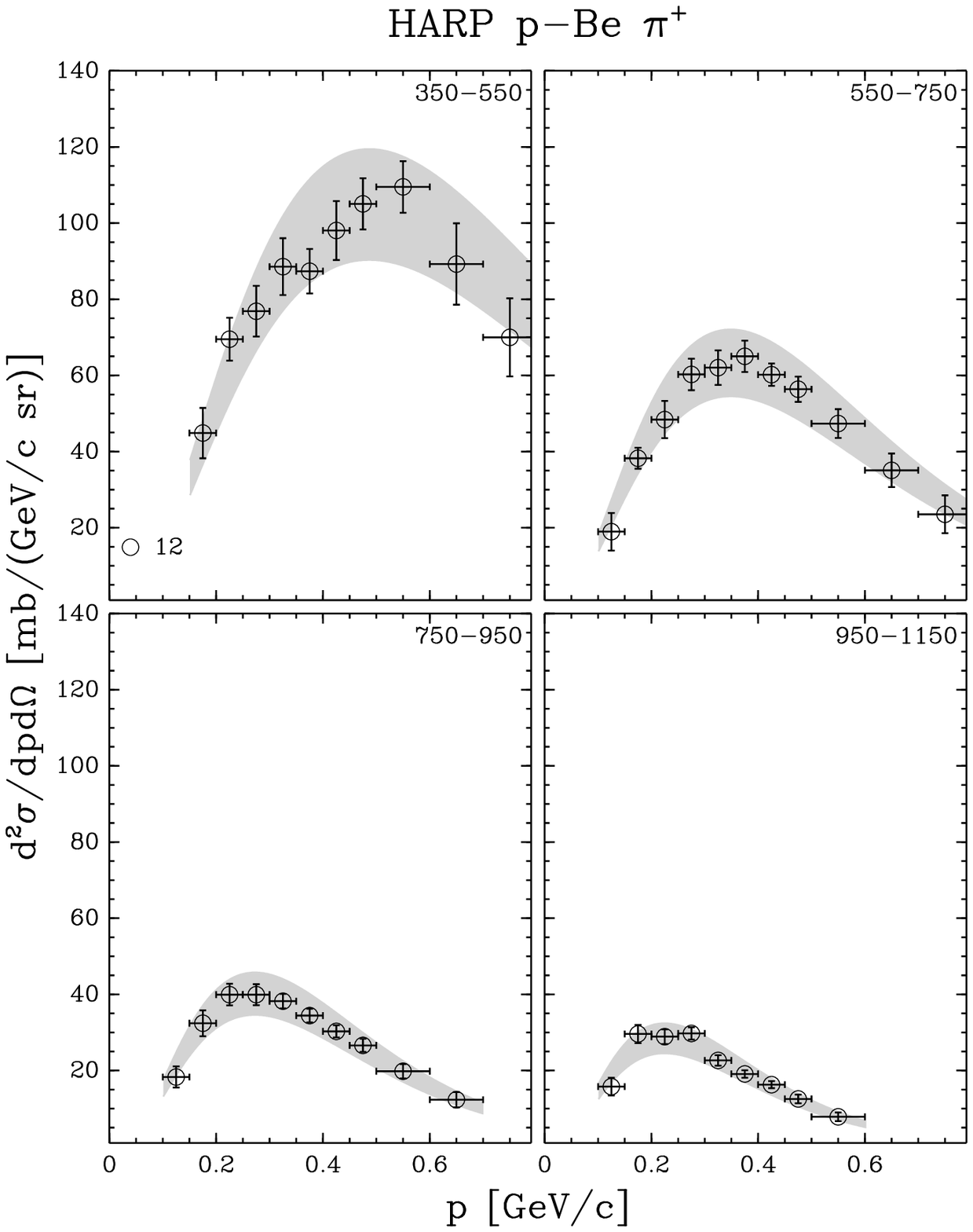,width=0.7\textwidth}
    \end{center}
   \end{minipage}
 ~
   \begin{minipage}[b]{0.47\textwidth}
    \begin{center}
     \epsfig{figure=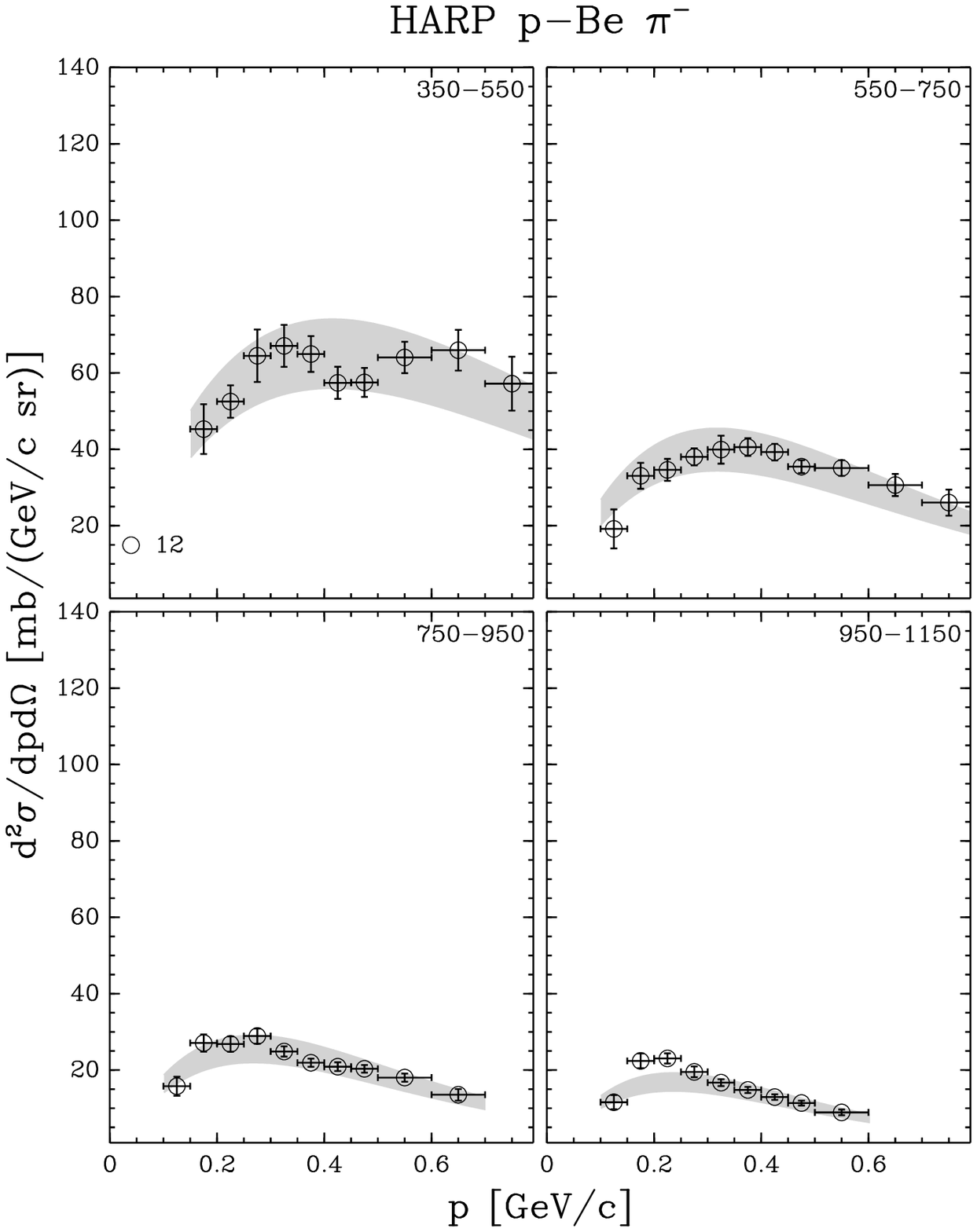,width=0.7\textwidth}
    \end{center}
   \end{minipage}

\end{center}
 \begin{center}
  \epsfig{figure=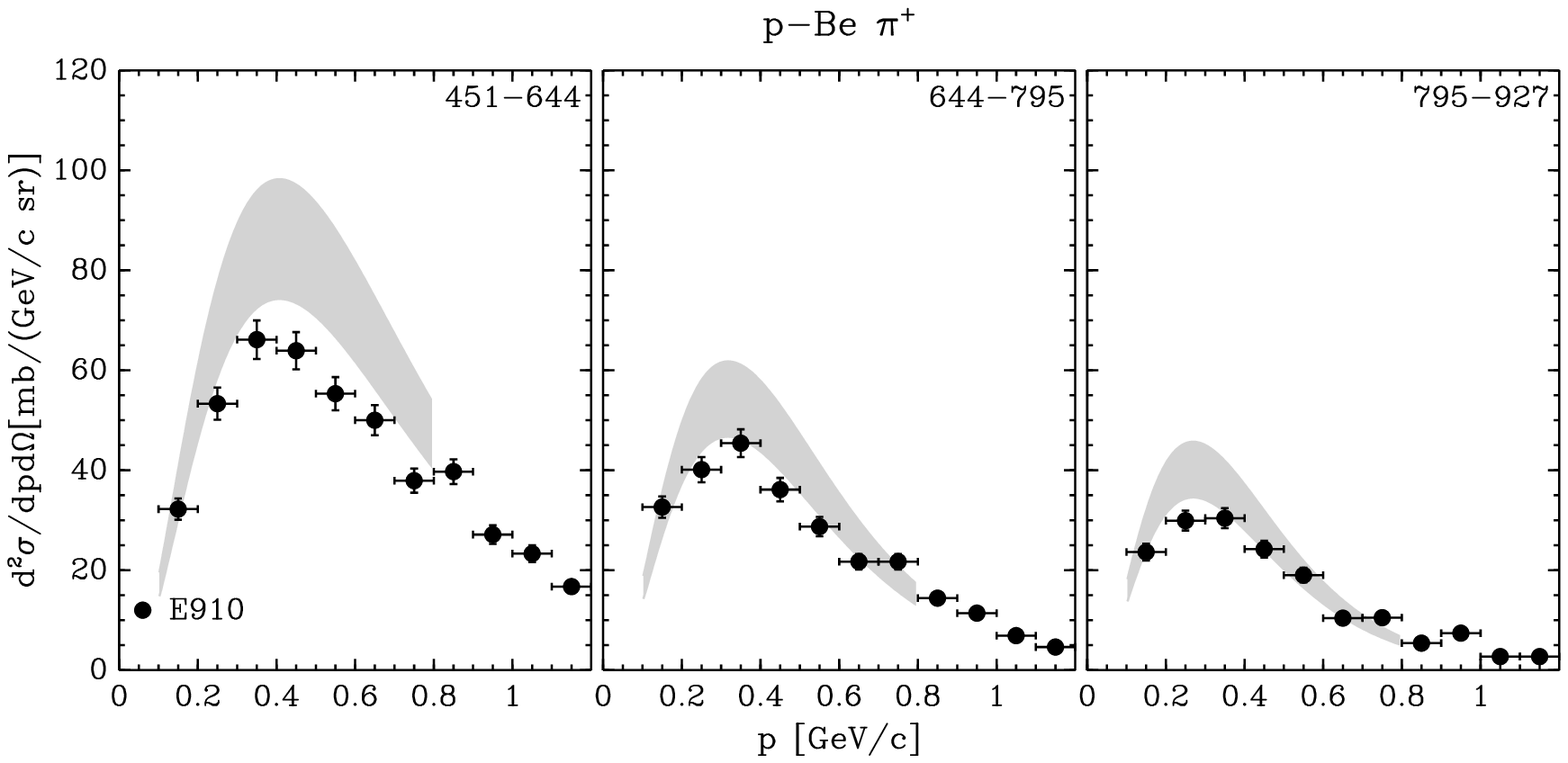,width=0.48\textwidth}
  ~
  \epsfig{figure=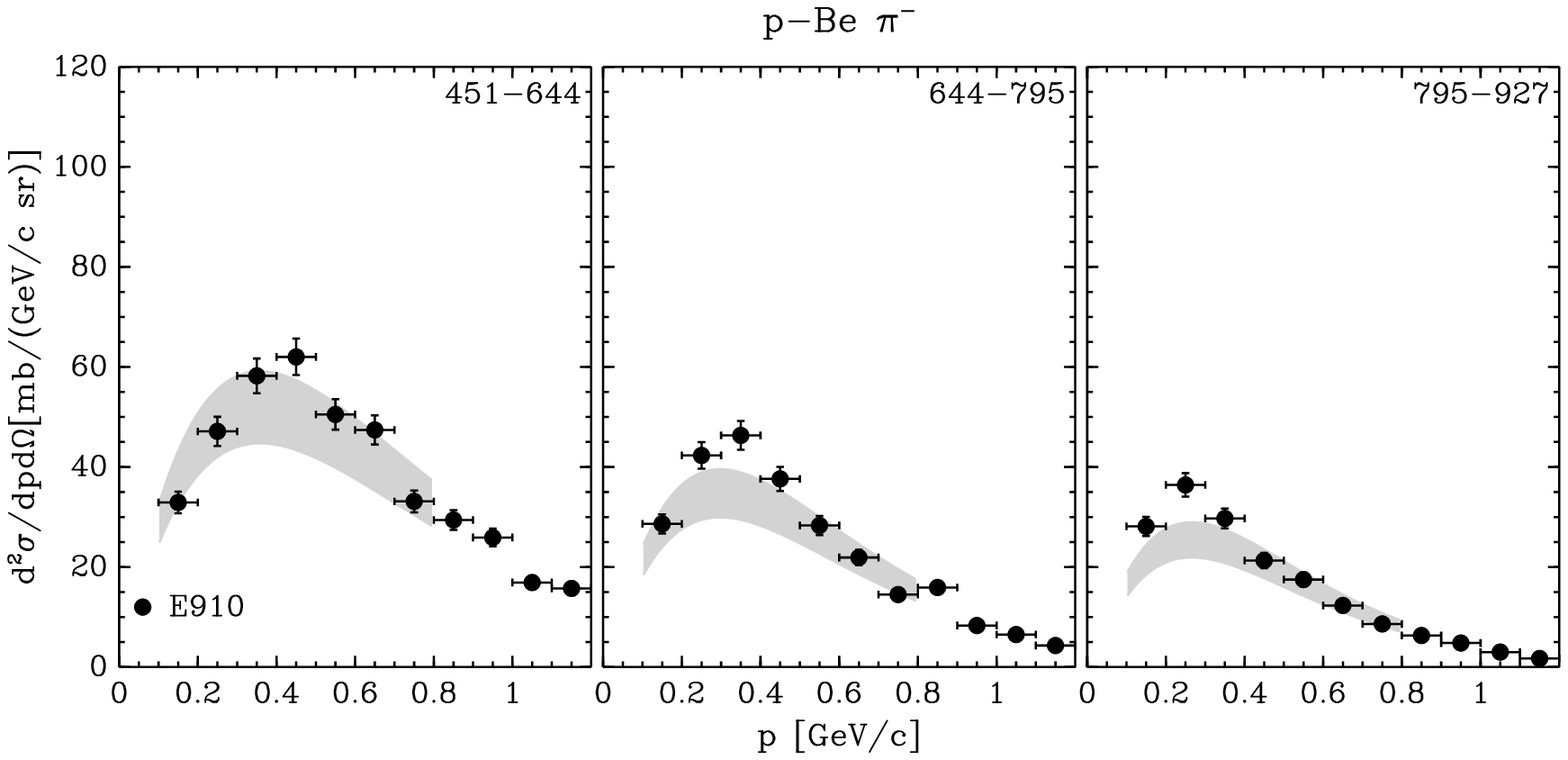,width=0.48\textwidth}
 \end{center}
\caption{
Comparison of the HARP Be data with \pip and \pim production data from
 Ref.~\cite{ref:E910} taken with 12.3~\GeVc protons.
The top panels show a parametrization of the \pip (left) and \pim(right)
 production  data described in this paper.
The data have been normalized to represent
$
{{\mathrm{d}^2 \sigma^{\pi}}}/{{\mathrm{d}p\mathrm{d}\Omega }} \ .
$
The shaded band represents the area between two parametrization which
 contain the data points.
The bottom panels show the comparison of the same parametrization,
 now binned according to the E910 data.  The bottom left (right) panel shows the
 \pip (\pim) production  data of Ref.~\cite{ref:E910}.
The angular regions are indicated in \mrad in the upper right-hand corner of each plot.
}
\label{fig:e910}
\end{figure}

Available data to be compared with are very scarce and in general suffer from large 
systematic and statistical uncertainties, 
except for the following two examples.

The p--Be data at 12.3 GeV/c from the E910 experiment \cite{ref:E910} are in
reasonable agreement with our results as shown in
Fig.~\ref{fig:e910}.
In order to take into account the different angular binnings which
prevent a direct comparison, a Sanford-Wang parametrization \cite{ref:SW} is
fitted to our data. 
The fit is performed to the data redefined as
$
{{\mathrm{d}^2 \sigma^{\pi}}}/{{\mathrm{d}p\mathrm{d}\Omega }} .
$
As the Sanford-Wang parametrization does not fit perfectly
our data, a $\pm 15 \%$ band which contains fully our 
experimental data points has been chosen for the comparison as shown in
Fig.~\ref{fig:e910} (top panels). 
%An area between two parametrizations is defined which contains our data
%points. 
The same parametrizations are then displayed in the binning of E910.
While the shape of the distributions are similar for both \pip and \pim
in the HARP and E910 data sets, the absolute cross-sections disagree by
up to 15\% for the \pip data and agree well for the \pim data. 
(The parametrization shows similar difficulties to fit both \pim
data sets). 
One should note that the range of the systematic errors of the HARP data
is 5\% to 10\% and similar for the E910 data, such that the disagreement
is not much larger than one standard deviation. 
%The effects are opposite in \pip and \pim, giving a 15\%
%difference in the \pip/\pim ratio between the two experiments. 
The difference in the \pip/\pim ratio between the two experiments is
about 15\%, which is more significant given the expected correlations
between the uncertainties in the measurements of the \pip and \pim
spectra. 
This effect may point to an underestimation of systematic effects on the
absolute normalization, efficiencies or background subtractions.
Owing to the symmetry of the HARP TPC, including its trigger counter, we
do not expect a large systematic error in the HARP data between \pip and
\pim production cross-sections\footnote{As a side remark, E910 data shows
a step around 800 MeV/c in momentum (inverse for \pip, \pim) that may 
explain also the observed discrepancy.}. 

\begin{figure}[tbp]
\epsfig{figure=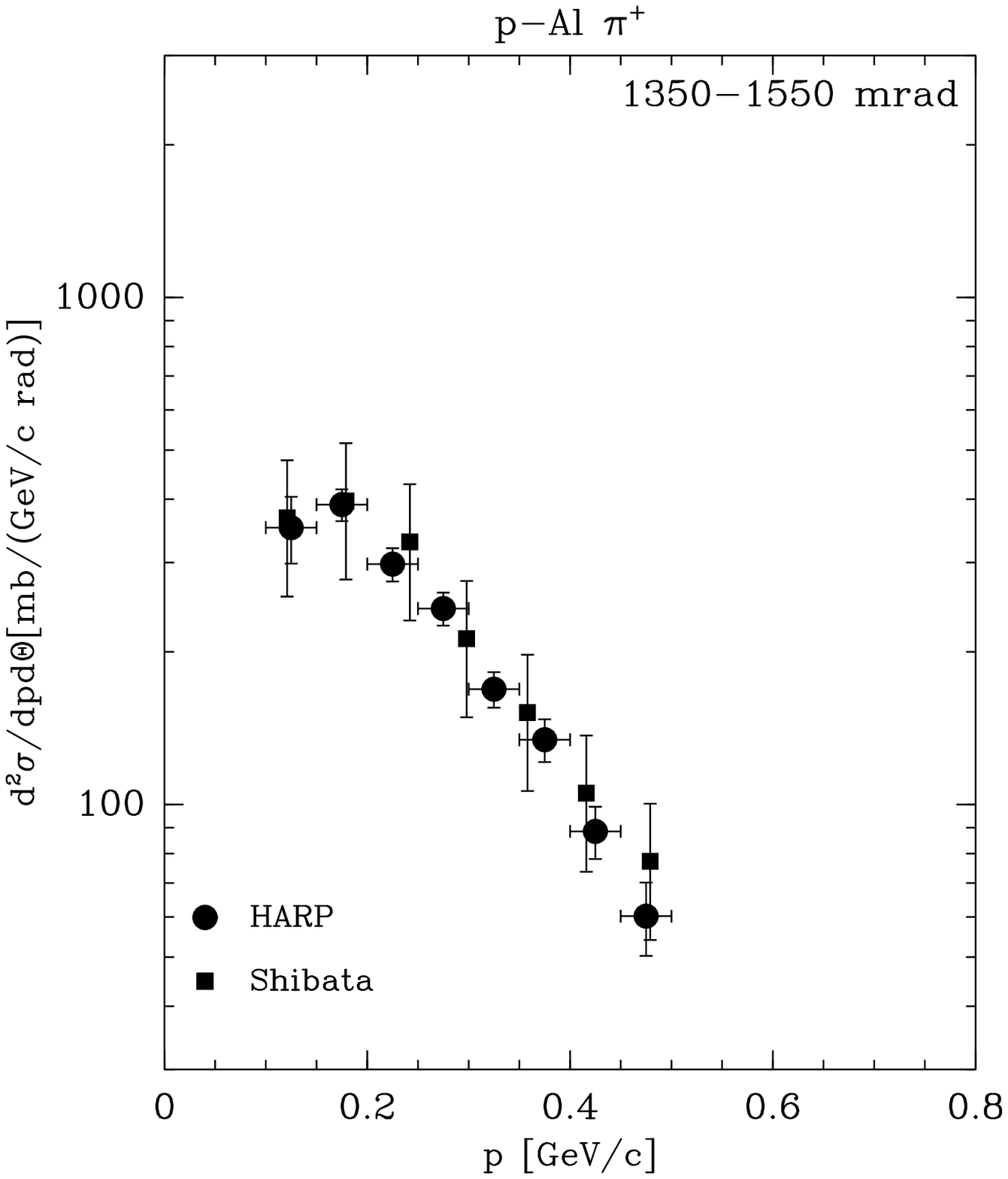,width=0.47\textwidth}
~
\epsfig{figure=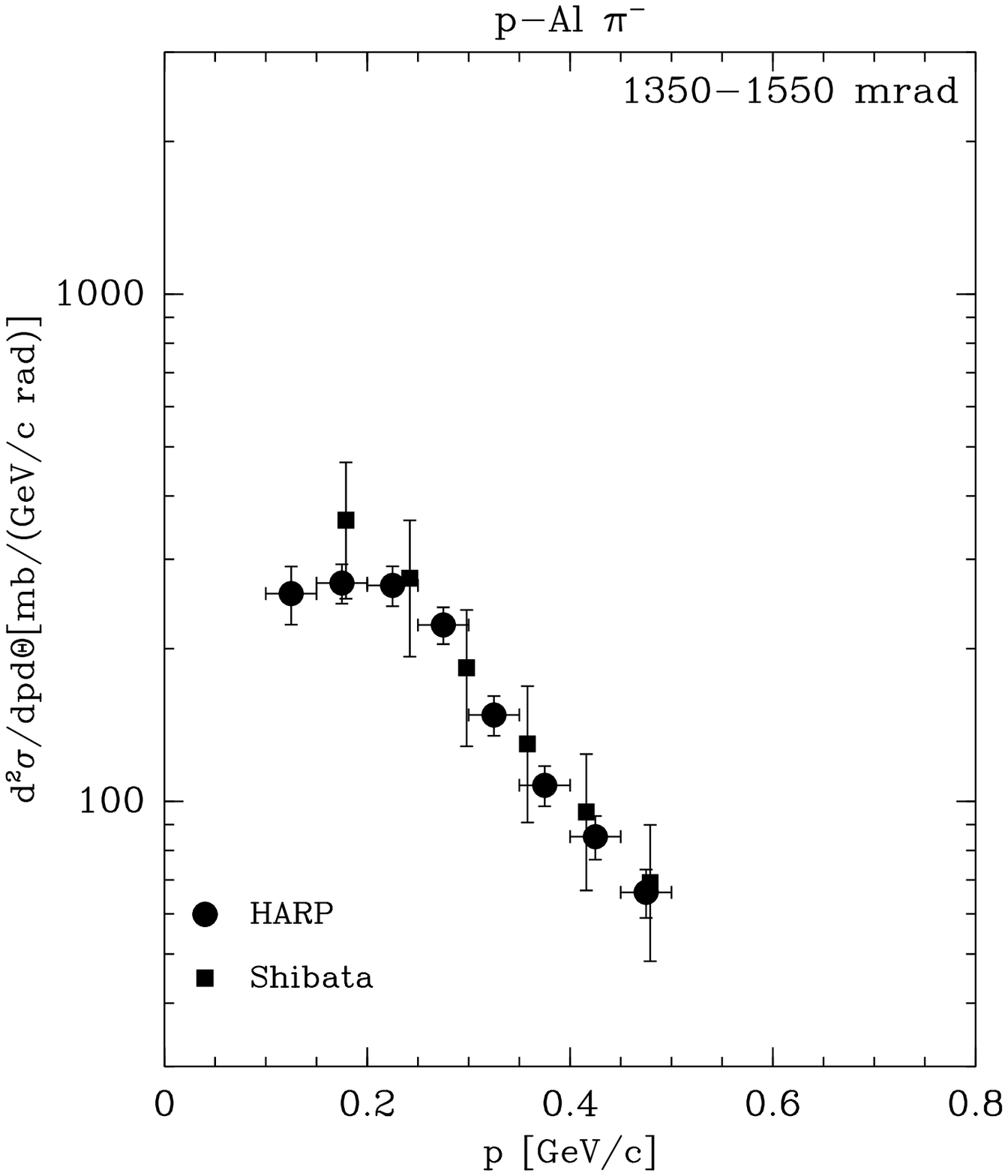,width=0.47\textwidth}
\caption{
Comparison of HARP p--Al data with \pip and \pim production data at 90
 degrees from Ref.~\cite{ref:shibata} taken with 12~\GeVc incident protons.
The left panel shows the comparison of the \pip production
 data of Ref.~\cite{ref:shibata} with the data
reported here; the right panel shows instead the comparison with the \pim production
data.
The data sets compares well with our results
(filled circles) in the angular region 1.35~$\leq \ \theta \
<$1.55~\rad.
}
\label{fig:shibata}
\end{figure}

Our p--Al data have  been compared
with \pip and \pim production measurements
at 12~\GeVc incident proton momentum from Shibata {\it et al.}~\cite{ref:shibata}.
Their data were taken with a magnetic spectrometer and only measurements
at 90 degrees from the initial proton direction are available.
The statistical point--to--point errors are quoted to be 3\%, while the
overall normalization has a 30\% uncertainty due to the knowledge of the
acceptance.
In Fig.~\ref{fig:shibata} their data are shown together with the results
reported in this paper.
Their data set compares well with the data described in this paper
(filled circles) in the angular region 1.35~\rad~$\leq \ \theta \
<$1.55~\rad at the same proton beam momentum\footnote{In this comparison data are compared with their proper 
experimental normalization factors, while in the previous published 
comparison of our carbon and copper target data with their data sets 
normalization factors  0.72 (0.91)   were used, still compatible
with their overall quoted normalization uncertainty
of 30\%.}.
 
\section{Summary and Conclusions}
\label{sec:summary}

The analysis of the production of charged pions at
large angles with respect to the beam direction for protons of
3~\GeVc, 5~\GeVc, 8~\GeVc, 8.9~\GeVc (Be only), 12~\GeVc  and 
12.9~\GeVc (Al only) impinging on  thin
(5\% nuclear interaction length) beryllium, aluminium  and lead  targets is presented. 
The secondary pion yields are measured in a large angular and momentum
range and double-differential cross-sections are obtained.
A detailed error estimation is discussed.
Results on the dependence of charged pion production on the target atomic number A
are also presented.

The use of a single detector for a range of beam momenta makes 
possible to measure the dependence of the pion yield on the secondary
particle momentum and emission angle $\theta$ with high precision.
The $A$ dependence of the cross-section can be studied using the
combination of the present data with the data obtained with 
carbon, copper, tin~\cite{ref:harp:cacotin} 
and tantalum~\cite{ref:harp:tantalum} targets in the same experiment. 
The yields integrated over relatively large angular and momentum regions
show a smooth trend in their $A$ and beam-momentum dependence. 

The data taken with the lead target is important for the design studies
for a neutrino factory.
These data show a similar behaviour as the previously reported data with
a tantalum target.

%Very few pion production measurements in this energy range are reported
%in the literature.
%The only comparable results found in the literature agrees with the
%analysis described in this paper.
%Hadronic production models describing this energy range can now be
%compared with our new results and, if needed, improved.
%Data taken with different target materials and beam momenta
%will be presented in subsequent papers.

\section{Acknowledgements}

We gratefully acknowledge the help and support of the PS beam staff
and of the numerous technical collaborators who contributed to the
detector design, construction, commissioning and operation.  
In particular, we would like to thank
G.~Barichello,
R.~Brocard,
K.~Burin,
V.~Carassiti,
F.~Chignoli,
D.~Conventi,
G.~Decreuse,
M.~Delattre,
C.~Detraz,  
A.~Domeniconi,
M.~Dwuznik,   
F.~Evangelisti,
B.~Friend,
A.~Iaciofano,
I.~Krasin, 
D.~Lacroix,
J.-C.~Legrand,
M.~Lobello, 
M.~Lollo,
J.~Loquet,
F.~Marinilli,
J.~Mulon,
L.~Musa,
R.~Nicholson,
A.~Pepato,
P.~Petev, 
X.~Pons,
I.~Rusinov,
M.~Scandurra,
E.~Usenko,
and
R.~van der Vlugt,
for their support in the construction of the detector.
The collaboration acknowledges the major contributions and advice of
M.~Baldo-Ceolin, 
L.~Linssen, 
M.T.~Muciaccia and A. Pullia
during the construction of the experiment.
The collaboration is indebted to 
V.~Ableev,
P.~Arce,   %DR
%%%M.~Baldo~Ceolin,
F.~Bergsma,
P.~Binko,
E.~Boter,
C.~Buttar,  %DR
M.~Calvi, 
M.~Campanelli, %DR
C.~Cavion, 
A.~Chukanov, 
A.~De~Min,    %DR
M.~Doucet,
D.~D\"{u}llmann,
R.~Engel,   %DR
V.~Ermilova, 
W.~Flegel,
P.~Gruber,   %DR
Y.~Hayato,
P.~Hodgson,  %DR
A.~Ichikawa,
A.~Ivanchenko,
I.~Kato,  %DR
O.~Klimov,
T.~Kobayashi,
%S.~Kotov,
D.~Kustov,
M.~Laveder,  
%%%L.~Linssen,
M.~Mass,
H.~Meinhard,
%%%M.T.~Muciaccia, 
T.~Nakaya,
K.~Nishikawa,
M.~Paganoni,     %DR
F.~Paleari,  %DR
M.~Pasquali,
J.~Pasternak,   %DR
C.~Pattison,    %DR
M.~Placentino,
%I.~Potrap, 
%%%A.~Pullia,
S.~Robbins,   %DR
G.~Santin,  %DR
S.~Simone,
A.~Tornero,   %DR
S.~Troquereau,
S.~Ueda, 
A.~Valassi,
F.~Vannucci   %DR
and
K.~Zuber   %DR
for their contributions to the experiment and to P. Dini for his
contribution to MC production.

We acknowledge the contributions of 
V.~Ammosov,
G.~Chelkov,
D.~Dedovich,
F.~Dydak,
M.~Gostkin,
A.~Guskov, 
D.~Khartchenko, 
V.~Koreshev,
Z.~Kroumchtein,
I.~Nefedov,
A.~Semak, 
J.~Wotschack,
V.~Zaets and
A.~Zhemchugov
to the work described in this paper.

 The experiment was made possible by grants from
the Institut Interuniversitaire des Sciences Nucl\'eair\-es and the
Interuniversitair Instituut voor Kernwetenschappen (Belgium), 
Ministerio de Educacion y Ciencia, Grant FPA2003-06921-c02-02 and
Generalitat Valenciana, grant GV00-054-1,
CERN (Geneva, Switzerland), 
the German Bundesministerium f\"ur Bildung und Forschung (Germany), 
the Istituto Na\-zio\-na\-le di Fisica Nucleare (Italy), 
INR RAS (Moscow) and the Particle Physics and Astronomy Research Council (UK).
We gratefully acknowledge their support.
This work was supported in part by the Swiss National Science Foundation
and the Swiss Agency for Development and Cooperation in the framework of
the programme SCOPES - Scientific co-operation between Eastern Europe
and Switzerland.

\clearpage

\begin{appendix}

\section{Cross-section data}\label{app:data}
\begin{table}[hp!] 
\begin{center}
  \caption{\label{tab:xsec-p-be}
    HARP results for the double-differential $\pi^+$ production
    cross-section in the laboratory system,
    $d^2\sigma^{\pi^+}/(dpd\theta)$ for beryllium. Each row refers to a
    different $(p_{\hbox{\small min}} \le p<p_{\hbox{\small max}},
    \theta_{\hbox{\small min}} \le \theta<\theta_{\hbox{\small max}})$ bin,
    where $p$ and $\theta$ are the pion momentum and polar angle, respectively.
    The central value as well as the square-root of the diagonal elements
    of the covariance matrix are given.
 The overall normalization has an uncertainty
 of 2\%, and is not reported in the table.
}
\vspace{2mm}
\small{
%\begin{tabular}{rrrr|r@{$\pm$}lr{$\pm$}lr{$\pm$}lr{$\pm$}l} 
\begin{tabular}{rrrr|r@{$\pm$}lr@{$\pm$}lr@{$\pm$}lr@{$\pm$}lr@{$\pm$}l} 
\hline
$\theta_{\hbox{\small min}}$ &
$\theta_{\hbox{\small max}}$ &
$p_{\hbox{\small min}}$ &
$p_{\hbox{\small max}}$ &
\multicolumn{10}{c}{$d^2\sigma^{\pi^+}/(dpd\theta)$} 
\\
(rad) & (rad) & (\GeVc) & (\GeVc) &
\multicolumn{10}{c}{(barn/(\GeVc rad))}
\\
  &  &  & 
&\multicolumn{2}{c}{$ \bf{3 \ \GeVc}$} 
&\multicolumn{2}{c}{$ \bf{5 \ \GeVc}$} 
&\multicolumn{2}{c}{$ \bf{8 \ \GeVc}$} 
&\multicolumn{2}{c}{$ \bf{8.9 \ \GeVc}$} 
&\multicolumn{2}{c}{$ \bf{12 \ \GeVc}$} 
\\ 
\hline
 0.35 & 0.55 & 0.15 & 0.20& 0.063 &  0.015& 0.107 &  0.017& 0.121 &  0.016& 0.130 &  0.015& 0.126 &  0.019\\ 
      &      & 0.20 & 0.25& 0.090 &  0.015& 0.131 &  0.014& 0.161 &  0.014& 0.174 &  0.014& 0.195 &  0.016\\ 
      &      & 0.25 & 0.30& 0.104 &  0.015& 0.191 &  0.023& 0.199 &  0.018& 0.213 &  0.018& 0.215 &  0.019\\ 
      &      & 0.30 & 0.35& 0.151 &  0.025& 0.229 &  0.021& 0.236 &  0.019& 0.232 &  0.022& 0.248 &  0.021\\ 
      &      & 0.35 & 0.40& 0.178 &  0.018& 0.237 &  0.016& 0.237 &  0.016& 0.266 &  0.018& 0.245 &  0.016\\ 
      &      & 0.40 & 0.45& 0.163 &  0.016& 0.222 &  0.014& 0.254 &  0.018& 0.261 &  0.014& 0.275 &  0.022\\ 
      &      & 0.45 & 0.50& 0.187 &  0.018& 0.245 &  0.020& 0.281 &  0.014& 0.281 &  0.019& 0.294 &  0.019\\ 
      &      & 0.50 & 0.60& 0.185 &  0.018& 0.232 &  0.016& 0.274 &  0.017& 0.284 &  0.018& 0.307 &  0.019\\ 
      &      & 0.60 & 0.70& 0.117 &  0.020& 0.214 &  0.022& 0.285 &  0.026& 0.249 &  0.024& 0.250 &  0.030\\ 
      &      & 0.70 & 0.80& 0.083 &  0.016& 0.150 &  0.029& 0.208 &  0.037& 0.187 &  0.029& 0.196 &  0.029\\ 
\hline \\ 
 0.55 & 0.75 & 0.10 & 0.15& 0.050 &  0.016& 0.076 &  0.017& 0.081 &  0.018& 0.092 &  0.015& 0.074 &  0.019\\ 
      &      & 0.15 & 0.20& 0.108 &  0.016& 0.111 &  0.013& 0.145 &  0.012& 0.132 &  0.010& 0.149 &  0.011\\ 
      &      & 0.20 & 0.25& 0.126 &  0.019& 0.171 &  0.017& 0.202 &  0.014& 0.197 &  0.015& 0.189 &  0.019\\ 
      &      & 0.25 & 0.30& 0.181 &  0.020& 0.218 &  0.020& 0.210 &  0.017& 0.218 &  0.016& 0.235 &  0.016\\ 
      &      & 0.30 & 0.35& 0.167 &  0.020& 0.214 &  0.016& 0.212 &  0.014& 0.220 &  0.015& 0.242 &  0.018\\ 
      &      & 0.35 & 0.40& 0.179 &  0.017& 0.196 &  0.013& 0.217 &  0.014& 0.222 &  0.013& 0.253 &  0.016\\ 
      &      & 0.40 & 0.45& 0.156 &  0.015& 0.183 &  0.013& 0.224 &  0.012& 0.209 &  0.010& 0.235 &  0.011\\ 
      &      & 0.45 & 0.50& 0.139 &  0.013& 0.175 &  0.011& 0.206 &  0.011& 0.211 &  0.011& 0.220 &  0.013\\ 
      &      & 0.50 & 0.60& 0.103 &  0.013& 0.149 &  0.013& 0.185 &  0.013& 0.179 &  0.014& 0.185 &  0.015\\ 
      &      & 0.60 & 0.70& 0.050 &  0.014& 0.102 &  0.016& 0.130 &  0.018& 0.131 &  0.017& 0.137 &  0.017\\ 
      &      & 0.70 & 0.80& 0.023 &  0.006& 0.057 &  0.013& 0.079 &  0.018& 0.086 &  0.016& 0.092 &  0.019\\ 
\hline \\ 
 0.75 & 0.95 & 0.10 & 0.15& 0.079 &  0.017& 0.067 &  0.013& 0.079 &  0.013& 0.087 &  0.013& 0.089 &  0.013\\ 
      &      & 0.15 & 0.20& 0.121 &  0.017& 0.143 &  0.015& 0.162 &  0.012& 0.157 &  0.012& 0.157 &  0.016\\ 
      &      & 0.20 & 0.25& 0.136 &  0.017& 0.173 &  0.016& 0.193 &  0.015& 0.191 &  0.013& 0.193 &  0.014\\ 
      &      & 0.25 & 0.30& 0.125 &  0.014& 0.167 &  0.012& 0.188 &  0.012& 0.197 &  0.013& 0.193 &  0.013\\ 
      &      & 0.30 & 0.35& 0.120 &  0.014& 0.146 &  0.012& 0.175 &  0.011& 0.192 &  0.012& 0.185 &  0.009\\ 
      &      & 0.35 & 0.40& 0.103 &  0.012& 0.150 &  0.011& 0.171 &  0.009& 0.175 &  0.008& 0.167 &  0.009\\ 
      &      & 0.40 & 0.45& 0.078 &  0.009& 0.129 &  0.010& 0.146 &  0.009& 0.137 &  0.007& 0.147 &  0.008\\ 
      &      & 0.45 & 0.50& 0.072 &  0.009& 0.096 &  0.010& 0.126 &  0.007& 0.121 &  0.006& 0.129 &  0.008\\ 
      &      & 0.50 & 0.60& 0.044 &  0.008& 0.065 &  0.008& 0.097 &  0.009& 0.095 &  0.008& 0.096 &  0.009\\ 
      &      & 0.60 & 0.70& 0.021 &  0.006& 0.039 &  0.007& 0.059 &  0.010& 0.058 &  0.010& 0.060 &  0.010\\ 
\hline \\ 
 0.95 & 1.15 & 0.10 & 0.15& 0.105 &  0.017& 0.079 &  0.014& 0.093 &  0.014& 0.089 &  0.012& 0.088 &  0.013\\ 
      &      & 0.15 & 0.20& 0.127 &  0.018& 0.154 &  0.015& 0.159 &  0.011& 0.166 &  0.011& 0.165 &  0.013\\ 
      &      & 0.20 & 0.25& 0.165 &  0.018& 0.167 &  0.013& 0.151 &  0.010& 0.168 &  0.008& 0.161 &  0.011\\ 
      &      & 0.25 & 0.30& 0.119 &  0.014& 0.163 &  0.012& 0.158 &  0.009& 0.155 &  0.009& 0.166 &  0.009\\ 
      &      & 0.30 & 0.35& 0.115 &  0.013& 0.106 &  0.009& 0.131 &  0.008& 0.122 &  0.007& 0.127 &  0.008\\ 
      &      & 0.35 & 0.40& 0.074 &  0.010& 0.086 &  0.007& 0.107 &  0.006& 0.107 &  0.006& 0.107 &  0.006\\ 
      &      & 0.40 & 0.45& 0.054 &  0.008& 0.067 &  0.006& 0.085 &  0.006& 0.088 &  0.004& 0.091 &  0.005\\ 
      &      & 0.45 & 0.50& 0.034 &  0.006& 0.051 &  0.005& 0.066 &  0.006& 0.068 &  0.006& 0.070 &  0.006\\ 
      &      & 0.50 & 0.60& 0.018 &  0.004& 0.032 &  0.005& 0.042 &  0.006& 0.044 &  0.006& 0.044 &  0.006\\ 
%%%%%%%%%%%%%%%%%%%%%%%%%%%%%%%%%%%%%%%%%%%%%%%%%%%%%%%%%%%%%%%%%%%%%%%%%%%%%
\hline
\end{tabular}
}
\end{center}
\end{table}

\begin{table}[hp!] 
\begin{center}
\small{
\begin{tabular}{rrrr|r@{$\pm$}lr@{$\pm$}lr@{$\pm$}lr@{$\pm$}lr@{$\pm$}l} 
\hline
$\theta_{\hbox{\small min}}$ &
$\theta_{\hbox{\small max}}$ &
$p_{\hbox{\small min}}$ &
$p_{\hbox{\small max}}$ &
\multicolumn{10}{c}{$d^2\sigma^{\pi^+}/(dpd\theta)$} 
\\
(rad) & (rad) & (\GeVc) & (\GeVc) &
\multicolumn{10}{c}{(barn/(\GeVc rad))}
\\
  &  &  & 
&\multicolumn{2}{c}{$ \bf{3 \ \GeVc}$} 
&\multicolumn{2}{c}{$ \bf{5 \ \GeVc}$} 
&\multicolumn{2}{c}{$ \bf{8 \ \GeVc}$} 
&\multicolumn{2}{c}{$ \bf{8.9 \ \GeVc}$} 
&\multicolumn{2}{c}{$ \bf{12 \ \GeVc}$} 
\\ 
\hline
%%%%%%%%%%%%%%%%%%%%%%%%%%%%%%%%%%%%%%%%%%%%%%%%%%%%%%%%%%%%%%%%%%%%%%%%%%%%%
%\hline \\ 
 1.15 & 1.35 & 0.10 & 0.15& 0.098 &  0.016& 0.102 &  0.014& 0.094 &  0.012& 0.105 &  0.013& 0.093 &  0.014\\ 
      &      & 0.15 & 0.20& 0.114 &  0.017& 0.144 &  0.013& 0.155 &  0.011& 0.149 &  0.008& 0.136 &  0.011\\ 
      &      & 0.20 & 0.25& 0.134 &  0.016& 0.127 &  0.010& 0.141 &  0.008& 0.145 &  0.007& 0.133 &  0.008\\ 
      &      & 0.25 & 0.30& 0.084 &  0.011& 0.110 &  0.009& 0.099 &  0.007& 0.103 &  0.007& 0.116 &  0.008\\ 
      &      & 0.30 & 0.35& 0.049 &  0.009& 0.076 &  0.007& 0.076 &  0.006& 0.075 &  0.004& 0.082 &  0.005\\ 
      &      & 0.35 & 0.40& 0.026 &  0.005& 0.057 &  0.005& 0.063 &  0.004& 0.066 &  0.003& 0.072 &  0.004\\ 
      &      & 0.40 & 0.45& 0.022 &  0.004& 0.041 &  0.006& 0.048 &  0.004& 0.048 &  0.003& 0.053 &  0.005\\ 
      &      & 0.45 & 0.50& 0.017 &  0.004& 0.024 &  0.004& 0.036 &  0.004& 0.034 &  0.004& 0.037 &  0.005\\ 
\hline \\ 
 1.35 & 1.55 & 0.10 & 0.15& 0.084 &  0.015& 0.109 &  0.015& 0.086 &  0.012& 0.097 &  0.012& 0.089 &  0.013\\ 
      &      & 0.15 & 0.20& 0.107 &  0.014& 0.120 &  0.011& 0.128 &  0.011& 0.145 &  0.010& 0.137 &  0.009\\ 
      &      & 0.20 & 0.25& 0.079 &  0.012& 0.124 &  0.011& 0.117 &  0.009& 0.118 &  0.006& 0.105 &  0.007\\ 
      &      & 0.25 & 0.30& 0.070 &  0.011& 0.081 &  0.009& 0.067 &  0.007& 0.083 &  0.006& 0.081 &  0.006\\ 
      &      & 0.30 & 0.35& 0.036 &  0.007& 0.049 &  0.006& 0.050 &  0.004& 0.055 &  0.003& 0.057 &  0.005\\ 
      &      & 0.35 & 0.40& 0.023 &  0.005& 0.032 &  0.004& 0.041 &  0.003& 0.044 &  0.003& 0.045 &  0.004\\ 
      &      & 0.40 & 0.45& 0.012 &  0.004& 0.018 &  0.003& 0.030 &  0.004& 0.029 &  0.003& 0.028 &  0.004\\ 
      &      & 0.45 & 0.50& 0.007 &  0.003& 0.010 &  0.002& 0.018 &  0.003& 0.020 &  0.003& 0.018 &  0.003\\ 
\hline \\ 
 1.55 & 1.75 & 0.10 & 0.15& 0.067 &  0.015& 0.088 &  0.014& 0.080 &  0.012& 0.084 &  0.011& 0.088 &  0.011\\ 
      &      & 0.15 & 0.20& 0.104 &  0.014& 0.130 &  0.011& 0.115 &  0.008& 0.114 &  0.006& 0.103 &  0.007\\ 
      &      & 0.20 & 0.25& 0.058 &  0.010& 0.073 &  0.009& 0.079 &  0.006& 0.082 &  0.005& 0.079 &  0.006\\ 
      &      & 0.25 & 0.30& 0.040 &  0.008& 0.044 &  0.005& 0.062 &  0.005& 0.060 &  0.004& 0.051 &  0.005\\ 
      &      & 0.30 & 0.35& 0.023 &  0.006& 0.030 &  0.004& 0.042 &  0.004& 0.039 &  0.003& 0.031 &  0.003\\ 
      &      & 0.35 & 0.40& 0.010 &  0.003& 0.015 &  0.003& 0.028 &  0.003& 0.024 &  0.003& 0.021 &  0.002\\ 
      &      & 0.40 & 0.45& 0.006 &  0.003& 0.008 &  0.002& 0.015 &  0.002& 0.015 &  0.002& 0.012 &  0.002\\ 
      &      & 0.45 & 0.50& 0.002 &  0.002& 0.004 &  0.002& 0.009 &  0.002& 0.009 &  0.001& 0.008 &  0.001\\ 
\hline \\ 
 1.75 & 1.95 & 0.10 & 0.15& 0.048 &  0.011& 0.072 &  0.011& 0.081 &  0.011& 0.085 &  0.010& 0.068 &  0.009\\ 
      &      & 0.15 & 0.20& 0.091 &  0.015& 0.097 &  0.009& 0.091 &  0.006& 0.100 &  0.006& 0.098 &  0.008\\ 
      &      & 0.20 & 0.25& 0.058 &  0.011& 0.053 &  0.007& 0.067 &  0.006& 0.061 &  0.004& 0.058 &  0.007\\ 
      &      & 0.25 & 0.30& 0.018 &  0.007& 0.033 &  0.005& 0.031 &  0.004& 0.043 &  0.003& 0.036 &  0.003\\ 
      &      & 0.30 & 0.35& 0.008 &  0.003& 0.016 &  0.003& 0.022 &  0.002& 0.024 &  0.003& 0.025 &  0.002\\ 
      &      & 0.35 & 0.40& 0.003 &  0.002& 0.009 &  0.002& 0.014 &  0.002& 0.016 &  0.002& 0.017 &  0.003\\ 
      &      & 0.40 & 0.45& 0.001 &  0.001& 0.004 &  0.001& 0.007 &  0.002& 0.008 &  0.002& 0.007 &  0.002\\ 
      &      & 0.45 & 0.50& 0.002 &  0.002& 0.002 &  0.001& 0.003 &  0.001& 0.004 &  0.001& 0.003 &  0.001\\ 
\hline \\ 
 1.95 & 2.15 & 0.10 & 0.15& 0.055 &  0.014& 0.062 &  0.010& 0.056 &  0.008& 0.068 &  0.008& 0.054 &  0.007\\ 
      &      & 0.15 & 0.20& 0.062 &  0.011& 0.082 &  0.008& 0.083 &  0.006& 0.068 &  0.005& 0.072 &  0.006\\ 
      &      & 0.20 & 0.25& 0.042 &  0.009& 0.045 &  0.006& 0.045 &  0.005& 0.048 &  0.004& 0.049 &  0.005\\ 
      &      & 0.25 & 0.30& 0.015 &  0.005& 0.019 &  0.005& 0.026 &  0.003& 0.027 &  0.002& 0.020 &  0.004\\ 
      &      & 0.30 & 0.35& 0.007 &  0.003& 0.007 &  0.002& 0.017 &  0.002& 0.018 &  0.002& 0.010 &  0.001\\ 
      &      & 0.35 & 0.40& 0.004 &  0.003& 0.007 &  0.002& 0.008 &  0.002& 0.008 &  0.002& 0.007 &  0.001\\ 
      &      & 0.40 & 0.45& \multicolumn{2}{c}{}& 0.004 &  0.002& 0.003 &  0.001& 0.004 &  0.001& 0.004 &  0.001\\ 
      &      & 0.45 & 0.50& \multicolumn{2}{c}{}& 0.002 &  0.002& 0.002 &  0.001& 0.003 &  0.001& 0.002 &  0.001\\ 
%%%%%%%%%%%%%%%%%%%%%%%%%%%%%%%%%%%%%%%%%%%%%%%%%%%%%%%%%%%%%%%%%%%%%%%%%%%%%
\hline
\end{tabular}
}
\end{center}
\end{table}

\begin{table}[hp!] 
\begin{center}
  \caption{\label{tab:xsec-n-be}
    HARP results for the double-differential $\pi^-$ production
    cross-section in the laboratory system,
    $d^2\sigma^{\pi^-}/(dpd\theta)$ for beryllium. Each row refers to a
    different $(p_{\hbox{\small min}} \le p<p_{\hbox{\small max}},
    \theta_{\hbox{\small min}} \le \theta<\theta_{\hbox{\small max}})$ bin,
    where $p$ and $\theta$ are the pion momentum and polar angle, respectively.
    The central value as well as the square-root of the diagonal elements
    of the covariance matrix are given.
 The overall normalization has an uncertainty
 of 2\%, and is not reported in the table.
}
\vspace{2mm}
\small{
\begin{tabular}{rrrr|r@{$\pm$}lr@{$\pm$}lr@{$\pm$}lr@{$\pm$}lr@{$\pm$}l} 
\hline
$\theta_{\hbox{\small min}}$ &
$\theta_{\hbox{\small max}}$ &
$p_{\hbox{\small min}}$ &
$p_{\hbox{\small max}}$ &
\multicolumn{10}{c}{$d^2\sigma^{\pi^-}/(dpd\theta)$} 
\\
(rad) & (rad) & (\GeVc) & (\GeVc) &
\multicolumn{10}{c}{(barn/(\GeVc rad))}
\\
  &  &  & 
&\multicolumn{2}{c}{$ \bf{3 \ \GeVc}$} 
&\multicolumn{2}{c}{$ \bf{5 \ \GeVc}$} 
&\multicolumn{2}{c}{$ \bf{8 \ \GeVc}$} 
&\multicolumn{2}{c}{$ \bf{8.9 \ \GeVc}$} 
&\multicolumn{2}{c}{$ \bf{12 \ \GeVc}$} 
\\ 
\hline
%%%%%%%%%%%%%%%%%%%%%%%%%%%%%%%%%%%%%%%%%%%%%%%%%%%%%%%%%%%%%%%%%%%%%%%%%%%%%
 0.35 & 0.55 & 0.15 & 0.20& 0.047 &  0.012& 0.078 &  0.013& 0.104 &  0.015& 0.114 &  0.014& 0.127 &  0.018\\ 
      &      & 0.20 & 0.25& 0.057 &  0.013& 0.095 &  0.012& 0.132 &  0.011& 0.146 &  0.012& 0.147 &  0.012\\ 
      &      & 0.25 & 0.30& 0.070 &  0.012& 0.147 &  0.017& 0.155 &  0.013& 0.164 &  0.011& 0.181 &  0.019\\ 
      &      & 0.30 & 0.35& 0.058 &  0.010& 0.130 &  0.011& 0.157 &  0.012& 0.170 &  0.012& 0.188 &  0.015\\ 
      &      & 0.35 & 0.40& 0.059 &  0.010& 0.109 &  0.009& 0.161 &  0.011& 0.164 &  0.008& 0.182 &  0.013\\ 
      &      & 0.40 & 0.45& 0.067 &  0.010& 0.122 &  0.013& 0.155 &  0.010& 0.157 &  0.009& 0.161 &  0.012\\ 
      &      & 0.45 & 0.50& 0.063 &  0.009& 0.122 &  0.010& 0.148 &  0.009& 0.166 &  0.011& 0.161 &  0.011\\ 
      &      & 0.50 & 0.60& 0.049 &  0.007& 0.120 &  0.009& 0.150 &  0.010& 0.166 &  0.010& 0.179 &  0.011\\ 
      &      & 0.60 & 0.70& 0.050 &  0.009& 0.101 &  0.012& 0.145 &  0.012& 0.164 &  0.013& 0.185 &  0.015\\ 
      &      & 0.70 & 0.80& 0.035 &  0.009& 0.081 &  0.010& 0.137 &  0.016& 0.140 &  0.017& 0.160 &  0.020\\ 
\hline \\ 
 0.55 & 0.75 & 0.10 & 0.15& 0.032 &  0.011& 0.056 &  0.015& 0.055 &  0.015& 0.069 &  0.016& 0.075 &  0.020\\ 
      &      & 0.15 & 0.20& 0.042 &  0.010& 0.101 &  0.012& 0.121 &  0.012& 0.127 &  0.009& 0.129 &  0.013\\ 
      &      & 0.20 & 0.25& 0.058 &  0.011& 0.131 &  0.011& 0.155 &  0.012& 0.152 &  0.011& 0.135 &  0.011\\ 
      &      & 0.25 & 0.30& 0.050 &  0.010& 0.110 &  0.010& 0.148 &  0.011& 0.155 &  0.010& 0.148 &  0.009\\ 
      &      & 0.30 & 0.35& 0.070 &  0.011& 0.106 &  0.010& 0.135 &  0.008& 0.154 &  0.008& 0.156 &  0.014\\ 
      &      & 0.35 & 0.40& 0.066 &  0.009& 0.108 &  0.009& 0.134 &  0.009& 0.146 &  0.009& 0.158 &  0.009\\ 
      &      & 0.40 & 0.45& 0.060 &  0.009& 0.099 &  0.007& 0.133 &  0.008& 0.146 &  0.007& 0.153 &  0.008\\ 
      &      & 0.45 & 0.50& 0.062 &  0.008& 0.103 &  0.010& 0.125 &  0.007& 0.137 &  0.006& 0.138 &  0.007\\ 
      &      & 0.50 & 0.60& 0.054 &  0.007& 0.101 &  0.008& 0.116 &  0.007& 0.126 &  0.006& 0.137 &  0.008\\ 
      &      & 0.60 & 0.70& 0.040 &  0.008& 0.075 &  0.009& 0.106 &  0.008& 0.103 &  0.010& 0.120 &  0.011\\ 
      &      & 0.70 & 0.80& 0.024 &  0.006& 0.056 &  0.011& 0.094 &  0.014& 0.079 &  0.012& 0.102 &  0.013\\ 
\hline \\ 
 0.75 & 0.95 & 0.10 & 0.15& 0.033 &  0.010& 0.043 &  0.009& 0.078 &  0.010& 0.071 &  0.009& 0.076 &  0.012\\ 
      &      & 0.15 & 0.20& 0.077 &  0.014& 0.097 &  0.011& 0.114 &  0.012& 0.132 &  0.009& 0.131 &  0.011\\ 
      &      & 0.20 & 0.25& 0.094 &  0.013& 0.120 &  0.011& 0.144 &  0.010& 0.149 &  0.010& 0.130 &  0.010\\ 
      &      & 0.25 & 0.30& 0.071 &  0.010& 0.107 &  0.009& 0.126 &  0.008& 0.140 &  0.007& 0.140 &  0.010\\ 
      &      & 0.30 & 0.35& 0.061 &  0.009& 0.096 &  0.007& 0.112 &  0.008& 0.121 &  0.006& 0.120 &  0.006\\ 
      &      & 0.35 & 0.40& 0.062 &  0.009& 0.087 &  0.007& 0.114 &  0.006& 0.111 &  0.005& 0.106 &  0.005\\ 
      &      & 0.40 & 0.45& 0.052 &  0.008& 0.077 &  0.006& 0.096 &  0.005& 0.098 &  0.005& 0.101 &  0.006\\ 
      &      & 0.45 & 0.50& 0.037 &  0.006& 0.066 &  0.005& 0.094 &  0.005& 0.088 &  0.004& 0.098 &  0.005\\ 
      &      & 0.50 & 0.60& 0.036 &  0.006& 0.058 &  0.005& 0.083 &  0.005& 0.078 &  0.004& 0.087 &  0.005\\ 
      &      & 0.60 & 0.70& 0.023 &  0.005& 0.045 &  0.006& 0.063 &  0.008& 0.063 &  0.006& 0.065 &  0.007\\ 
\hline \\ 
 0.95 & 1.15 & 0.10 & 0.15& 0.040 &  0.010& 0.055 &  0.010& 0.067 &  0.008& 0.069 &  0.008& 0.065 &  0.010\\ 
      &      & 0.15 & 0.20& 0.070 &  0.012& 0.116 &  0.015& 0.099 &  0.009& 0.131 &  0.010& 0.125 &  0.011\\ 
      &      & 0.20 & 0.25& 0.048 &  0.008& 0.115 &  0.010& 0.116 &  0.008& 0.131 &  0.006& 0.129 &  0.007\\ 
      &      & 0.25 & 0.30& 0.052 &  0.009& 0.090 &  0.008& 0.113 &  0.007& 0.120 &  0.006& 0.109 &  0.008\\ 
      &      & 0.30 & 0.35& 0.077 &  0.013& 0.076 &  0.007& 0.094 &  0.005& 0.099 &  0.004& 0.093 &  0.005\\ 
      &      & 0.35 & 0.40& 0.054 &  0.010& 0.070 &  0.006& 0.083 &  0.004& 0.082 &  0.004& 0.083 &  0.005\\ 
      &      & 0.40 & 0.45& 0.033 &  0.005& 0.052 &  0.005& 0.069 &  0.004& 0.071 &  0.003& 0.072 &  0.004\\ 
      &      & 0.45 & 0.50& 0.028 &  0.005& 0.038 &  0.004& 0.059 &  0.004& 0.060 &  0.003& 0.064 &  0.004\\ 
      &      & 0.50 & 0.60& 0.017 &  0.004& 0.030 &  0.003& 0.045 &  0.004& 0.045 &  0.004& 0.050 &  0.004\\ 
%%%%%%%%%%%%%%%%%%%%%%%%%%%%%%%%%%%%%%%%%%%%%%%%%%%%%%%%%%%%%%%%%%%%%%%%%%%%%
\hline
\end{tabular}
}
\end{center}
\end{table}

\begin{table}[hp!] 
\begin{center}
\small{
\begin{tabular}{rrrr|r@{$\pm$}lr@{$\pm$}lr@{$\pm$}lr@{$\pm$}lr@{$\pm$}l} 
\hline
$\theta_{\hbox{\small min}}$ &
$\theta_{\hbox{\small max}}$ &
$p_{\hbox{\small min}}$ &
$p_{\hbox{\small max}}$ &
\multicolumn{10}{c}{$d^2\sigma^{\pi^-}/(dpd\theta)$} 
\\
(rad) & (rad) & (\GeVc) & (\GeVc) &
\multicolumn{10}{c}{(barn/(\GeVc rad))}
\\
  &  &  & 
&\multicolumn{2}{c}{$ \bf{3 \ \GeVc}$} 
&\multicolumn{2}{c}{$ \bf{5 \ \GeVc}$} 
&\multicolumn{2}{c}{$ \bf{8 \ \GeVc}$} 
&\multicolumn{2}{c}{$ \bf{8.9 \ \GeVc}$} 
&\multicolumn{2}{c}{$ \bf{12 \ \GeVc}$} 
\\ 
\hline
%%%%%%%%%%%%%%%%%%%%%%%%%%%%%%%%%%%%%%%%%%%%%%%%%%%%%%%%%%%%%%%%%%%%%%%%%%%%%
 1.15 & 1.35 & 0.10 & 0.15& 0.023 &  0.007& 0.075 &  0.011& 0.081 &  0.008& 0.075 &  0.008& 0.080 &  0.010\\ 
      &      & 0.15 & 0.20& 0.069 &  0.012& 0.107 &  0.010& 0.098 &  0.008& 0.117 &  0.008& 0.124 &  0.009\\ 
      &      & 0.20 & 0.25& 0.087 &  0.015& 0.086 &  0.008& 0.118 &  0.008& 0.114 &  0.005& 0.118 &  0.007\\ 
      &      & 0.25 & 0.30& 0.060 &  0.010& 0.070 &  0.006& 0.091 &  0.006& 0.087 &  0.005& 0.088 &  0.006\\ 
      &      & 0.30 & 0.35& 0.035 &  0.007& 0.049 &  0.005& 0.074 &  0.005& 0.068 &  0.004& 0.070 &  0.004\\ 
      &      & 0.35 & 0.40& 0.020 &  0.004& 0.040 &  0.004& 0.055 &  0.004& 0.056 &  0.003& 0.051 &  0.004\\ 
      &      & 0.40 & 0.45& 0.016 &  0.003& 0.035 &  0.003& 0.046 &  0.003& 0.046 &  0.002& 0.041 &  0.003\\ 
      &      & 0.45 & 0.50& 0.013 &  0.003& 0.032 &  0.003& 0.037 &  0.003& 0.037 &  0.002& 0.033 &  0.003\\ 
\hline \\ 
 1.35 & 1.55 & 0.10 & 0.15& 0.045 &  0.011& 0.075 &  0.011& 0.057 &  0.008& 0.086 &  0.009& 0.079 &  0.010\\ 
      &      & 0.15 & 0.20& 0.049 &  0.009& 0.099 &  0.010& 0.111 &  0.009& 0.103 &  0.006& 0.107 &  0.008\\ 
      &      & 0.20 & 0.25& 0.074 &  0.012& 0.097 &  0.009& 0.102 &  0.007& 0.101 &  0.005& 0.100 &  0.007\\ 
      &      & 0.25 & 0.30& 0.041 &  0.008& 0.072 &  0.008& 0.069 &  0.005& 0.075 &  0.005& 0.081 &  0.007\\ 
      &      & 0.30 & 0.35& 0.027 &  0.005& 0.041 &  0.006& 0.050 &  0.004& 0.053 &  0.003& 0.050 &  0.003\\ 
      &      & 0.35 & 0.40& 0.016 &  0.004& 0.024 &  0.003& 0.039 &  0.003& 0.036 &  0.003& 0.038 &  0.003\\ 
      &      & 0.40 & 0.45& 0.012 &  0.003& 0.021 &  0.003& 0.031 &  0.003& 0.028 &  0.002& 0.027 &  0.003\\ 
      &      & 0.45 & 0.50& 0.009 &  0.003& 0.019 &  0.003& 0.022 &  0.002& 0.020 &  0.002& 0.020 &  0.002\\ 
\hline \\ 
 1.55 & 1.75 & 0.10 & 0.15& 0.042 &  0.010& 0.067 &  0.009& 0.068 &  0.008& 0.072 &  0.009& 0.072 &  0.010\\ 
      &      & 0.15 & 0.20& 0.084 &  0.014& 0.082 &  0.008& 0.088 &  0.007& 0.098 &  0.006& 0.102 &  0.008\\ 
      &      & 0.20 & 0.25& 0.046 &  0.010& 0.066 &  0.007& 0.078 &  0.005& 0.079 &  0.005& 0.081 &  0.006\\ 
      &      & 0.25 & 0.30& 0.021 &  0.006& 0.054 &  0.006& 0.050 &  0.004& 0.053 &  0.004& 0.051 &  0.005\\ 
      &      & 0.30 & 0.35& 0.011 &  0.003& 0.034 &  0.005& 0.044 &  0.004& 0.040 &  0.003& 0.042 &  0.003\\ 
      &      & 0.35 & 0.40& 0.009 &  0.003& 0.021 &  0.003& 0.034 &  0.004& 0.028 &  0.002& 0.030 &  0.003\\ 
      &      & 0.40 & 0.45& 0.009 &  0.003& 0.015 &  0.002& 0.020 &  0.002& 0.019 &  0.002& 0.021 &  0.002\\ 
      &      & 0.45 & 0.50& 0.008 &  0.003& 0.011 &  0.002& 0.014 &  0.002& 0.013 &  0.001& 0.015 &  0.002\\ 
\hline \\ 
 1.75 & 1.95 & 0.10 & 0.15& 0.042 &  0.011& 0.064 &  0.009& 0.059 &  0.008& 0.064 &  0.007& 0.058 &  0.008\\ 
      &      & 0.15 & 0.20& 0.049 &  0.010& 0.067 &  0.008& 0.070 &  0.006& 0.080 &  0.005& 0.090 &  0.008\\ 
      &      & 0.20 & 0.25& 0.028 &  0.007& 0.060 &  0.007& 0.056 &  0.004& 0.059 &  0.004& 0.058 &  0.006\\ 
      &      & 0.25 & 0.30& 0.019 &  0.006& 0.035 &  0.005& 0.039 &  0.004& 0.038 &  0.003& 0.044 &  0.004\\ 
      &      & 0.30 & 0.35& 0.006 &  0.003& 0.022 &  0.003& 0.023 &  0.003& 0.025 &  0.002& 0.029 &  0.003\\ 
      &      & 0.35 & 0.40& 0.008 &  0.004& 0.017 &  0.003& 0.014 &  0.002& 0.017 &  0.001& 0.021 &  0.003\\ 
      &      & 0.40 & 0.45& 0.005 &  0.003& 0.008 &  0.002& 0.012 &  0.001& 0.011 &  0.001& 0.012 &  0.002\\ 
      &      & 0.45 & 0.50& \multicolumn{2}{c}{}& 0.004 &  0.001& 0.008 &  0.001& 0.008 &  0.001& 0.006 &  0.001\\ 
\hline \\ 
 1.95 & 2.15 & 0.10 & 0.15& 0.038 &  0.010& 0.047 &  0.008& 0.060 &  0.006& 0.058 &  0.006& 0.053 &  0.006\\ 
      &      & 0.15 & 0.20& 0.036 &  0.009& 0.053 &  0.007& 0.063 &  0.005& 0.065 &  0.004& 0.066 &  0.006\\ 
      &      & 0.20 & 0.25& 0.023 &  0.007& 0.052 &  0.007& 0.042 &  0.004& 0.051 &  0.003& 0.045 &  0.005\\ 
      &      & 0.25 & 0.30& 0.013 &  0.005& 0.023 &  0.004& 0.028 &  0.004& 0.032 &  0.003& 0.023 &  0.003\\ 
      &      & 0.30 & 0.35& 0.007 &  0.004& 0.020 &  0.003& 0.016 &  0.002& 0.017 &  0.002& 0.019 &  0.002\\ 
      &      & 0.35 & 0.40& \multicolumn{2}{c}{}& 0.013 &  0.003& 0.011 &  0.002& 0.012 &  0.001& 0.010 &  0.002\\ 
      &      & 0.40 & 0.45& \multicolumn{2}{c}{}& 0.007 &  0.002& 0.010 &  0.001& 0.008 &  0.001& 0.009 &  0.001\\ 
      &      & 0.45 & 0.50& \multicolumn{2}{c}{}& 0.004 &  0.001& 0.007 &  0.001& 0.006 &  0.001& 0.007 &  0.001\\ 
%%%%%%%%%%%%%%%%%%%%%%%%%%%%%%%%%%%%%%%%%%%%%%%%%%%%%%%%%%%%%%%%%%%%%%%%%%%%%
\hline
\end{tabular}
}
\end{center}
\end{table}

\begin{table}[hp!]
\begin{center}
  \caption{\label{tab:xsec-p-al}
    HARP results for the double-differential $\pi^+$ production
    cross-section in the laboratory system,
    $d^2\sigma^{\pi^+}/(dpd\theta)$ for aluminium. Each row refers to a
    different $(p_{\hbox{\small min}} \le p<p_{\hbox{\small max}},
    \theta_{\hbox{\small min}} \le \theta<\theta_{\hbox{\small max}})$ bin,
    where $p$ and $\theta$ are the pion momentum and polar angle, respectively.
    The central value as well as the square-root of the diagonal elements
    of the covariance matrix are given.
 The overall normalization has an uncertainty
 of 2\% for Be and Al and 3\% for Pb, and is not reported in the table.
}
\vspace{2mm}
%\begin{tabular}{rrrr|r@{$\pm$}lr{$\pm$}lr{$\pm$}lr{$\pm$}l}
\small{
\begin{tabular}{rrrr|r@{$\pm$}lr@{$\pm$}lr@{$\pm$}lr@{$\pm$}lr@{$\pm$}l}
\hline
$\theta_{\hbox{\small min}}$ &
$\theta_{\hbox{\small max}}$ &
$p_{\hbox{\small min}}$ &
$p_{\hbox{\small max}}$ &
\multicolumn{10}{c}{$d^2\sigma^{\pi^+}/(dpd\theta)$}
\\
(rad) & (rad) & (\GeVc) & (\GeVc) &
\multicolumn{10}{c}{(barn/(\GeVc rad))}
\\
  &  &  &
&\multicolumn{2}{c}{$ \bf{3 \ \GeVc}$}
&\multicolumn{2}{c}{$ \bf{5 \ \GeVc}$}
&\multicolumn{2}{c}{$ \bf{8 \ \GeVc}$}
&\multicolumn{2}{c}{$ \bf{12 \ \GeVc}$}
&\multicolumn{2}{c}{$ \bf{12.9 \ \GeVc}$}
\\
\hline
  0.35 & 0.55& 0.15 & 0.20& 0.121 &  0.027& 0.267 &  0.047& 0.321 &  0.050& 0.376 &  0.062& 0.402 &  0.063\\ 
      &      & 0.20 & 0.25& 0.151 &  0.026& 0.345 &  0.038& 0.485 &  0.048& 0.494 &  0.044& 0.533 &  0.035\\ 
      &      & 0.25 & 0.30& 0.196 &  0.028& 0.451 &  0.046& 0.517 &  0.039& 0.597 &  0.056& 0.582 &  0.037\\ 
      &      & 0.30 & 0.35& 0.280 &  0.042& 0.450 &  0.036& 0.625 &  0.059& 0.614 &  0.043& 0.666 &  0.054\\ 
      &      & 0.35 & 0.40& 0.300 &  0.030& 0.398 &  0.030& 0.645 &  0.038& 0.667 &  0.055& 0.682 &  0.038\\ 
      &      & 0.40 & 0.45& 0.305 &  0.029& 0.409 &  0.036& 0.643 &  0.041& 0.717 &  0.048& 0.722 &  0.037\\ 
      &      & 0.45 & 0.50& 0.270 &  0.025& 0.455 &  0.032& 0.675 &  0.039& 0.759 &  0.051& 0.698 &  0.032\\ 
      &      & 0.50 & 0.60& 0.281 &  0.027& 0.500 &  0.036& 0.626 &  0.037& 0.695 &  0.046& 0.706 &  0.035\\ 
      &      & 0.60 & 0.70& 0.197 &  0.032& 0.366 &  0.047& 0.621 &  0.060& 0.709 &  0.074& 0.673 &  0.068\\ 
      &      & 0.70 & 0.80& 0.095 &  0.024& 0.274 &  0.045& 0.492 &  0.077& 0.569 &  0.087& 0.526 &  0.088\\ 
\hline \\ 
 0.55 & 0.75 & 0.10 & 0.15& 0.160 &  0.048& 0.227 &  0.061& 0.229 &  0.053& 0.265 &  0.065& 0.253 &  0.067\\ 
      &      & 0.15 & 0.20& 0.251 &  0.035& 0.367 &  0.036& 0.353 &  0.036& 0.378 &  0.041& 0.417 &  0.035\\ 
      &      & 0.20 & 0.25& 0.327 &  0.037& 0.427 &  0.039& 0.553 &  0.050& 0.546 &  0.044& 0.592 &  0.036\\ 
      &      & 0.25 & 0.30& 0.349 &  0.037& 0.420 &  0.047& 0.626 &  0.051& 0.557 &  0.040& 0.612 &  0.043\\ 
      &      & 0.30 & 0.35& 0.280 &  0.030& 0.449 &  0.033& 0.601 &  0.032& 0.568 &  0.040& 0.620 &  0.032\\ 
      &      & 0.35 & 0.40& 0.261 &  0.028& 0.376 &  0.030& 0.544 &  0.030& 0.599 &  0.039& 0.648 &  0.032\\ 
      &      & 0.40 & 0.45& 0.265 &  0.024& 0.383 &  0.028& 0.491 &  0.028& 0.615 &  0.033& 0.604 &  0.022\\ 
      &      & 0.45 & 0.50& 0.260 &  0.026& 0.371 &  0.026& 0.505 &  0.028& 0.537 &  0.035& 0.551 &  0.019\\ 
      &      & 0.50 & 0.60& 0.178 &  0.028& 0.297 &  0.026& 0.485 &  0.032& 0.464 &  0.033& 0.474 &  0.028\\ 
      &      & 0.60 & 0.70& 0.098 &  0.020& 0.193 &  0.030& 0.336 &  0.048& 0.373 &  0.052& 0.368 &  0.046\\ 
      &      & 0.70 & 0.80& 0.046 &  0.014& 0.123 &  0.027& 0.234 &  0.043& 0.234 &  0.049& 0.250 &  0.047\\ 
\hline \\ 
 0.75 & 0.95 & 0.10 & 0.15& 0.195 &  0.039& 0.267 &  0.049& 0.265 &  0.049& 0.217 &  0.044& 0.307 &  0.053\\ 
      &      & 0.15 & 0.20& 0.369 &  0.041& 0.348 &  0.034& 0.464 &  0.034& 0.439 &  0.047& 0.532 &  0.034\\ 
      &      & 0.20 & 0.25& 0.346 &  0.033& 0.394 &  0.030& 0.536 &  0.037& 0.538 &  0.040& 0.593 &  0.037\\ 
      &      & 0.25 & 0.30& 0.375 &  0.039& 0.438 &  0.042& 0.521 &  0.032& 0.569 &  0.036& 0.548 &  0.026\\ 
      &      & 0.30 & 0.35& 0.244 &  0.033& 0.367 &  0.028& 0.442 &  0.024& 0.451 &  0.024& 0.519 &  0.021\\ 
      &      & 0.35 & 0.40& 0.190 &  0.020& 0.291 &  0.021& 0.411 &  0.022& 0.399 &  0.022& 0.468 &  0.017\\ 
      &      & 0.40 & 0.45& 0.176 &  0.018& 0.284 &  0.022& 0.346 &  0.018& 0.375 &  0.021& 0.403 &  0.014\\ 
      &      & 0.45 & 0.50& 0.141 &  0.017& 0.268 &  0.020& 0.307 &  0.018& 0.347 &  0.021& 0.359 &  0.016\\ 
      &      & 0.50 & 0.60& 0.091 &  0.013& 0.193 &  0.023& 0.256 &  0.023& 0.272 &  0.025& 0.294 &  0.024\\ 
      &      & 0.60 & 0.70& 0.055 &  0.010& 0.110 &  0.022& 0.161 &  0.028& 0.179 &  0.029& 0.173 &  0.028\\ 
\hline \\ 
 0.95 & 1.15 & 0.10 & 0.15& 0.257 &  0.043& 0.274 &  0.043& 0.264 &  0.039& 0.289 &  0.052& 0.312 &  0.046\\ 
      &      & 0.15 & 0.20& 0.327 &  0.034& 0.449 &  0.037& 0.489 &  0.040& 0.457 &  0.038& 0.477 &  0.027\\ 
      &      & 0.20 & 0.25& 0.314 &  0.031& 0.417 &  0.034& 0.513 &  0.029& 0.508 &  0.034& 0.520 &  0.028\\ 
      &      & 0.25 & 0.30& 0.221 &  0.026& 0.357 &  0.029& 0.396 &  0.023& 0.405 &  0.025& 0.438 &  0.021\\ 
      &      & 0.30 & 0.35& 0.160 &  0.024& 0.267 &  0.021& 0.337 &  0.020& 0.338 &  0.020& 0.368 &  0.014\\ 
      &      & 0.35 & 0.40& 0.107 &  0.013& 0.191 &  0.019& 0.293 &  0.016& 0.277 &  0.017& 0.322 &  0.010\\ 
      &      & 0.40 & 0.45& 0.081 &  0.010& 0.122 &  0.015& 0.231 &  0.014& 0.225 &  0.015& 0.262 &  0.015\\ 
      &      & 0.45 & 0.50& 0.061 &  0.011& 0.078 &  0.010& 0.184 &  0.014& 0.183 &  0.015& 0.196 &  0.017\\ 
      &      & 0.50 & 0.60& 0.032 &  0.007& 0.058 &  0.009& 0.126 &  0.016& 0.128 &  0.016& 0.127 &  0.014\\ 
\hline \\ 
%%%%%%%%%%%%%%%%%%%%%%%%%%%%%%%%%%%%%%%%%%%%%%%%%%%%%%%%%%%%%%%%%%%%%%%%%%%%%
\end{tabular}
}
\end{center}
\end{table}
\begin{table}[hp!]
\begin{center}
\vspace{2mm}
%\begin{tabular}{rrrr|r@{$\pm$}lr{$\pm$}lr{$\pm$}lr{$\pm$}l}
\small{
\begin{tabular}{rrrr|r@{$\pm$}lr@{$\pm$}lr@{$\pm$}lr@{$\pm$}lr@{$\pm$}l}
\hline
$\theta_{\hbox{\small min}}$ &
$\theta_{\hbox{\small max}}$ &
$p_{\hbox{\small min}}$ &
$p_{\hbox{\small max}}$ &
\multicolumn{10}{c}{$d^2\sigma^{\pi^+}/(dpd\theta)$}
\\
(rad) & (rad) & (\GeVc) & (\GeVc) &
\multicolumn{10}{c}{(barn/(\GeVc rad))}
\\
  &  &  &
&\multicolumn{2}{c}{$ \bf{3 \ \GeVc}$}
&\multicolumn{2}{c}{$ \bf{5 \ \GeVc}$}
&\multicolumn{2}{c}{$ \bf{8 \ \GeVc}$}
&\multicolumn{2}{c}{$ \bf{12 \ \GeVc}$}
&\multicolumn{2}{c}{$ \bf{12.9 \ \GeVc}$}
\\
\hline
 1.15 & 1.35 & 0.10 & 0.15& 0.208 &  0.031& 0.271 &  0.040& 0.308 &  0.043& 0.257 &  0.038& 0.332 &  0.048\\ 
      &      & 0.15 & 0.20& 0.323 &  0.037& 0.487 &  0.043& 0.422 &  0.029& 0.486 &  0.048& 0.468 &  0.024\\ 
      &      & 0.20 & 0.25& 0.249 &  0.027& 0.332 &  0.031& 0.390 &  0.023& 0.431 &  0.029& 0.436 &  0.019\\ 
      &      & 0.25 & 0.30& 0.190 &  0.023& 0.234 &  0.021& 0.337 &  0.020& 0.336 &  0.027& 0.336 &  0.015\\ 
      &      & 0.30 & 0.35& 0.123 &  0.019& 0.193 &  0.017& 0.258 &  0.016& 0.292 &  0.021& 0.253 &  0.013\\ 
      &      & 0.35 & 0.40& 0.068 &  0.010& 0.142 &  0.014& 0.186 &  0.015& 0.188 &  0.019& 0.189 &  0.010\\ 
      &      & 0.40 & 0.45& 0.044 &  0.008& 0.104 &  0.014& 0.138 &  0.011& 0.142 &  0.013& 0.133 &  0.008\\ 
      &      & 0.45 & 0.50& 0.030 &  0.007& 0.067 &  0.011& 0.097 &  0.011& 0.095 &  0.013& 0.102 &  0.009\\ 
\hline \\ 
 1.35 & 1.55 & 0.10 & 0.15& 0.195 &  0.034& 0.269 &  0.038& 0.353 &  0.045& 0.351 &  0.053& 0.343 &  0.048\\ 
      &      & 0.15 & 0.20& 0.357 &  0.041& 0.343 &  0.032& 0.416 &  0.027& 0.390 &  0.028& 0.427 &  0.023\\ 
      &      & 0.20 & 0.25& 0.231 &  0.039& 0.266 &  0.023& 0.303 &  0.020& 0.298 &  0.022& 0.357 &  0.015\\ 
      &      & 0.25 & 0.30& 0.121 &  0.016& 0.174 &  0.016& 0.215 &  0.015& 0.243 &  0.018& 0.246 &  0.014\\ 
      &      & 0.30 & 0.35& 0.077 &  0.013& 0.163 &  0.016& 0.174 &  0.013& 0.169 &  0.014& 0.175 &  0.010\\ 
      &      & 0.35 & 0.40& 0.052 &  0.009& 0.096 &  0.014& 0.119 &  0.011& 0.134 &  0.013& 0.128 &  0.008\\ 
      &      & 0.40 & 0.45& 0.034 &  0.007& 0.057 &  0.009& 0.073 &  0.008& 0.088 &  0.010& 0.087 &  0.009\\ 
      &      & 0.45 & 0.50& 0.021 &  0.005& 0.036 &  0.007& 0.045 &  0.008& 0.060 &  0.010& 0.055 &  0.008\\ 
\hline \\ 
 1.55 & 1.75 & 0.10 & 0.15& 0.224 &  0.037& 0.278 &  0.044& 0.288 &  0.036& 0.267 &  0.044& 0.324 &  0.041\\ 
      &      & 0.15 & 0.20& 0.272 &  0.030& 0.399 &  0.033& 0.331 &  0.021& 0.384 &  0.030& 0.382 &  0.020\\ 
      &      & 0.20 & 0.25& 0.159 &  0.020& 0.233 &  0.023& 0.221 &  0.016& 0.280 &  0.028& 0.268 &  0.014\\ 
      &      & 0.25 & 0.30& 0.130 &  0.019& 0.118 &  0.015& 0.159 &  0.012& 0.169 &  0.017& 0.165 &  0.010\\ 
      &      & 0.30 & 0.35& 0.044 &  0.012& 0.094 &  0.012& 0.111 &  0.010& 0.095 &  0.012& 0.112 &  0.006\\ 
      &      & 0.35 & 0.40& 0.022 &  0.006& 0.061 &  0.009& 0.070 &  0.007& 0.063 &  0.006& 0.081 &  0.007\\ 
      &      & 0.40 & 0.45& 0.014 &  0.005& 0.035 &  0.007& 0.047 &  0.006& 0.051 &  0.007& 0.049 &  0.006\\ 
      &      & 0.45 & 0.50& 0.007 &  0.003& 0.018 &  0.005& 0.027 &  0.005& 0.034 &  0.007& 0.029 &  0.004\\ 
\hline \\ 
 1.75 & 1.95 & 0.10 & 0.15& 0.263 &  0.038& 0.222 &  0.031& 0.279 &  0.034& 0.297 &  0.038& 0.293 &  0.034\\ 
      &      & 0.15 & 0.20& 0.232 &  0.026& 0.284 &  0.027& 0.329 &  0.028& 0.299 &  0.024& 0.309 &  0.014\\ 
      &      & 0.20 & 0.25& 0.128 &  0.020& 0.216 &  0.022& 0.204 &  0.020& 0.175 &  0.018& 0.197 &  0.009\\ 
      &      & 0.25 & 0.30& 0.081 &  0.013& 0.090 &  0.016& 0.125 &  0.011& 0.111 &  0.012& 0.117 &  0.009\\ 
      &      & 0.30 & 0.35& 0.052 &  0.013& 0.051 &  0.008& 0.064 &  0.009& 0.067 &  0.009& 0.076 &  0.005\\ 
      &      & 0.35 & 0.40& 0.015 &  0.007& 0.040 &  0.007& 0.035 &  0.004& 0.049 &  0.007& 0.051 &  0.005\\ 
      &      & 0.40 & 0.45& 0.004 &  0.003& 0.025 &  0.006& 0.025 &  0.004& 0.032 &  0.007& 0.030 &  0.004\\ 
      &      & 0.45 & 0.50& 0.002 &  0.002& 0.014 &  0.005& 0.013 &  0.003& 0.016 &  0.005& 0.016 &  0.003\\ 
\hline \\ 
 1.95 & 2.15 & 0.10 & 0.15& 0.203 &  0.029& 0.193 &  0.027& 0.249 &  0.033& 0.202 &  0.031& 0.218 &  0.029\\ 
      &      & 0.15 & 0.20& 0.179 &  0.024& 0.248 &  0.028& 0.255 &  0.018& 0.242 &  0.021& 0.254 &  0.012\\ 
      &      & 0.20 & 0.25& 0.079 &  0.019& 0.161 &  0.028& 0.132 &  0.017& 0.124 &  0.015& 0.147 &  0.011\\ 
      &      & 0.25 & 0.30& 0.048 &  0.011& 0.046 &  0.008& 0.066 &  0.007& 0.058 &  0.009& 0.077 &  0.006\\ 
      &      & 0.30 & 0.35& 0.018 &  0.008& 0.033 &  0.007& 0.043 &  0.006& 0.036 &  0.006& 0.045 &  0.004\\ 
      &      & 0.35 & 0.40& 0.007 &  0.003& 0.022 &  0.006& 0.022 &  0.004& 0.029 &  0.006& 0.031 &  0.003\\ 
      &      & 0.40 & 0.45& 0.004 &  0.003& 0.007 &  0.003& 0.015 &  0.003& 0.013 &  0.004& 0.016 &  0.003\\ 
      &      & 0.45 & 0.50& 0.002 &  0.002& 0.003 &  0.002& 0.006 &  0.002& 0.007 &  0.002& 0.009 &  0.002\\ 

%%%%%%%%%%%%%%%%%%%%%%%%%%%%%%%%%%%%%%%%%%%%%%%%%%%%%%%%%%%%%%%%%%%%%%%%%%%%%
\hline
\end{tabular}
}
\end{center}
\end{table}
\begin{table}[hp!]
\begin{center}
  \caption{\label{tab:xsec-n-al}
    HARP results for the double-differential $\pi^-$ production
    cross-section in the laboratory system,
    $d^2\sigma^{\pi^+}/(dpd\theta)$ for aluminium. Each row refers to a
    different $(p_{\hbox{\small min}} \le p<p_{\hbox{\small max}},
    \theta_{\hbox{\small min}} \le \theta<\theta_{\hbox{\small max}})$ bin,
    where $p$ and $\theta$ are the pion momentum and polar angle, respectively.
    The central value as well as the square-root of the diagonal elements
    of the covariance matrix are given.
 The overall normalization has an uncertainty
 of 2\%, and is not reported in the table.
}
\vspace{2mm}
%\begin{tabular}{rrrr|r@{$\pm$}lr{$\pm$}lr{$\pm$}lr{$\pm$}l}
\small{
\begin{tabular}{rrrr|r@{$\pm$}lr@{$\pm$}lr@{$\pm$}lr@{$\pm$}lr@{$\pm$}l}
\hline
$\theta_{\hbox{\small min}}$ &
$\theta_{\hbox{\small max}}$ &
$p_{\hbox{\small min}}$ &
$p_{\hbox{\small max}}$ &
\multicolumn{10}{c}{$d^2\sigma^{\pi^+}/(dpd\theta)$}
\\
(rad) & (rad) & (\GeVc) & (\GeVc) &
\multicolumn{10}{c}{(barn/(\GeVc rad))}
\\
  &  &  &
&\multicolumn{2}{c}{$ \bf{3 \ \GeVc}$}
&\multicolumn{2}{c}{$ \bf{5 \ \GeVc}$}
&\multicolumn{2}{c}{$ \bf{8 \ \GeVc}$}
&\multicolumn{2}{c}{$ \bf{12 \ \GeVc}$}
&\multicolumn{2}{c}{$ \bf{12.9 \ \GeVc}$}
\\
\hline
 0.35 & 0.55 & 0.15 & 0.20& 0.107 &  0.029& 0.231 &  0.046& 0.356 &  0.056& 0.321 &  0.055& 0.378 &  0.062\\ 
      &      & 0.20 & 0.25& 0.099 &  0.022& 0.249 &  0.031& 0.371 &  0.035& 0.410 &  0.044& 0.498 &  0.036\\ 
      &      & 0.25 & 0.30& 0.106 &  0.025& 0.300 &  0.034& 0.455 &  0.030& 0.485 &  0.037& 0.537 &  0.028\\ 
      &      & 0.30 & 0.35& 0.180 &  0.026& 0.313 &  0.030& 0.355 &  0.023& 0.454 &  0.037& 0.510 &  0.019\\ 
      &      & 0.35 & 0.40& 0.164 &  0.022& 0.250 &  0.024& 0.383 &  0.035& 0.478 &  0.035& 0.459 &  0.020\\ 
      &      & 0.40 & 0.45& 0.114 &  0.015& 0.230 &  0.024& 0.432 &  0.032& 0.442 &  0.027& 0.468 &  0.022\\ 
      &      & 0.45 & 0.50& 0.127 &  0.019& 0.279 &  0.028& 0.422 &  0.024& 0.415 &  0.028& 0.453 &  0.020\\ 
      &      & 0.50 & 0.60& 0.137 &  0.018& 0.261 &  0.023& 0.367 &  0.023& 0.411 &  0.028& 0.469 &  0.025\\ 
      &      & 0.60 & 0.70& 0.088 &  0.017& 0.193 &  0.030& 0.350 &  0.028& 0.401 &  0.034& 0.459 &  0.037\\ 
      &      & 0.70 & 0.80& 0.060 &  0.014& 0.135 &  0.020& 0.333 &  0.041& 0.406 &  0.046& 0.395 &  0.049\\ 
\hline \\ 
 0.55 & 0.75 & 0.10 & 0.15& 0.083 &  0.030& 0.123 &  0.047& 0.232 &  0.062& 0.246 &  0.071& 0.257 &  0.066\\ 
      &      & 0.15 & 0.20& 0.117 &  0.023& 0.286 &  0.032& 0.368 &  0.034& 0.376 &  0.035& 0.408 &  0.028\\ 
      &      & 0.20 & 0.25& 0.188 &  0.028& 0.233 &  0.028& 0.424 &  0.027& 0.396 &  0.032& 0.440 &  0.023\\ 
      &      & 0.25 & 0.30& 0.142 &  0.020& 0.285 &  0.029& 0.406 &  0.031& 0.389 &  0.030& 0.428 &  0.020\\ 
      &      & 0.30 & 0.35& 0.133 &  0.020& 0.263 &  0.023& 0.355 &  0.021& 0.454 &  0.037& 0.428 &  0.022\\ 
      &      & 0.35 & 0.40& 0.150 &  0.020& 0.214 &  0.019& 0.366 &  0.026& 0.336 &  0.018& 0.427 &  0.019\\ 
      &      & 0.40 & 0.45& 0.139 &  0.017& 0.225 &  0.023& 0.341 &  0.018& 0.368 &  0.027& 0.403 &  0.014\\ 
      &      & 0.45 & 0.50& 0.123 &  0.015& 0.238 &  0.020& 0.323 &  0.019& 0.384 &  0.024& 0.379 &  0.013\\ 
      &      & 0.50 & 0.60& 0.104 &  0.014& 0.212 &  0.018& 0.281 &  0.019& 0.338 &  0.024& 0.350 &  0.017\\ 
      &      & 0.60 & 0.70& 0.081 &  0.014& 0.155 &  0.022& 0.250 &  0.022& 0.282 &  0.029& 0.301 &  0.026\\ 
      &      & 0.70 & 0.80& 0.052 &  0.013& 0.110 &  0.020& 0.218 &  0.029& 0.254 &  0.031& 0.242 &  0.035\\ 
\hline \\ 
 0.75 & 0.95 & 0.10 & 0.15& 0.100 &  0.030& 0.176 &  0.035& 0.224 &  0.041& 0.258 &  0.052& 0.278 &  0.048\\ 
      &      & 0.15 & 0.20& 0.198 &  0.028& 0.319 &  0.032& 0.356 &  0.026& 0.418 &  0.032& 0.439 &  0.022\\ 
      &      & 0.20 & 0.25& 0.141 &  0.022& 0.283 &  0.027& 0.360 &  0.026& 0.374 &  0.030& 0.391 &  0.022\\ 
      &      & 0.25 & 0.30& 0.145 &  0.020& 0.272 &  0.025& 0.348 &  0.026& 0.393 &  0.031& 0.392 &  0.018\\ 
      &      & 0.30 & 0.35& 0.138 &  0.018& 0.230 &  0.020& 0.323 &  0.019& 0.348 &  0.022& 0.374 &  0.020\\ 
      &      & 0.35 & 0.40& 0.139 &  0.017& 0.210 &  0.018& 0.293 &  0.017& 0.283 &  0.017& 0.339 &  0.012\\ 
      &      & 0.40 & 0.45& 0.117 &  0.015& 0.209 &  0.018& 0.258 &  0.015& 0.269 &  0.016& 0.281 &  0.009\\ 
      &      & 0.45 & 0.50& 0.080 &  0.012& 0.169 &  0.017& 0.233 &  0.013& 0.245 &  0.015& 0.251 &  0.010\\ 
      &      & 0.50 & 0.60& 0.054 &  0.011& 0.115 &  0.013& 0.203 &  0.013& 0.230 &  0.017& 0.211 &  0.012\\ 
      &      & 0.60 & 0.70& 0.028 &  0.007& 0.079 &  0.012& 0.156 &  0.017& 0.178 &  0.019& 0.174 &  0.018\\ 
\hline \\ 
 0.95 & 1.15 & 0.10 & 0.15& 0.174 &  0.031& 0.213 &  0.032& 0.222 &  0.031& 0.215 &  0.037& 0.293 &  0.036\\ 
      &      & 0.15 & 0.20& 0.216 &  0.027& 0.259 &  0.029& 0.326 &  0.026& 0.338 &  0.030& 0.400 &  0.023\\ 
      &      & 0.20 & 0.25& 0.197 &  0.027& 0.280 &  0.025& 0.355 &  0.024& 0.354 &  0.027& 0.371 &  0.019\\ 
      &      & 0.25 & 0.30& 0.177 &  0.021& 0.201 &  0.019& 0.288 &  0.017& 0.331 &  0.025& 0.345 &  0.014\\ 
      &      & 0.30 & 0.35& 0.137 &  0.018& 0.172 &  0.018& 0.231 &  0.014& 0.263 &  0.020& 0.269 &  0.010\\ 
      &      & 0.35 & 0.40& 0.087 &  0.014& 0.118 &  0.011& 0.189 &  0.012& 0.201 &  0.013& 0.232 &  0.008\\ 
      &      & 0.40 & 0.45& 0.065 &  0.010& 0.118 &  0.013& 0.153 &  0.010& 0.192 &  0.013& 0.191 &  0.007\\ 
      &      & 0.45 & 0.50& 0.050 &  0.009& 0.096 &  0.009& 0.139 &  0.009& 0.158 &  0.012& 0.165 &  0.006\\ 
      &      & 0.50 & 0.60& 0.028 &  0.006& 0.074 &  0.008& 0.115 &  0.009& 0.126 &  0.011& 0.132 &  0.008\\ 
%%%%%%%%%%%%%%%%%%%%%%%%%%%%%%%%%%%%%%%%%%%%%%%%%%%%%%%%%%%%%%%%%%%%%%%%%%%%%
\hline
\end{tabular}
}
\end{center}
\end{table}
\begin{table}[hp!]
\begin{center}
\vspace{2mm}
%\begin{tabular}{rrrr|r@{$\pm$}lr{$\pm$}lr{$\pm$}lr{$\pm$}l}
\small{
\begin{tabular}{rrrr|r@{$\pm$}lr@{$\pm$}lr@{$\pm$}lr@{$\pm$}lr@{$\pm$}l}
\hline
$\theta_{\hbox{\small min}}$ &
$\theta_{\hbox{\small max}}$ &
$p_{\hbox{\small min}}$ &
$p_{\hbox{\small max}}$ &
\multicolumn{10}{c}{$d^2\sigma^{\pi^+}/(dpd\theta)$}
\\
(rad) & (rad) & (\GeVc) & (\GeVc) &
\multicolumn{10}{c}{(barn/(\GeVc rad))}
\\
  &  &  &
&\multicolumn{2}{c}{$ \bf{3 \ \GeVc}$}
&\multicolumn{2}{c}{$ \bf{5 \ \GeVc}$}
&\multicolumn{2}{c}{$ \bf{8 \ \GeVc}$}
&\multicolumn{2}{c}{$ \bf{12 \ \GeVc}$}
&\multicolumn{2}{c}{$ \bf{12.9 \ \GeVc}$}
\\
\hline
 1.15 & 1.35 & 0.10 & 0.15& 0.116 &  0.023& 0.127 &  0.021& 0.240 &  0.032& 0.249 &  0.045& 0.302 &  0.032\\ 
      &      & 0.15 & 0.20& 0.159 &  0.023& 0.242 &  0.033& 0.338 &  0.024& 0.371 &  0.028& 0.370 &  0.023\\ 
      &      & 0.20 & 0.25& 0.166 &  0.022& 0.271 &  0.026& 0.300 &  0.020& 0.308 &  0.023& 0.323 &  0.012\\ 
      &      & 0.25 & 0.30& 0.137 &  0.019& 0.165 &  0.016& 0.253 &  0.017& 0.271 &  0.020& 0.269 &  0.010\\ 
      &      & 0.30 & 0.35& 0.082 &  0.016& 0.161 &  0.016& 0.168 &  0.015& 0.202 &  0.017& 0.210 &  0.009\\ 
      &      & 0.35 & 0.40& 0.047 &  0.009& 0.110 &  0.012& 0.128 &  0.009& 0.136 &  0.011& 0.160 &  0.006\\ 
      &      & 0.40 & 0.45& 0.038 &  0.007& 0.086 &  0.009& 0.108 &  0.007& 0.111 &  0.010& 0.126 &  0.006\\ 
      &      & 0.45 & 0.50& 0.031 &  0.006& 0.071 &  0.008& 0.091 &  0.007& 0.088 &  0.008& 0.096 &  0.007\\ 
\hline \\ 
 1.35 & 1.55 & 0.10 & 0.15& 0.144 &  0.028& 0.208 &  0.029& 0.257 &  0.027& 0.257 &  0.034& 0.291 &  0.033\\ 
      &      & 0.15 & 0.20& 0.193 &  0.025& 0.226 &  0.024& 0.279 &  0.020& 0.269 &  0.024& 0.329 &  0.016\\ 
      &      & 0.20 & 0.25& 0.126 &  0.019& 0.165 &  0.019& 0.252 &  0.017& 0.266 &  0.024& 0.276 &  0.012\\ 
      &      & 0.25 & 0.30& 0.069 &  0.011& 0.104 &  0.012& 0.187 &  0.015& 0.223 &  0.019& 0.212 &  0.011\\ 
      &      & 0.30 & 0.35& 0.078 &  0.014& 0.101 &  0.012& 0.141 &  0.010& 0.148 &  0.013& 0.156 &  0.008\\ 
      &      & 0.35 & 0.40& 0.041 &  0.009& 0.086 &  0.010& 0.111 &  0.008& 0.108 &  0.010& 0.116 &  0.006\\ 
      &      & 0.40 & 0.45& 0.023 &  0.007& 0.062 &  0.008& 0.090 &  0.008& 0.085 &  0.008& 0.084 &  0.006\\ 
      &      & 0.45 & 0.50& 0.015 &  0.005& 0.041 &  0.008& 0.066 &  0.008& 0.066 &  0.007& 0.063 &  0.005\\ 
\hline \\ 
 1.55 & 1.75 & 0.10 & 0.15& 0.141 &  0.028& 0.239 &  0.032& 0.234 &  0.034& 0.280 &  0.038& 0.274 &  0.029\\ 
      &      & 0.15 & 0.20& 0.170 &  0.023& 0.231 &  0.023& 0.257 &  0.018& 0.274 &  0.023& 0.280 &  0.015\\ 
      &      & 0.20 & 0.25& 0.120 &  0.020& 0.177 &  0.020& 0.184 &  0.013& 0.233 &  0.021& 0.220 &  0.010\\ 
      &      & 0.25 & 0.30& 0.057 &  0.011& 0.092 &  0.016& 0.150 &  0.012& 0.148 &  0.018& 0.159 &  0.009\\ 
      &      & 0.30 & 0.35& 0.034 &  0.008& 0.058 &  0.008& 0.094 &  0.012& 0.085 &  0.011& 0.103 &  0.008\\ 
      &      & 0.35 & 0.40& 0.017 &  0.005& 0.055 &  0.008& 0.052 &  0.006& 0.057 &  0.006& 0.076 &  0.004\\ 
      &      & 0.40 & 0.45& 0.012 &  0.004& 0.043 &  0.007& 0.040 &  0.004& 0.043 &  0.005& 0.057 &  0.004\\ 
      &      & 0.45 & 0.50& 0.008 &  0.003& 0.028 &  0.006& 0.033 &  0.003& 0.041 &  0.005& 0.039 &  0.004\\ 
\hline \\ 
 1.75 & 1.95 & 0.10 & 0.15& 0.123 &  0.023& 0.174 &  0.023& 0.218 &  0.023& 0.223 &  0.028& 0.236 &  0.026\\ 
      &      & 0.15 & 0.20& 0.154 &  0.023& 0.206 &  0.022& 0.234 &  0.017& 0.229 &  0.020& 0.235 &  0.011\\ 
      &      & 0.20 & 0.25& 0.091 &  0.016& 0.123 &  0.016& 0.166 &  0.013& 0.149 &  0.016& 0.171 &  0.007\\ 
      &      & 0.25 & 0.30& 0.042 &  0.009& 0.088 &  0.012& 0.113 &  0.011& 0.093 &  0.010& 0.115 &  0.006\\ 
      &      & 0.30 & 0.35& 0.032 &  0.009& 0.054 &  0.009& 0.069 &  0.007& 0.088 &  0.010& 0.077 &  0.005\\ 
      &      & 0.35 & 0.40& 0.020 &  0.006& 0.041 &  0.007& 0.048 &  0.006& 0.061 &  0.009& 0.056 &  0.003\\ 
      &      & 0.40 & 0.45& 0.013 &  0.005& 0.023 &  0.005& 0.036 &  0.004& 0.034 &  0.006& 0.040 &  0.004\\ 
      &      & 0.45 & 0.50& 0.009 &  0.004& 0.016 &  0.004& 0.026 &  0.003& 0.024 &  0.004& 0.026 &  0.003\\ 
\hline \\ 
 1.95 & 2.15 & 0.10 & 0.15& 0.096 &  0.019& 0.097 &  0.016& 0.176 &  0.019& 0.128 &  0.022& 0.206 &  0.021\\ 
      &      & 0.15 & 0.20& 0.086 &  0.016& 0.172 &  0.021& 0.178 &  0.015& 0.211 &  0.021& 0.181 &  0.010\\ 
      &      & 0.20 & 0.25& 0.040 &  0.009& 0.088 &  0.016& 0.138 &  0.012& 0.117 &  0.014& 0.132 &  0.007\\ 
      &      & 0.25 & 0.30& 0.036 &  0.010& 0.046 &  0.008& 0.061 &  0.010& 0.070 &  0.009& 0.078 &  0.005\\ 
      &      & 0.30 & 0.35& 0.021 &  0.007& 0.048 &  0.009& 0.040 &  0.005& 0.055 &  0.008& 0.051 &  0.003\\ 
      &      & 0.35 & 0.40& 0.016 &  0.006& 0.030 &  0.007& 0.039 &  0.005& 0.034 &  0.006& 0.037 &  0.003\\ 
      &      & 0.40 & 0.45& 0.010 &  0.005& 0.013 &  0.005& 0.032 &  0.005& 0.019 &  0.005& 0.027 &  0.002\\ 
      &      & 0.45 & 0.50& 0.005 &  0.003& 0.006 &  0.003& 0.017 &  0.004& 0.012 &  0.003& 0.019 &  0.002\\ 
%%%%%%%%%%%%%%%%%%%%%%%%%%%%%%%%%%%%%%%%%%%%%%%%%%%%%%%%%%%%%%%%%%%%%%%%%%%%%
\hline
\end{tabular}
}
\end{center}
\end{table}
\begin{table}[hp!] 
\begin{center}
  \caption{\label{tab:xsec-p-pb}
    HARP results for the double-differential $\pi^+$ production
    cross-section in the laboratory system,
    $d^2\sigma^{\pi^+}/(dpd\theta)$ for lead. Each row refers to a
    different $(p_{\hbox{\small min}} \le p<p_{\hbox{\small max}},
    \theta_{\hbox{\small min}} \le \theta<\theta_{\hbox{\small max}})$ bin,
    where $p$ and $\theta$ are the pion momentum and polar angle, respectively.
    The central value as well as the square-root of the diagonal elements
    of the covariance matrix are given.
 The overall normalization has an uncertainty
 of 3\%, and is not reported in the table.
}
\vspace{2mm}
\small{
%\begin{tabular}{rrrr|r@{$\pm$}lr{$\pm$}lr{$\pm$}lr{$\pm$}l} 
\begin{tabular}{rrrr|r@{$\pm$}lr@{$\pm$}lr@{$\pm$}lr@{$\pm$}l} 
\hline
$\theta_{\hbox{\small min}}$ &
$\theta_{\hbox{\small max}}$ &
$p_{\hbox{\small min}}$ &
$p_{\hbox{\small max}}$ &
\multicolumn{8}{c}{$d^2\sigma^{\pi^+}/(dpd\theta)$} 
\\
(rad) & (rad) & (\GeVc) & (\GeVc) &
\multicolumn{8}{c}{(barn/(\GeVc rad))}
\\
  &  &  & 
&\multicolumn{2}{c}{$ \bf{3 \ \GeVc}$} 
&\multicolumn{2}{c}{$ \bf{5 \ \GeVc}$} 
&\multicolumn{2}{c}{$ \bf{8 \ \GeVc}$} 
&\multicolumn{2}{c}{$ \bf{12 \ \GeVc}$} 
\\ 
\hline
%%%%%%%%%%%%%%%%%%%%%%%%%%%%%%%%%%%%%%%%%%%%%%%%%%%%%%%%%%%%%%%%%%%%%%%%%%%%%
 0.35 & 0.55 & 0.15 & 0.20& 0.34 &  0.12& 0.80 &  0.21& 1.14 &  0.32& 1.49 &  0.40\\ 
      &      & 0.20 & 0.25& 0.29 &  0.09& 0.95 &  0.16& 1.86 &  0.24& 1.80 &  0.26\\ 
      &      & 0.25 & 0.30& 0.48 &  0.11& 1.26 &  0.13& 2.09 &  0.14& 2.42 &  0.26\\ 
      &      & 0.30 & 0.35& 0.75 &  0.11& 1.25 &  0.11& 1.99 &  0.15& 2.75 &  0.26\\ 
      &      & 0.35 & 0.40& 0.56 &  0.08& 1.31 &  0.13& 2.19 &  0.16& 2.69 &  0.17\\ 
      &      & 0.40 & 0.45& 0.53 &  0.07& 1.43 &  0.11& 2.21 &  0.13& 2.33 &  0.14\\ 
      &      & 0.45 & 0.50& 0.56 &  0.10& 1.29 &  0.09& 2.04 &  0.11& 2.32 &  0.16\\ 
      &      & 0.50 & 0.60& 0.45 &  0.07& 1.22 &  0.09& 2.07 &  0.13& 2.22 &  0.16\\ 
      &      & 0.60 & 0.70& 0.22 &  0.05& 0.84 &  0.12& 1.73 &  0.20& 2.03 &  0.21\\ 
      &      & 0.70 & 0.80& 0.12 &  0.03& 0.52 &  0.09& 1.09 &  0.19& 1.52 &  0.22\\ 
\hline \\ 
 0.55 & 0.75 & 0.10 & 0.15& 0.21 &  0.11& 0.66 &  0.24& 0.93 &  0.35& 0.91 &  0.41\\ 
      &      & 0.15 & 0.20& 0.47 &  0.11& 1.25 &  0.17& 1.81 &  0.21& 2.14 &  0.30\\ 
      &      & 0.20 & 0.25& 0.70 &  0.12& 1.23 &  0.13& 2.11 &  0.18& 3.03 &  0.24\\ 
      &      & 0.25 & 0.30& 0.88 &  0.12& 1.37 &  0.13& 2.26 &  0.15& 2.89 &  0.19\\ 
      &      & 0.30 & 0.35& 0.53 &  0.07& 1.12 &  0.09& 2.09 &  0.15& 2.62 &  0.16\\ 
      &      & 0.35 & 0.40& 0.55 &  0.07& 0.95 &  0.09& 2.09 &  0.14& 2.68 &  0.22\\ 
      &      & 0.40 & 0.45& 0.58 &  0.07& 1.03 &  0.10& 1.86 &  0.10& 2.69 &  0.15\\ 
      &      & 0.45 & 0.50& 0.54 &  0.08& 1.02 &  0.07& 1.63 &  0.09& 2.38 &  0.13\\ 
      &      & 0.50 & 0.60& 0.27 &  0.06& 0.74 &  0.08& 1.30 &  0.09& 1.78 &  0.16\\ 
      &      & 0.60 & 0.70& 0.16 &  0.03& 0.42 &  0.06& 0.89 &  0.11& 1.21 &  0.14\\ 
      &      & 0.70 & 0.80& 0.11 &  0.03& 0.28 &  0.06& 0.56 &  0.11& 0.80 &  0.14\\ 
\hline \\ 
 0.75 & 0.95 & 0.10 & 0.15& 0.44 &  0.13& 0.90 &  0.22& 1.21 &  0.31& 1.27 &  0.35\\ 
      &      & 0.15 & 0.20& 0.77 &  0.11& 1.55 &  0.14& 2.16 &  0.15& 2.54 &  0.21\\ 
      &      & 0.20 & 0.25& 0.86 &  0.10& 1.37 &  0.13& 2.27 &  0.14& 2.93 &  0.24\\ 
      &      & 0.25 & 0.30& 0.70 &  0.09& 1.22 &  0.13& 2.02 &  0.13& 2.75 &  0.18\\ 
      &      & 0.30 & 0.35& 0.54 &  0.07& 1.05 &  0.08& 1.85 &  0.12& 2.22 &  0.15\\ 
      &      & 0.35 & 0.40& 0.40 &  0.05& 0.96 &  0.07& 1.60 &  0.10& 2.09 &  0.13\\ 
      &      & 0.40 & 0.45& 0.41 &  0.06& 0.80 &  0.06& 1.39 &  0.08& 1.81 &  0.11\\ 
      &      & 0.45 & 0.50& 0.32 &  0.05& 0.67 &  0.05& 1.22 &  0.07& 1.52 &  0.09\\ 
      &      & 0.50 & 0.60& 0.18 &  0.03& 0.44 &  0.05& 0.83 &  0.08& 1.04 &  0.10\\ 
      &      & 0.60 & 0.70& 0.08 &  0.02& 0.25 &  0.04& 0.46 &  0.07& 0.63 &  0.08\\ 
\hline \\ 
 0.95 & 1.15 & 0.10 & 0.15& 0.73 &  0.18& 1.01 &  0.20& 1.47 &  0.26& 1.52 &  0.32\\ 
      &      & 0.15 & 0.20& 0.82 &  0.09& 1.64 &  0.12& 2.19 &  0.15& 2.60 &  0.21\\ 
      &      & 0.20 & 0.25& 0.71 &  0.10& 1.37 &  0.10& 1.97 &  0.12& 2.82 &  0.19\\ 
      &      & 0.25 & 0.30& 0.73 &  0.09& 1.10 &  0.10& 1.60 &  0.11& 2.13 &  0.16\\ 
      &      & 0.30 & 0.35& 0.41 &  0.07& 0.80 &  0.08& 1.31 &  0.10& 1.73 &  0.12\\ 
      &      & 0.35 & 0.40& 0.29 &  0.05& 0.66 &  0.06& 1.11 &  0.06& 1.64 &  0.11\\ 
      &      & 0.40 & 0.45& 0.29 &  0.05& 0.61 &  0.05& 0.88 &  0.05& 1.23 &  0.10\\ 
      &      & 0.45 & 0.50& 0.20 &  0.04& 0.48 &  0.05& 0.66 &  0.04& 0.86 &  0.08\\ 
      &      & 0.50 & 0.60& 0.12 &  0.02& 0.23 &  0.04& 0.46 &  0.04& 0.53 &  0.07\\ 
%%%%%%%%%%%%%%%%%%%%%%%%%%%%%%%%%%%%%%%%%%%%%%%%%%%%%%%%%%%%%%%%%%%%%%%%%%%%%
\hline
\end{tabular}
}
\end{center}
\end{table}

\begin{table}[hp!] 
\begin{center}
\small{
\begin{tabular}{rrrr|r@{$\pm$}lr@{$\pm$}lr@{$\pm$}lr@{$\pm$}l} 
\hline
$\theta_{\hbox{\small min}}$ &
$\theta_{\hbox{\small max}}$ &
$p_{\hbox{\small min}}$ &
$p_{\hbox{\small max}}$ &
\multicolumn{8}{c}{$d^2\sigma^{\pi^+}/(dpd\theta)$} 
\\
(rad) & (rad) & (\GeVc) & (\GeVc) &
\multicolumn{8}{c}{(barn/(\GeVc rad))}
\\
  &  &  & 
&\multicolumn{2}{c}{$ \bf{3 \ \GeVc}$} 
&\multicolumn{2}{c}{$ \bf{5 \ \GeVc}$} 
&\multicolumn{2}{c}{$ \bf{8 \ \GeVc}$} 
&\multicolumn{2}{c}{$ \bf{12 \ \GeVc}$} 
\\ 
\hline
%%%%%%%%%%%%%%%%%%%%%%%%%%%%%%%%%%%%%%%%%%%%%%%%%%%%%%%%%%%%%%%%%%%%%%%%%%%%%
 1.15 & 1.35 & 0.10 & 0.15& 0.79 &  0.18& 1.17 &  0.23& 1.82 &  0.30& 1.76 &  0.34\\ 
      &      & 0.15 & 0.20& 0.93 &  0.10& 1.51 &  0.11& 2.30 &  0.14& 2.53 &  0.22\\ 
      &      & 0.20 & 0.25& 0.65 &  0.08& 1.21 &  0.09& 1.86 &  0.11& 2.64 &  0.20\\ 
      &      & 0.25 & 0.30& 0.51 &  0.07& 0.91 &  0.08& 1.41 &  0.10& 1.84 &  0.17\\ 
      &      & 0.30 & 0.35& 0.33 &  0.05& 0.59 &  0.06& 0.94 &  0.08& 1.17 &  0.09\\ 
      &      & 0.35 & 0.40& 0.23 &  0.04& 0.38 &  0.04& 0.74 &  0.05& 0.93 &  0.07\\ 
      &      & 0.40 & 0.45& 0.17 &  0.03& 0.29 &  0.03& 0.55 &  0.04& 0.72 &  0.07\\ 
      &      & 0.45 & 0.50& 0.11 &  0.03& 0.21 &  0.03& 0.37 &  0.04& 0.45 &  0.06\\ 
\hline \\ 
 1.35 & 1.55 & 0.10 & 0.15& 0.85 &  0.17& 1.44 &  0.29& 1.84 &  0.39& 2.03 &  0.41\\ 
      &      & 0.15 & 0.20& 0.94 &  0.10& 1.41 &  0.12& 2.21 &  0.15& 2.50 &  0.19\\ 
      &      & 0.20 & 0.25& 0.71 &  0.08& 0.96 &  0.08& 1.57 &  0.11& 1.84 &  0.14\\ 
      &      & 0.25 & 0.30& 0.38 &  0.06& 0.64 &  0.06& 1.04 &  0.08& 1.48 &  0.12\\ 
      &      & 0.30 & 0.35& 0.22 &  0.04& 0.43 &  0.04& 0.74 &  0.05& 0.87 &  0.08\\ 
      &      & 0.35 & 0.40& 0.11 &  0.03& 0.35 &  0.04& 0.53 &  0.04& 0.63 &  0.06\\ 
      &      & 0.40 & 0.45& 0.04 &  0.01& 0.23 &  0.03& 0.38 &  0.03& 0.46 &  0.05\\ 
      &      & 0.45 & 0.50& 0.04 &  0.01& 0.15 &  0.02& 0.22 &  0.03& 0.28 &  0.04\\ 
\hline \\ 
 1.55 & 1.75 & 0.10 & 0.15& 0.66 &  0.15& 1.37 &  0.23& 1.54 &  0.30& 2.03 &  0.40\\ 
      &      & 0.15 & 0.20& 0.70 &  0.09& 1.40 &  0.11& 1.93 &  0.13& 2.27 &  0.18\\ 
      &      & 0.20 & 0.25& 0.53 &  0.07& 0.99 &  0.07& 1.34 &  0.09& 1.42 &  0.11\\ 
      &      & 0.25 & 0.30& 0.26 &  0.06& 0.58 &  0.06& 0.87 &  0.06& 0.95 &  0.08\\ 
      &      & 0.30 & 0.35& 0.13 &  0.03& 0.41 &  0.04& 0.53 &  0.04& 0.66 &  0.06\\ 
      &      & 0.35 & 0.40& 0.10 &  0.03& 0.26 &  0.03& 0.36 &  0.03& 0.47 &  0.05\\ 
      &      & 0.40 & 0.45& 0.06 &  0.02& 0.16 &  0.03& 0.25 &  0.02& 0.31 &  0.04\\ 
      &      & 0.45 & 0.50& 0.04 &  0.01& 0.09 &  0.02& 0.15 &  0.02& 0.17 &  0.03\\ 
\hline \\ 
 1.75 & 1.95 & 0.10 & 0.15& 0.64 &  0.14& 1.03 &  0.19& 1.31 &  0.20& 1.72 &  0.29\\ 
      &      & 0.15 & 0.20& 0.66 &  0.07& 1.14 &  0.09& 1.64 &  0.09& 1.85 &  0.13\\ 
      &      & 0.20 & 0.25& 0.36 &  0.05& 0.74 &  0.06& 1.02 &  0.07& 1.10 &  0.10\\ 
      &      & 0.25 & 0.30& 0.18 &  0.04& 0.34 &  0.05& 0.63 &  0.05& 0.56 &  0.06\\ 
      &      & 0.30 & 0.35& 0.08 &  0.02& 0.17 &  0.03& 0.34 &  0.04& 0.36 &  0.05\\ 
      &      & 0.35 & 0.40& 0.07 &  0.02& 0.10 &  0.02& 0.22 &  0.03& 0.18 &  0.03\\ 
      &      & 0.40 & 0.45& 0.05 &  0.02& 0.06 &  0.01& 0.14 &  0.02& 0.10 &  0.02\\ 
      &      & 0.45 & 0.50& 0.02 &  0.01& 0.03 &  0.01& 0.07 &  0.02& 0.07 &  0.01\\ 
\hline \\ 
 1.95 & 2.15 & 0.10 & 0.15& 0.79 &  0.15& 0.83 &  0.14& 1.14 &  0.18& 1.46 &  0.23\\ 
      &      & 0.15 & 0.20& 0.65 &  0.09& 0.89 &  0.08& 1.17 &  0.06& 1.36 &  0.11\\ 
      &      & 0.20 & 0.25& 0.36 &  0.07& 0.51 &  0.05& 0.75 &  0.05& 1.04 &  0.10\\ 
      &      & 0.25 & 0.30& 0.15 &  0.05& 0.28 &  0.05& 0.45 &  0.05& 0.40 &  0.07\\ 
      &      & 0.30 & 0.35& 0.05 &  0.02& 0.14 &  0.03& 0.20 &  0.03& 0.19 &  0.03\\ 
      &      & 0.35 & 0.40& 0.03 &  0.02& 0.09 &  0.02& 0.09 &  0.01& 0.13 &  0.02\\ 
      &      & 0.40 & 0.45& 0.02 &  0.01& 0.04 &  0.01& 0.06 &  0.01& 0.09 &  0.02\\ 
      &      & 0.45 & 0.50& 0.01 &  0.01& 0.02 &  0.01& 0.04 &  0.01& 0.07 &  0.02\\ 
%%%%%%%%%%%%%%%%%%%%%%%%%%%%%%%%%%%%%%%%%%%%%%%%%%%%%%%%%%%%%%%%%%%%%%%%%%%%%
\hline
\end{tabular}
}
\end{center}
\end{table}

\begin{table}[hp!] 
\begin{center}
  \caption{\label{tab:xsec-n-pb}
    HARP results for the double-differential $\pi^-$ production
    cross-section in the laboratory system,
    $d^2\sigma^{\pi^-}/(dpd\theta)$ for lead. Each row refers to a
    different $(p_{\hbox{\small min}} \le p<p_{\hbox{\small max}},
    \theta_{\hbox{\small min}} \le \theta<\theta_{\hbox{\small max}})$ bin,
    where $p$ and $\theta$ are the pion momentum and polar angle, respectively.
    The central value as well as the square-root of the diagonal elements
    of the covariance matrix are given.
 The overall normalization has an uncertainty
 of 3\%, and is not reported in the table.
}
\vspace{2mm}
\small{
\begin{tabular}{rrrr|r@{$\pm$}lr@{$\pm$}lr@{$\pm$}lr@{$\pm$}l} 
\hline
$\theta_{\hbox{\small min}}$ &
$\theta_{\hbox{\small max}}$ &
$p_{\hbox{\small min}}$ &
$p_{\hbox{\small max}}$ &
\multicolumn{8}{c}{$d^2\sigma^{\pi^-}/(dpd\theta)$} 
\\
(rad) & (rad) & (\GeVc) & (\GeVc) &
\multicolumn{8}{c}{(barn/(\GeVc rad))}
\\
  &  &  & 
&\multicolumn{2}{c}{$ \bf{3 \ \GeVc}$} 
&\multicolumn{2}{c}{$ \bf{5 \ \GeVc}$} 
&\multicolumn{2}{c}{$ \bf{8 \ \GeVc}$} 
&\multicolumn{2}{c}{$ \bf{12 \ \GeVc}$} 
\\ 
\hline
%%%%%%%%%%%%%%%%%%%%%%%%%%%%%%%%%%%%%%%%%%%%%%%%%%%%%%%%%%%%%%%%%%%%%%%%%%%%%
 0.35 & 0.55 & 0.15 & 0.20& 0.17 &  0.09& 0.58 &  0.22& 1.59 &  0.36& 1.92 &  0.47\\ 
      &      & 0.20 & 0.25& 0.33 &  0.11& 0.89 &  0.18& 2.00 &  0.27& 2.66 &  0.31\\ 
      &      & 0.25 & 0.30& 0.25 &  0.06& 0.98 &  0.12& 2.07 &  0.14& 2.54 &  0.24\\ 
      &      & 0.30 & 0.35& 0.49 &  0.11& 0.96 &  0.11& 1.99 &  0.13& 2.30 &  0.16\\ 
      &      & 0.35 & 0.40& 0.37 &  0.08& 0.88 &  0.09& 1.63 &  0.09& 2.09 &  0.12\\ 
      &      & 0.40 & 0.45& 0.30 &  0.06& 0.92 &  0.07& 1.63 &  0.08& 2.04 &  0.13\\ 
      &      & 0.45 & 0.50& 0.31 &  0.06& 0.79 &  0.06& 1.48 &  0.07& 1.89 &  0.13\\ 
      &      & 0.50 & 0.60& 0.34 &  0.06& 0.78 &  0.06& 1.38 &  0.08& 1.50 &  0.14\\ 
      &      & 0.60 & 0.70& 0.21 &  0.05& 0.64 &  0.07& 1.37 &  0.10& 1.32 &  0.12\\ 
      &      & 0.70 & 0.80& 0.11 &  0.04& 0.45 &  0.08& 1.06 &  0.15& 1.46 &  0.15\\ 
\hline \\ 
 0.55 & 0.75 & 0.10 & 0.15& 0.34 &  0.18& 1.03 &  0.37& 1.24 &  0.44& 1.70 &  0.49\\ 
      &      & 0.15 & 0.20& 0.40 &  0.10& 1.30 &  0.17& 2.09 &  0.22& 2.30 &  0.29\\ 
      &      & 0.20 & 0.25& 0.43 &  0.09& 1.21 &  0.12& 2.18 &  0.16& 2.45 &  0.17\\ 
      &      & 0.25 & 0.30& 0.54 &  0.09& 1.29 &  0.11& 2.19 &  0.14& 2.20 &  0.19\\ 
      &      & 0.30 & 0.35& 0.53 &  0.09& 1.01 &  0.08& 1.88 &  0.11& 2.29 &  0.14\\ 
      &      & 0.35 & 0.40& 0.40 &  0.08& 0.93 &  0.08& 1.51 &  0.08& 1.91 &  0.13\\ 
      &      & 0.40 & 0.45& 0.31 &  0.05& 0.89 &  0.07& 1.38 &  0.07& 1.79 &  0.11\\ 
      &      & 0.45 & 0.50& 0.37 &  0.06& 0.70 &  0.06& 1.26 &  0.07& 1.59 &  0.10\\ 
      &      & 0.50 & 0.60& 0.27 &  0.05& 0.62 &  0.05& 1.20 &  0.06& 1.35 &  0.09\\ 
      &      & 0.60 & 0.70& 0.16 &  0.04& 0.55 &  0.05& 0.95 &  0.09& 1.19 &  0.11\\ 
      &      & 0.70 & 0.80& 0.09 &  0.03& 0.41 &  0.07& 0.70 &  0.09& 0.95 &  0.12\\ 
\hline \\ 
 0.75 & 0.95 & 0.10 & 0.15& 0.52 &  0.16& 1.12 &  0.26& 1.71 &  0.38& 1.95 &  0.43\\ 
      &      & 0.15 & 0.20& 0.73 &  0.10& 1.47 &  0.15& 2.28 &  0.17& 2.74 &  0.23\\ 
      &      & 0.20 & 0.25& 0.61 &  0.10& 1.17 &  0.10& 2.04 &  0.12& 2.78 &  0.17\\ 
      &      & 0.25 & 0.30& 0.60 &  0.09& 1.04 &  0.09& 1.88 &  0.11& 2.25 &  0.16\\ 
      &      & 0.30 & 0.35& 0.45 &  0.07& 0.94 &  0.07& 1.54 &  0.08& 2.29 &  0.14\\ 
      &      & 0.35 & 0.40& 0.29 &  0.05& 0.76 &  0.07& 1.26 &  0.06& 1.67 &  0.12\\ 
      &      & 0.40 & 0.45& 0.19 &  0.04& 0.53 &  0.05& 1.11 &  0.06& 1.43 &  0.09\\ 
      &      & 0.45 & 0.50& 0.18 &  0.04& 0.44 &  0.04& 0.95 &  0.05& 1.25 &  0.08\\ 
      &      & 0.50 & 0.60& 0.19 &  0.04& 0.45 &  0.04& 0.78 &  0.04& 0.99 &  0.08\\ 
      &      & 0.60 & 0.70& 0.10 &  0.03& 0.36 &  0.04& 0.61 &  0.06& 0.67 &  0.09\\ 
\hline \\ 
 0.95 & 1.15 & 0.10 & 0.15& 0.44 &  0.10& 1.28 &  0.24& 2.17 &  0.36& 2.58 &  0.45\\ 
      &      & 0.15 & 0.20& 0.77 &  0.11& 1.53 &  0.13& 2.30 &  0.16& 2.93 &  0.19\\ 
      &      & 0.20 & 0.25& 0.47 &  0.07& 1.13 &  0.08& 1.85 &  0.11& 2.59 &  0.19\\ 
      &      & 0.25 & 0.30& 0.40 &  0.06& 0.99 &  0.08& 1.50 &  0.10& 2.19 &  0.15\\ 
      &      & 0.30 & 0.35& 0.39 &  0.06& 0.85 &  0.07& 1.15 &  0.07& 1.77 &  0.12\\ 
      &      & 0.35 & 0.40& 0.30 &  0.05& 0.61 &  0.06& 0.99 &  0.06& 1.44 &  0.09\\ 
      &      & 0.40 & 0.45& 0.28 &  0.05& 0.42 &  0.04& 0.80 &  0.05& 1.11 &  0.08\\ 
      &      & 0.45 & 0.50& 0.18 &  0.04& 0.34 &  0.03& 0.69 &  0.04& 0.82 &  0.06\\ 
      &      & 0.50 & 0.60& 0.08 &  0.03& 0.26 &  0.03& 0.52 &  0.04& 0.65 &  0.05\\ 
%%%%%%%%%%%%%%%%%%%%%%%%%%%%%%%%%%%%%%%%%%%%%%%%%%%%%%%%%%%%%%%%%%%%%%%%%%%%%
\hline
\end{tabular}
}
\end{center}
\end{table}

\begin{table}[hp!] 
\begin{center}
\small{
\begin{tabular}{rrrr|r@{$\pm$}lr@{$\pm$}lr@{$\pm$}lr@{$\pm$}l} 
\hline
$\theta_{\hbox{\small min}}$ &
$\theta_{\hbox{\small max}}$ &
$p_{\hbox{\small min}}$ &
$p_{\hbox{\small max}}$ &
\multicolumn{8}{c}{$d^2\sigma^{\pi^-}/(dpd\theta)$} 
\\
(rad) & (rad) & (\GeVc) & (\GeVc) &
\multicolumn{8}{c}{(barn/(\GeVc rad))}
\\
  &  &  & 
&\multicolumn{2}{c}{$ \bf{3 \ \GeVc}$} 
&\multicolumn{2}{c}{$ \bf{5 \ \GeVc}$} 
&\multicolumn{2}{c}{$ \bf{8 \ \GeVc}$} 
&\multicolumn{2}{c}{$ \bf{12 \ \GeVc}$} 
\\ 
\hline
%%%%%%%%%%%%%%%%%%%%%%%%%%%%%%%%%%%%%%%%%%%%%%%%%%%%%%%%%%%%%%%%%%%%%%%%%%%%%
 1.15 & 1.35 & 0.10 & 0.15& 0.43 &  0.10& 1.45 &  0.26& 2.40 &  0.37& 3.22 &  0.57\\ 
      &      & 0.15 & 0.20& 0.65 &  0.09& 1.56 &  0.10& 2.19 &  0.14& 3.14 &  0.21\\ 
      &      & 0.20 & 0.25& 0.37 &  0.06& 0.99 &  0.08& 1.70 &  0.11& 2.32 &  0.16\\ 
      &      & 0.25 & 0.30& 0.25 &  0.04& 0.83 &  0.07& 1.23 &  0.08& 1.74 &  0.14\\ 
      &      & 0.30 & 0.35& 0.16 &  0.03& 0.63 &  0.06& 0.92 &  0.07& 1.16 &  0.09\\ 
      &      & 0.35 & 0.40& 0.13 &  0.03& 0.42 &  0.04& 0.75 &  0.05& 0.90 &  0.06\\ 
      &      & 0.40 & 0.45& 0.11 &  0.03& 0.35 &  0.03& 0.59 &  0.04& 0.71 &  0.06\\ 
      &      & 0.45 & 0.50& 0.09 &  0.02& 0.26 &  0.03& 0.46 &  0.03& 0.52 &  0.05\\ 
\hline \\ 
 1.35 & 1.55 & 0.10 & 0.15& 0.60 &  0.13& 1.39 &  0.26& 2.34 &  0.45& 3.46 &  0.64\\ 
      &      & 0.15 & 0.20& 0.70 &  0.08& 1.29 &  0.11& 2.06 &  0.13& 2.89 &  0.20\\ 
      &      & 0.20 & 0.25& 0.41 &  0.06& 0.84 &  0.08& 1.60 &  0.10& 1.78 &  0.13\\ 
      &      & 0.25 & 0.30& 0.34 &  0.06& 0.55 &  0.06& 1.04 &  0.07& 1.28 &  0.09\\ 
      &      & 0.30 & 0.35& 0.24 &  0.05& 0.41 &  0.04& 0.69 &  0.05& 0.89 &  0.08\\ 
      &      & 0.35 & 0.40& 0.16 &  0.03& 0.32 &  0.03& 0.52 &  0.03& 0.66 &  0.05\\ 
      &      & 0.40 & 0.45& 0.08 &  0.02& 0.27 &  0.03& 0.41 &  0.03& 0.47 &  0.05\\ 
      &      & 0.45 & 0.50& 0.06 &  0.02& 0.20 &  0.03& 0.29 &  0.03& 0.31 &  0.04\\ 
\hline \\ 
 1.55 & 1.75 & 0.10 & 0.15& 0.73 &  0.16& 1.17 &  0.25& 2.09 &  0.40& 3.05 &  0.58\\ 
      &      & 0.15 & 0.20& 0.65 &  0.08& 1.20 &  0.10& 1.76 &  0.12& 2.45 &  0.16\\ 
      &      & 0.20 & 0.25& 0.39 &  0.06& 0.73 &  0.06& 1.29 &  0.08& 1.30 &  0.11\\ 
      &      & 0.25 & 0.30& 0.22 &  0.05& 0.43 &  0.05& 0.82 &  0.07& 0.86 &  0.07\\ 
      &      & 0.30 & 0.35& 0.16 &  0.04& 0.34 &  0.04& 0.48 &  0.05& 0.69 &  0.06\\ 
      &      & 0.35 & 0.40& 0.10 &  0.03& 0.26 &  0.03& 0.35 &  0.03& 0.49 &  0.04\\ 
      &      & 0.40 & 0.45& 0.07 &  0.02& 0.16 &  0.02& 0.26 &  0.02& 0.38 &  0.04\\ 
      &      & 0.45 & 0.50& 0.04 &  0.02& 0.10 &  0.02& 0.19 &  0.02& 0.26 &  0.03\\ 
\hline \\ 
 1.75 & 1.95 & 0.10 & 0.15& 0.72 &  0.15& 1.13 &  0.20& 1.78 &  0.25& 2.36 &  0.44\\ 
      &      & 0.15 & 0.20& 0.52 &  0.07& 1.09 &  0.09& 1.44 &  0.07& 1.84 &  0.12\\ 
      &      & 0.20 & 0.25& 0.32 &  0.06& 0.65 &  0.06& 0.92 &  0.06& 1.10 &  0.10\\ 
      &      & 0.25 & 0.30& 0.11 &  0.03& 0.36 &  0.04& 0.56 &  0.05& 0.58 &  0.05\\ 
      &      & 0.30 & 0.35& 0.11 &  0.03& 0.25 &  0.03& 0.30 &  0.03& 0.41 &  0.05\\ 
      &      & 0.35 & 0.40& 0.09 &  0.03& 0.15 &  0.03& 0.23 &  0.02& 0.31 &  0.03\\ 
      &      & 0.40 & 0.45& 0.07 &  0.03& 0.11 &  0.02& 0.20 &  0.02& 0.30 &  0.04\\ 
      &      & 0.45 & 0.50& 0.04 &  0.02& 0.09 &  0.02& 0.14 &  0.02& 0.21 &  0.04\\ 
\hline \\ 
 1.95 & 2.15 & 0.10 & 0.15& 0.69 &  0.14& 1.08 &  0.20& 1.52 &  0.20& 1.84 &  0.31\\ 
      &      & 0.15 & 0.20& 0.43 &  0.07& 0.84 &  0.07& 1.11 &  0.06& 1.35 &  0.10\\ 
      &      & 0.20 & 0.25& 0.23 &  0.05& 0.40 &  0.06& 0.68 &  0.05& 0.90 &  0.10\\ 
      &      & 0.25 & 0.30& 0.08 &  0.03& 0.21 &  0.03& 0.42 &  0.04& 0.51 &  0.06\\ 
      &      & 0.30 & 0.35& 0.05 &  0.02& 0.11 &  0.03& 0.24 &  0.03& 0.34 &  0.06\\ 
      &      & 0.35 & 0.40& 0.03 &  0.02& 0.07 &  0.01& 0.17 &  0.02& 0.19 &  0.03\\ 
      &      & 0.40 & 0.45& 0.02 &  0.02& 0.08 &  0.02& 0.12 &  0.02& 0.14 &  0.03\\ 
      &      & 0.45 & 0.50& 0.02 &  0.02& 0.08 &  0.02& 0.08 &  0.01& 0.09 &  0.02\\ 
%%%%%%%%%%%%%%%%%%%%%%%%%%%%%%%%%%%%%%%%%%%%%%%%%%%%%%%%%%%%%%%%%%%%%%%%%%%%%
\hline
\end{tabular}
}
\end{center}
\end{table}
%

%%%%%%%%%%%%%%%%%%%%%%%%%%%%%%%%%%

\end{appendix}

\clearpage

%\section*{References}

%%%%%%%%%%%

\end{document}